\documentclass[preprint]{aastex631}
\usepackage{makecell}
\usepackage{lineno}

\shorttitle{Calibration of PAH as a SFR Indicator in AKARI NEP}
\shortauthors{Kim et al.}

\graphicspath{{./}{figures/}}

\begin{document}

\title{The Calibration of Polycyclic Aromatic Hydrocarbon Dust Emission as a Star Formation Rate Indicator in the AKARI NEP Survey}

\correspondingauthor{Helen Kyung Kim}
\email{helenkkim@ucla.edu}

\author[0000-0003-2352-8636]{Helen Kyung Kim}
\affiliation{Department of Physics and Astronomy \\
University of California, Los Angeles \\
475 Portola Plaza, Los Angeles, CA 90095}

\author[0000-0001-6919-1237]{Matthew A. Malkan}
\affiliation{Department of Physics and Astronomy \\
University of California, Los Angeles \\
475 Portola Plaza, Los Angeles, CA 90095}

\author{Toshinobu Takagi}
\affiliation{Institute of Space and Astronautical Science, JAXA, \\
Sagamihara, Kanagawa 252-5210, Japan}

\author[0000-0002-4686-4985]{Nagisa Oi}
\affiliation{Faculty of Science Division II, Tokyo University of Science, \\
Kagurazaka, Shinjuku-ku, Tokyo 162-8601, Japan
}

\author[0000-0002-4193-2539]{Denis Burgarella}
\affiliation{Aix Marseille Univ, CNRS, CNES, LAM, Marseille, France}

\author[0000-0002-7562-485X]{Takamitsu Miyaji}
\affiliation{Instituto de Astronom\'ia sede Ensenada \\
Universidad Nacional Aut\'onoma de M\'exico \\
Km 107, Carret. Tij.-Ens., Ensenada, 22060, BC, Mexico
}

\author[0000-0002-4179-2628]{Hyunjin Shim}
\affiliation{Department of Earth Science Education, Kyungpook National University \\ Daegu 41566, Republic of Korea}

\author{Hideo Matsuhara}
\affiliation{Institute of Space and Astronautical Science, JAXA, \\
Sagamihara, Kanagawa 252-5210, Japan}

\author[0000-0002-6821-8669]{Tomotsugu Goto}
\affiliation{Institute of Astronomy, National Tsing Hua University \\
101, Section 2. Kuang-Fu Road, Hsinchu, 30013, Taiwan (R.O.C.)}

\author[0000-0001-9490-3582]{Yoichi Ohyama}
\affiliation{Institute of Astronomy and Astrophysics, Academia Sinica, \\
11F of Astronomy-Mathematics Building, No.1, Sec. 4, \\
Roosevelt Road,
Taipei 10617, Taiwan (R.O.C.)
}

\author[0000-0003-3441-903X]{Veronique Buat}
\affiliation{Aix Marseille Univ, CNRS, CNES, LAM, Marseille, France}

\author[0000-0001-9970-8145]{Seong Jin Kim}
\affiliation{Institute of Astronomy, National Tsing Hua University \\
101, Section 2. Kuang-Fu Road, Hsinchu, 30013, Taiwan (R.O.C.)}

\begin{abstract}

Polycyclic aromatic hydrocarbon (PAH) dust emission has been proposed as an effective extinction-independent star formation rate (SFR) indicator in the mid-infrared (MIR), but this may depend on conditions in the interstellar medium. The coverage of the \textit{AKARI}/Infrared Camera (IRC) allows us to study the effects of metallicity, starburst intensity, and active galactic nuclei on PAH emission in galaxies with $f_{\nu}(L18W)\lesssim 19$ AB mag. Observations include follow-up, rest-frame optical spectra of 443 galaxies within the \textit{AKARI} North Ecliptic Pole survey that have IRC detections from 7-24 $\mu$m. We use optical emission line diagnostics to infer SFR based on H$\alpha$ and [O II]$\lambda\lambda 3726,3729$ emission line luminosities. The PAH 6.2 $\mu$m and PAH 7.7 $\mu$m luminosities ($L(PAH\ 6.2\ \mu m)$ and $L(PAH\ 7.7\ \mu m)$, respectively) derived using multi-wavelength model fits are consistent with those derived from slitless spectroscopy within 0.2 dex. $L(PAH\ 6.2\ \mu m)$ and $L(PAH\ 7.7\ \mu m)$ correlate linearly with the 24 $\mu$m-dust corrected H$\alpha$ luminosity only for normal, star-forming ``main-sequence" galaxies. Assuming multi-linear correlations, we quantify the additional dependencies on metallicity and starburst intensity, which we use to correct our PAH SFR calibrations at $0<z<1.2$ for the first time. We derive the cosmic star formation rate density (SFRD) per comoving volume from $0.15 \lesssim z \lesssim 1$. The PAH SFRD is consistent with that of the far-infrared and reaches an order of magnitude higher than that of uncorrected UV observations at $z\sim1$. Starburst galaxies contribute $\gtrsim 0.7$ of the total SFRD at $z\sim1$ compared to main-sequence galaxies.

\end{abstract}

\keywords{Star formation (1569), Polycyclic aromatic hydrocarbons (1280), Interstellar dust (836), Infrared galaxies (790), Infrared photometry (792), Spectral energy distribution (2129), Spectroscopy (1558), Photometry (1234), Galaxy evolution (594), Starburst galaxies (1570), Active galaxies (17)}

\section{Introduction} \label{sec:intro}

The star formation rate (SFR) is one of the key parameters that describes a galaxy's growth and evolution, tracing the conversion of baryonic matter in the interstellar medium (ISM) into stellar radiation. The calibration of various SFR indicators across the electromagnetic spectrum has been extensively studied for the past three decades \citep{KennicuttEvans, Calzetti2013}. The Balmer recombination lines, most notably H$\alpha$ $\lambda$6563, and the rest-frame UV continuum emission from massive stars have been established as standard indicators at $z=0$ and $z\gtrsim1$, respectively. However, the SFRs derived from rest-frame optical/UV lines are prone to large uncertainties in the presence of interstellar dust, which absorbs stellar photons and re-radiates in the infrared. The total infrared luminosity, which measures the integrated dust continuum ($\sim8-1000$ $\mu$m) heated by O, B, A, and F-type stars, is also commonly used to trace star formation, but may overestimate the SFR in post-starburst galaxies \citep{Hayward2014}. Therefore, multi-wavelength data are crucial for providing a ``big picture" view of the energy generation in galaxies and for overcoming the limitations of relying on a single SFR indicator.

\begin{figure*}[ht!]
	\centering
	\includegraphics[scale=0.6]{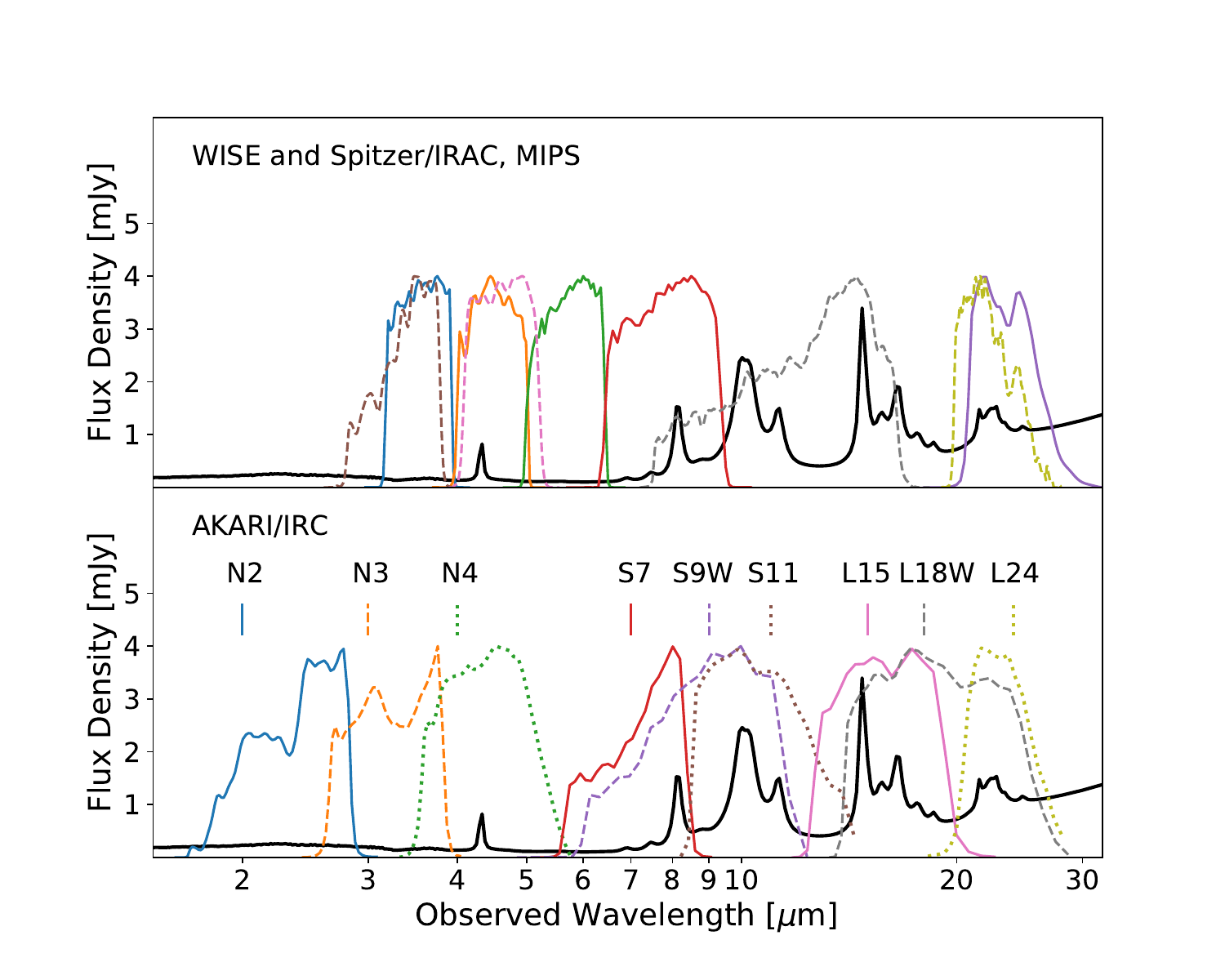}
	\vspace*{-5mm}
	\caption{Filter transmission curves from \textit{WISE} and \textit{Spitzer}/IRAC and MIPS (top) and \textit{AKARI}/IRC (bottom) superimposed on a modeled mid-IR spectrum of a star-forming galaxy at $z=0.308$.
	\label{filter_comparison}}
\end{figure*}

The brightness of the cosmic infrared background suggests at least half the luminous energy generated by stars has been reprocessed into the IR by dust \citep{Lagache, 2016ApJ...827....6S}, and that dust-obscured star formation was more important at higher redshifts than today \citep{Buat2007}. At $z$$\sim$1, infrared luminous galaxies ($L(8-1000\ \mu m) \geq 10^{11}\ L_{\odot}$) contribute $\sim$70\% of the cosmic infrared luminosity density \citep{2005ApJ...632..169L}, but the individual dusty galaxies producing this IR light are not well understood, partly due to the requirement of deep wide-field mid-IR surveys to find them, and then the extreme difficulty of studying their dust spectra. The only mission able to accomplish these tasks was the \textit{AKARI} space telescope with its Infrared Camera (IRC), which devoted a substantial fraction of its entire lifetime to intensive mapping of the 0.5 deg$^2$ area of North Ecliptic Pole (NEP) Deep field (orange scalloped circle in Figure 1 of \citealp{Takeuchi,Murata2013}). \textit{AKARI}-NEP's $>$6000 faint galaxy mid-IR 9-point spectra are a 1.5 order-of-magnitude increase over all other samples (e.g., \textit{Spitzer}/IRS) combined. The NEP-Deep field is in a prime location that has been observed very intensively at all other wavelengths from the X-ray and UV to radio wavelengths (e.g., \citealp{Matsuhara2006, SJKim2021}). Figure \ref{filter_comparison} highlights the continuous mid-infrared coverage of \textit{AKARI}/IRC photometry from 2-24 $\mu$m, compared to \textit{WISE}, \textit{Spitzer}/IRAC and MIPS. The superimposed spectrum is that of a modeled star-forming galaxy at the median redshift of our spectroscopic sample ($z=0.308$).

The luminosities of the 3.3, 6.2, 7.7, 8.6, and 11.3 $\mu$m PAH emission features have been proposed as extinction-independent SFR indicators \citep{Peeters, Brandl, Houck, Castro2014, Lai2020}. The wavelength coverage from the IRC-derived spectral energy distributions (SEDs) is uniquely suited to study the strongest of these: the PAH 7.7 $\mu$m blend composed of emission features at 7.42, 7.60, and 7.85 $\mu$m, which can contribute $\sim42$\% of the total PAH luminosity in star-forming galaxies \citep{Shipley2013}. However, studies of ultra-luminous infrared galaxies (ULIRGs: L(8-1000 $\mu$m) $\geq$ $10^{12}$ L$_{\odot}$) at $z<0.9$ targeted by \textit{Spitzer}/IRS have diverse mid-infrared spectra, with the majority of ULIRGs having depressed PAH 6.2 and 7.7 $\mu$m equivalent widths, which is likely due to UV photons from buried AGN or compact star formation destroying small dust grains \citep{Spoon, Imanishi, Desai}. But on the other hand, at $z\sim 1$, some ULIRGs have been observed with ``enhanced" peak PAH 7.7 $\mu$m luminosities per total infrared luminosity relative to local ULIRGs \citep{Takagi2010}.

The strength of PAH emission is also believed to depend on the gas-phase metallicity of the interstellar medium (ISM). \citet{Engelbracht} found decreased 8 $\mu$m-to-24 $\mu$m flux density ratios in local star-forming galaxies with 12+log(O/H) $\lesssim$ 8.2, possibly due to decreased PAH contribution to the 8 $\mu$m bandpass. \citet{Draine07} determined that the PAH index, defined as the fraction of the total dust mass in PAH grains with fewer than $10^{3}$ carbon atoms ($q_{PAH}$), is lower for galaxies with 12+log(O/H) $\leq$ 8.1 observed in the Spitzer Nearby Galaxy Survey (SINGS). Some possible physical explanations for reduced PAH strength include: UV dissociation of PAH molecules due to less dust shielding, increased destruction within shock-heated gas, and decreased efficiency in PAH production from carbon stars and planetary nebulae \citep{Draine07}. 

Given the effects of metallicity and radiation from AGN and starburst galaxies, observations of rest-frame UV/optical emission lines (e.g., ([O II], H$\beta$, [O III], H$\alpha$, [N II]) in mid-IR-detected sources are crucial for calibrating PAH dust emission as a star formation rate indicator. Emission line diagnostics provide the tools to classify sources, measure gas-phase metallicity and dust extinction, and measure accurate spectroscopic redshifts for SEDs.

In this work, we present SFR calibrations based on PAH luminosity, corrected for metallicity and starburst intensity. In Section \ref{sec:obs}, we describe our spectroscopic and photometric observations, data reduction processes, emission line measurements, galaxy classification, and SED fitting procedure. In section \ref{sec:pahmeasurement}, we present our methodology for measuring PAH luminosities. We confirm their validity by direct comparison with mid-infrared spectra of the same galaxies. In Section \ref{sec:analysis}, we discuss the effects of starburst intensity, metallicity, and AGN strength on PAH luminosity; we derive SFR prediction equations as a function of PAH luminosity calibrated against dust-corrected H$\alpha$ and [O II]$\lambda\lambda3726,3729$ line luminosities. In Section \ref{sec:sfrd}, we apply our SFR equations to estimate the cosmic star formation rate density (SFRD) from $0\lesssim z\lesssim1.2$ and compare our results to UV and FIR-based calibrations. In Section \ref{sec:disc} we discuss our PAH SFR results in the context of other studies, and how our calibrations may be used to analyze future \textit{JWST} observations. Lastly, in Section \ref{sec:summary}, we provide a summary of conclusions.

\section{Follow-up and Supporting Observations} \label{sec:obs}
To measure key physical properties of \textit{AKARI}/IRC-detected sources, we obtained follow-up optical and near-infrared spectra in the \textit{AKARI} NEP-Deep and NEP-Wide fields. The main goals of our spectroscopic analysis include: measuring reliable spectroscopic redshifts, classifying sources as star-forming galaxies or AGN, measuring metallicity, measuring dust extinction with the Balmer recombination lines, and calibrating the dust-corrected H$\alpha$ and [O II]$\lambda\lambda3726,3729$ SFR against the PAH luminosity. In this section, we present our multi-object spectroscopy from \textit{Keck II}/DEIMOS and \textit{Keck I}/MOSFIRE spectrographs, as well as supplementary spectra taken by the \textit{AKARI} collaboration from \textit{MMT}/Hectospec, \textit{WIYN}/Hydra, and \textit{Subaru}/FMOS.

\subsection{Keck II/DEIMOS spectroscopic sample}
From 2008-2016, we observed 19 slitmasks with \textit{Keck II}/DEep Imaging Multi-Object Spectrograph (DEIMOS; \citealp{Faber}) targeting the \textit{AKARI} NEP-Deep field. When selecting targets for each slit in the later years, we prioritized mid-infrared \textit{AKARI}/IRC sources with detected flux measurements at 9 and 18 $\mu$m, and assigned secondary priorities to high-redshift galaxies (photometric redshift $z_{phot} \sim1$; \citealp{Oi2014}) and optical counterparts of \textit{Chandra} X-ray sources \citep{Krumpe2015}. The spectrograph was configured with the 600ZD grating and GG495 order blocking filter (central wavelength $\lambda_{c} = 7500$ {\AA}) for the two slitmasks observed on August 24, 2014, and  the 900ZD grating and GG455 order blocking filter ($\lambda_{c} = 6800$ {\AA}) for the remaining four slitmasks observed in 2015-2016. A typical slitmask contained $\sim 80$ slits that were 1${''}$ in width. The spectral range covered approximately 4850 {\AA} $\lesssim\lambda\lesssim10050$ {\AA}, depending on the location of the slit. The spectral resolution was $R\sim2000$, which corresponds to full-width at half-maximum (FWHM) $\Delta \lambda\sim 3.75$ {\AA} at the central wavelength $\lambda_{c}=7500$ {\AA}, as confirmed by the observed widths of night-sky lines. Prior to observing each night, we acquired one set of Ne, Ar, Kr, and Xe arc lamp spectra and three flat field exposures of the detectors illuminated by an internal incandescent lamp. In addition, we observed standard star Feige 15 (RA $01^{h}49^{m}09.5^{s}$, Dec $+13^{\circ}33^{'}11.4^{''}$) in long-slit mode for flux calibration. We reduced the raw data from 2014-2016 using the UCB DEEP2 spec2d reduction pipeline \citep{Cooper2012, Newman2013}, which processed the flatfields and calibration arcs, fit a polynomial wavelength solution to the pixel locations of arc lines, subtracted night-sky emission, and extracted one- and two-dimensional spectra. The extended flux along the slit was included, but most spectra were unresolved in the spatial direction.

The \textit{AKARI} NEP collaboration observed eight DEIMOS slitmasks in July 2011, with the focus of targeting galaxy cluster candidates with \textit{AKARI}/IRC MIR detections in the NEP-Deep field (PI: Tomotsugu Goto; \citealp{Shogaki}). 553 targets were selected based on \textit{Subaru}/Suprime-Cam \citep{Imai2007, Wada2008} R-band and Z-band brightness cuts such that $Z < 23$ mag and $R > 21.5$ mag. Slitmasks contained $\sim50-80$ slits of 1${''}$ width. The instrument configuration consisted of the 600ZD grating and GG455 order blocking filter, yielding a spectral resolution of $R\sim2000$. A set of calibration images consisting of a Ne, Ar, Kr, Xe arc lamp image and five flat field lamp images was acquired for each slitmask. Bright spectrophotometric standard stars BD+$33^{\circ}2642$ and Feige 110 were observed for 90 seconds each in long-slit mode. Weather conditions were clear on both nights, with 0.6$''$ seeing. The data were reduced using the spec2d pipeline described above \citep{Shogaki}.

We then flux-calibrated the one-dimensional object spectra from 2011-2016 using standard IRAF routines as follows. The raw frames were overscan-trimmed and bias-subtracted using the \texttt{mscred.ccdproc} routine. A normalized flat field was created using \texttt{mscred.flatcombine}, \texttt{noao.twodspec.longslit.response}, and texttt{mscred.mscjoin}, and the standard star spectra were flat-fielded with \texttt{mscred.ccdproc}. A wavelength-calibrated standard star image was produced using \texttt{noao.twodspec.longslit.identify}, \texttt{reidentify}, \texttt{fitcoords}, and \texttt{transform} on the arc images to determine the wavelength solution and \texttt{noao.imred.kpnoslit.doslit} on the standard star. Then, the sensitivity function was derived using \texttt{noao.twodspec.longslit.standard}, \texttt{sensfunc}, and \texttt{calibrate} on the standard star. Lastly, the sensitivity function was input into the routine \texttt{noao.onedspec.calibrate} to apply extinction corrections and flux calibrations to the reduced one-dimensional object spectra. 

To estimate the photometric accuracy of the flux-calibrated spectra, we calculated the average flux density of the spectra within the \textit{CFHT}/MegaCam r-band filter (5690-6930 {\AA}):
\begin{equation}
    f_{\nu}\ [erg/s/cm^2/Hz] = \frac{1}{c}\frac{\int FT\lambda \,d\lambda}{\int T\lambda \, d\lambda},
\end{equation}
where $F$ is the galaxy spectrum in erg/s/cm$^2$/{\AA}, $T$ is the filter response function, $\lambda$ is the wavelength in {\AA}, and $c$ is the speed of light in {\AA}/s. We compared our measurements to observed photometric \textit{CFHT} r-band flux densities and applied a typical offset correction $\sim0.1-0.25$ dex to account for slit losses. After applying the correction, the emission line fluxes have an uncertainty of $\sim10$\%.

The sample contains 68 H$\alpha$ and 259 [O II] emission line candidates with at least four MIR detections/upper limits in the 7-24 $\mu$m filters. The ranges in redshift, stellar mass, and TIR luminosity of the combined unique 296 sources are $0.05<z<1.36$, $2.1\times10^{8}$ $M_{\odot} < M_{*} < 9.3\times10^{11}$ $M_{\odot}$, and $1.1\times10^{9}$ $L_{\odot} < L(TIR) < 2.2\times10^{12}$ $L_{\odot}$. Table \ref{tab:obs} summarizes the newly reduced and calibrated multi-slit spectroscopy from this work.

\begin{deluxetable*}{cccccccc}
    \tablenum{1}
    \tablecaption{Summary of observations using the \textit{Keck I} and \textit{Keck II} telescopes\label{tab:obs}}
    \tablewidth{0pt}
    \tablehead{
        \colhead{Instrument} & \colhead{UT Date} & \colhead{RA (J200)} & \colhead{Dec (J200)} & \colhead{PA} & \colhead{FWHM} & \colhead{$t_{exp}$} & \colhead{$N_{t}$} \\
        \colhead{} & \colhead{} & \colhead{(hh:mm:ss)} & \colhead{(dd:mm:ss)} & \colhead{(deg)} & \colhead{($''$)} & \colhead{(min)} & \colhead{}
        }
    \startdata
    DEIMOS & 2008-08-01 & 17:57:21.3 & +66:28:45.7 & 0.00 & 0.5 & 51.0 & 81 \\
    DEIMOS & 2008-08-01 & 17:56:24.6 & +66:32:26.1 & -10.00 & 0.6 & 51.0 & 80 \\
    DEIMOS & 2008-08-01 & 17:55:08.4 & +66:42:34.5 & -20.00 & 0.8 & 51.0 & 81 \\
    DEIMOS & 2008-08-01 & 17:56:26.7 & +66:45:40.0 & -30.00 & 0.9 & 60.0 & 82 \\
    DEIMOS & 2008-08-01 & 17:54:34.3 & +66:29:37.4 & -35.00 & 0.7 & 80.0 & 79 \\
    DEIMOS & 2011-07-03 & 17:53:51.9 & +66:44:55.6 & 359.82 & 0.6 & 120.0 & 66 \\
    DEIMOS & 2011-07-03 & 17:55:49.7 & +66:45:30.1 & 359.82 & 0.6 & 120.0 & 61 \\
    DEIMOS & 2011-07-03 & 17:56:28.2 & +66:41:00.1 & 359.83 & 0.6 & 120.0 & 55 \\
    DEIMOS & 2011-07-04 & 17:57:01.8 & +66:30:00.1 & 359.82 & 0.6 & 90.0 & 65 \\
    DEIMOS & 2011-07-04 & 17:56:12.8 & +66:30:05.8 & 359.80 & 0.6 & 90.0 & 64 \\
    DEIMOS & 2011-07-04 & 17:54:54.9 & +66:29:39.9 & 359.80 & 0.6 & 90.0 & 65 \\
    DEIMOS & 2011-07-04 & 17:53:44.2 & +66:27:54.6 & 359.82 & 0.6 & 90.0 & 83 \\
    DEIMOS & 2011-07-04 & 17:57:08.4 & +66:43:24.0 & 37.85 & 0.6 & 105.0 & 66 \\
    DEIMOS & 2014-08-24 & 17:55:15.7 & +66:38:36.3 & 130.07 & 0.6 & 120.0 & 83 \\
    DEIMOS & 2014-08-24 & 17:54:37.8 & +66:29:14.9 & 160.17 & 0.6 & 120.0 & 83 \\
    DEIMOS & 2015-09-14 & 17:52:59.5 & +66:30:37.7 & 145.04 & 0.8 & 120.0 & 78 \\
    DEIMOS & 2015-09-15 & 17:55:01.9 & +66:43:32.3 & 156.56 & 0.9 & 120.0 & 88 \\
    DEIMOS & 2016-09-09 & 17:55:39.0 & +66:29:59.1 & 150.03 & 0.9 & 80.0 & 77 \\
    DEIMOS & 2016-09-10 & 17:53:26.0 & +66:41:40.9 & 94.02 & 0.9 & 80.0 & 78 \\
    MOSFIRE & 2017-08-02 & 17:55:46.3 & +66:37:26.1 & 300.03 & 1.2 & 61.6 & 22 \\
    MOSFIRE & 2017-08-02 & 17:55:35.4 & +66:37:20.6 & 20.00 & 0.4 & 87.5 & 26 \\
    MOSFIRE & 2017-08-03 & 17:53:43.9 & +66:27:12.0 & 340.22 & 0.5 & 69.6 & 23 \\
    MOSFIRE & 2017-08-03 & 17:52:39.6 & +66:34:16.1 & 15.00 & 0.5 & 63.6 & 25 \\
    MOSFIRE & 2017-08-03 & 17:55:40.8 & +66:36:59.8 & 335.00 & 0.5 & 45.7 & 23 \\
    \enddata
    \tablecomments{The exposure time only includes reduced frames. $N_{t}$ is the number of targeted science objects (excluding alignment stars).}
\end{deluxetable*}

\subsection{Keck I/MOSFIRE sample}
In 2017, we targeted higher redshift galaxies in the \textit{AKARI} NEP-Deep field, with the Multi-Object Spectrometer For Infra-Red Exploration (MOSFIRE; \citealp{MOSFIRE, Mclean}) on the \textit{Keck I} telescope.  Based on photometric redshifts provided by the \textit{AKARI}-NEP team \citep{Oi2014}, we observed four slitmasks in the J-band (1.153--1.352 $\mu$m) and one slitmask in the Y-band (0.972--1.125 $\mu$m) to target H$\alpha$/[NII] in \textit{AKARI}/IRC sources at $0.7<z<1.1$ and $0.5<z<0.8$, respectively. Each mask had $\sim 25$ slits with slit width of 0.7${''}$, and resolution of $R\sim 3300$. Weather conditions were clear and dry on August 2nd and mostly clear with some high clouds on August 3rd. Calibration images consisted of Ne, Ar arc lamps (1.5 sec each) and internal flat fields (11 sec each). Slitmasks were observed in Multiple Correlated Double Sampling (MCDS) mode with 16 reads, and a dither position offset of $\pm$1.5${''}$ was used. Standard stars FS 139 (RA $16^{h}33^{m}52.9^{s}$, Dec $+54^{\circ}28{'}23{''}$) and HD 162208 were observed in long-slit mode. 

We used the standard MOSFIRE Data Reduction Pipeline (DRP) to reduce the spectroscopic data. The pipeline creates a pixel flatfield, performs slit edge tracing, calibrates the wavelength scale, performs background subtraction, rectifies the slits, and extracts the 1D spectra. We flux-calibrated the 1D spectra by using the standard star. First we smoothed the standard star spectrum and interpolated over telluric absorption lines. Then we modeled the star as a blackbody based on its effective temperature, and normalized the curve to the star's Ks-band magnitude from the Two Micron All Sky Survey (2MASS), factoring in the filter response function. The factor to convert from counts/second to erg/s/cm$^2$/{\AA} is then proportional to the ratio of the normalized blackbody function to the calibration star spectrum. We accounted for slit losses by comparing each galaxy's average flux density to \textit{CFHT}/WIRCAM J-band photometry.

Our MOSFIRE sample contains 14 H$\alpha$ and 9 [O II] emission-line candidates with at least four MIR detections/upper limits in the 7--24 $\mu$m filters. The ranges in redshift, stellar mass, and TIR luminosity of the combined unique 15 sources are $0.54<z<1.04$, $8.5\times10^{9}$ $M_{\odot} < M_{*} < 1.1\times10^{11}$ $M_{\odot}$, and $5.5\times10^{10}$ $L_{\odot} < L(TIR) < 8.6\times10^{11}$ $L_{\odot}$.

\subsection{Takagi et al. (2010) sample}
In addition to our 2011-2016 \textit{Keck II}/DEIMOS observations, we obtained reduced DEIMOS spectra and emission line flux measurements from observations in semester 2008A through the \textit{AKARI}-NEP collaboration (PI: Toshinobu Takagi). Targets included 242 \textit{AKARI}-NEP mid-IR-selected sources with \textit{Subaru}/Suprime-Cam \textit{R}-band magnitude $\lesssim$ 24 mag. Five slitmasks were observed using the 600ZD grating and GG495 order blocking filter. Standard stars HZ 44 
and Feige 110 were observed with a long slit (1.5${''}$ slit width) using the same grating and filter setup as the targets. Data were reduced using the spec2d pipeline, and then flux-calibrated with standard IRAF routines. [O II]$\lambda\lambda3726,3729$, H$\beta$, [O III]$\lambda5007$, H$\alpha$, and [N II]$\lambda6584$ emission lines were fit using Gaussian models with the IDL/MPFIT routine. Double Gaussians were used to fit the resolved [O II] doublet, H$\beta$ emission with underlying absorption, and H$\alpha$ with broad and narrow line components in AGN. The continuum was modeled with a linear fit. H$\alpha$ and [N II]$\lambda\lambda6548,6584$ were fit simultaneously with three Gaussians. The Takagi et al. sample contains 48 H$\alpha$ and 126 [O II] emission line candidates with at least four MIR detections in the 7--24 $\mu$m filters. The ranges in redshift, stellar mass, and TIR luminosity of the combined unique 152 sources are $0.09<z<1.35$, $9.5\times10^{8}$ $M_{\odot} < M_{*} < 3.1\times10^{11}$ $M_{\odot}$, and $8.3\times10^{8}$ $L_{\odot} < L(TIR) < 3.1\times10^{12}$ $L_{\odot}$.

\subsection{Shim et al. (2013) sample}
We also include spectroscopic redshifts and emission line fluxes from the Shim et al. (2013) sample, which consists of spectroscopy of hundreds of low-redshift galaxies and AGN ($z\lesssim0.4$)  in the \textit{AKARI} NEP-Wide field.  This provides rest-frame UV/optical emission line detections from \textit{MMT}/Hectospec and \textit{WIYN}/HYDRA multi-object spectrographs. The authors primarily selected targets with MIR detections at 11 $\mu$m and 15 $\mu$m. Secondary target selection criteria are summarized in their Table 1, and include AGN candidates, \textit{Herschel} FIR sources, and PAH-luminous galaxies. The Shim et al. sample contains 498 H$\alpha$ and 571 [O II] emission line galaxieses with at least four MIR detections/upper limits in the 7--24 $\mu$m filters and secure redshifts (i.e., quality flag of 4). The ranges in redshift, stellar mass, and TIR luminosity of the combined unique 711 sources are $0.03<z<1.29$, $1.9\times10^{8}$ $M_{\odot} < M_{*} < 9.9\times10^{11}$ $M_{\odot}$, and $4.3\times10^{8}$ $L_{\odot} < L(TIR) < 3.2\times10^{12}$ $L_{\odot}$.

\subsection{Oi et al. (2017) sample}
We supplement our high-redshift MOSFIRE data with near-infrared H$\alpha$ detections from \citet{Oi2017}, who studied the mass-metallicity relation in luminous infrared galaxies at $z\sim0.9$. Their sample consists of  \textit{Subaru}/FMOS H$\alpha$/[N II] multi-object spectroscopy in the J-long band (1.11--1.35 $\mu$m), of \textit{AKARI} mid-infrared sources in the NEP-Deep field. Target galaxies were selected to have detections at 11 $\mu$m, 15 $\mu$m, and/or 18 $\mu$m, and estimated photometric redshifts $z_{phot} \sim 1$. H$\alpha$/[N II] fluxes are given for 28 secure H$\alpha$ emitters and H$\alpha$ fluxes are given for 36 ``non-secure" H$\alpha$ detections (i.e., objects where a single emission line was observed and assumed to be H$\alpha$.) Some of the non-secure H$\alpha$ objects have since been re-observed in optical wavelengths by \textit{AKARI} NEP collaborators using \textit{MMT}/Hectospec, \textit{WIYN}/Hydra, \textit{Keck II}/DEIMOS, and \textit{GTC}/OSIRIS, allowing us to confirm that 10/36 are indeed H$\alpha$ detections based on the detection of other strong emission lines at shorter wavelengths (e.g., [OII], H$\gamma$, H$\beta$, [OIII]). In addition, we find that the emission line from object 61016583, assumed by the Oi et al. to be H$\alpha$, is [O III]$\lambda$5007 at $z=1.357$. For reference, we list the updated secure H$\alpha$ objects in Table \ref{tab:fmos} in the Appendix. Our SFR calibration sample includes 37 FMOS objects with H$\alpha$ detections and at least four MIR detections/upper limits in the 7--24 $\mu$m filters. The ranges in redshift, stellar mass, and TIR luminosity are $0.70<z<1.03$, $2.1\times10^9$ $M_{\odot}<M_{*}<1.5\times10^{11}$ $M_{\odot}$, and $8.1\times10^{10}$  $L_{\odot}<L(TIR)<1.5\times10^{12}$ $L_{\odot}$.

\subsection{Validation sample: Ohyama et al. (2018) SPICY galaxies}
\begin{figure*}[ht!]
	\centering
	\includegraphics[scale=0.6]{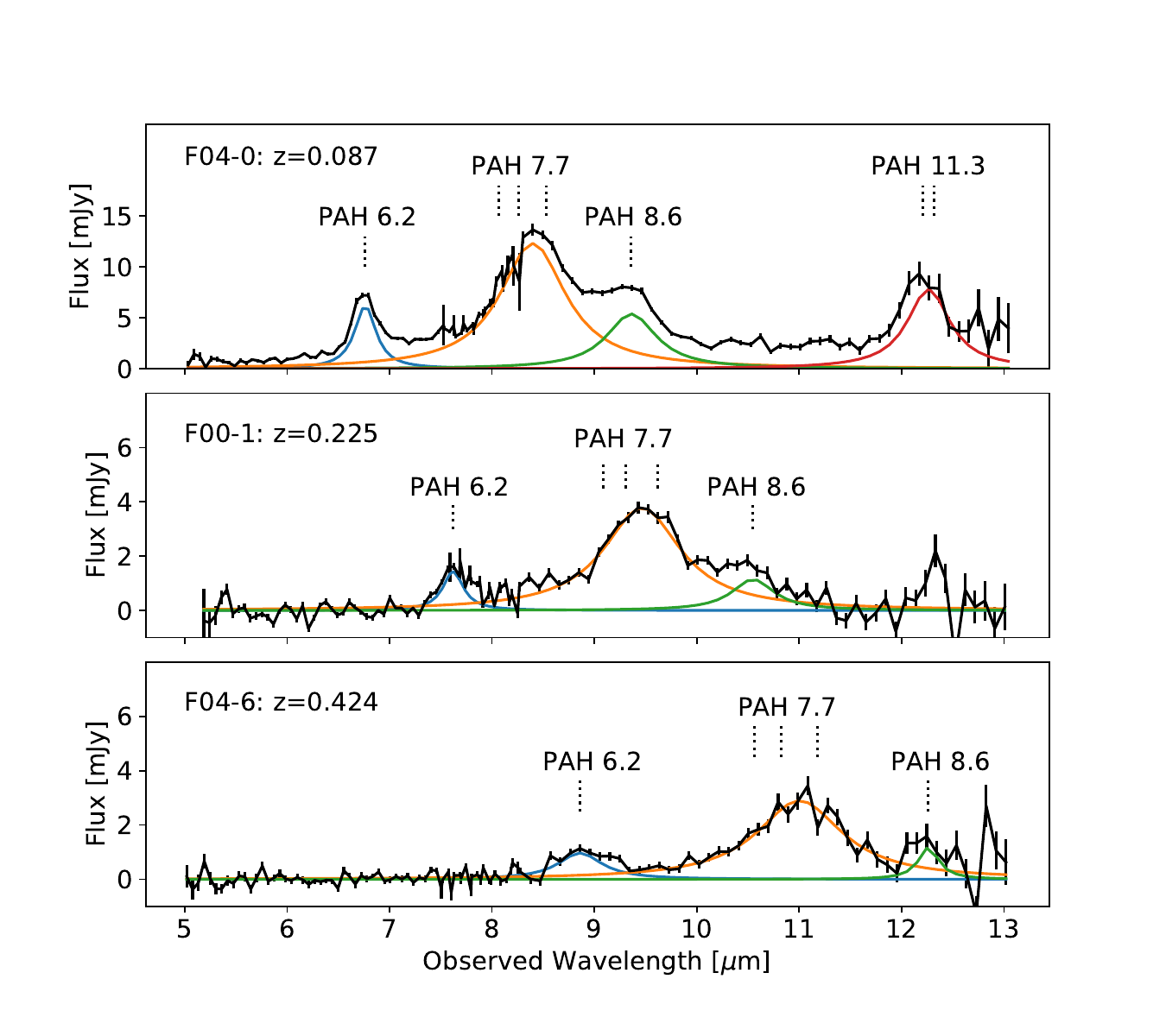}
	\vspace*{-5mm}
	\caption{Examples of \textit{AKARI} SPICY spectra of star-forming galaxies from \citet{Ohyama2018} with PAH emission detections. Underlying curves represent Lorentzian fits to the PAH 6.2, 7.7, 8.6, and 11.3 $\mu$m emission features \citep{Ohyama2018}.
	\label{spicy_examples}}
\end{figure*}

The \textit{AKARI}/IRC slitless SpectroscoPIC surveY (SPICY) obtained infrared spectra of all sources with 9 $\mu$m flux density brighter than 0.3 mJy in 14 NEP fields of $10{'}\times10{'}$. The primary goal was to study PAH emission in galaxies at $z\lesssim0.5$ \citep{Ohyama2018} within the NEP-Deep and Wide fields. Low-resolution spectra ($R\sim50$) were acquired with the short-wavelength MIR camera on the IRC, which covers a wavelength range of $5-13\ \mu$m. Ohyama et al. identified 48 galaxies with detectable PAH 6.2, 7.7, and 8.6 $\mu$m features. All 48 galaxies have \textit{AKARI}/IRC photometry, with 83\% having detections in the \textit{N2}, \textit{N3}, \textit{N4}, \textit{S7}, \textit{S9W}, \textit{S11}, \textit{L15}, and \textit{L18W} filters ($\sim2-18\ \mu$m). Of the 48 PAH galaxies in the sample, 41 galaxies have optical spectroscopic detections and 33 galaxies have detected H$\alpha$ emission from \textit{Keck II}/DEIMOS (this work) or \textit{MMT}/Hectospec (PI: Ho Seong Hwang). Figure \ref{spicy_examples} illustrates typical SPICY spectra of star-forming galaxies from \citet{Ohyama2018}. We use these 41 galaxies as our validation sample, to test the consistency and accuracy of our photometrically-derived PAH luminosity measurements. The ranges in redshift, stellar mass, and TIR luminosity of the 41 sources are $0.06<z<0.49$, $2\times10^{9}$ $M_{\odot} < M_{*} < 2\times10^{11}$ $M_{\odot}$, and $2.1\times10^{9}$ $L_{\odot} < L(TIR) < 3.9\times10^{11}$ $L_{\odot}$. 

\subsection{Emission line measurements}
\begin{figure*}[ht!]
	\centering
	\includegraphics[scale=0.8]{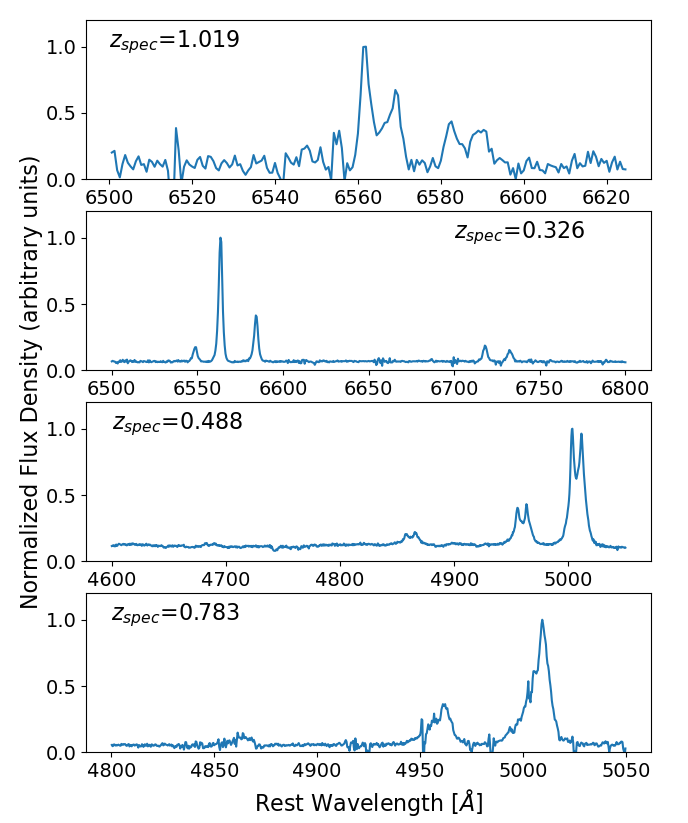}
	\caption{Example  observations illustrating the quality and variety of selected  sections of our high resolution spectra. The galaxy in the top panel was observed with \textit{Keck I}/MOSFIRE J-band, and the bottom three were observed with \textit{Keck II}/DEIMOS. The H$\alpha$/[N II] complex is clearly detected in the top two spectra, while the H$\beta$/[O III] complex is seen in the bottom two.}
	\label{fig:spec}
\end{figure*}

Figure \ref{fig:spec} shows examples of the quality of our DEIMOS and MOSFIRE spectra. The top panel shows a MOSFIRE galaxy detected in the J-band that has red- and blue-shifted lines in H$\alpha$ and [N II] due to rotational kinematics. The second panel is an example of a typical star-forming galaxy observed with DEIMOS with H$\alpha$, [N II], and [S II] emission lines. The third panel shows a Seyfert galaxy with double-peaked emission in H$\beta$ and [O III]. The signal-to-noise is also high enough to resolve the doublet in H$\gamma$ (not shown). The fourth panel shows broad blue wings in [O III] emission, suggestive of outflows from an AGN.
	
We used the Specpro software \citep{MastersCapak} to visually inspect each spectrum and to determine an initial redshift guess based on a template of emission and absorption lines. This initial redshift served as the seed for our python routine that used non-linear least-squares fitting. We used Gaussian fits to measure redshifts, emission line fluxes, and equivalent widths for [O II]$\lambda \lambda$3726,3729, H$\gamma$, H$\beta$, [O III]$\lambda \lambda$4959,5007, H$\alpha$, [N II]$\lambda \lambda$6548,6584, and [S II]$\lambda \lambda$6716,6731. In general, a linear local continuum was added. For H$\gamma$ and H$\beta$, the continuum was modeled with a Voigt profile in order to account for underlying stellar absorption. 

For multi-component Gaussian fits, the following constraints were used in order to limit the number of free parameters. When resolved, the $[$O II$]$$\lambda \lambda$3726,3729 doublet was fit with a double Gaussian, with the central wavelength and width of [O II]$\lambda$3729 fixed relative to [O II]$\lambda$3726. The $[$O III$]\lambda \lambda$4959,5007 lines were fit with a double Gaussian such that the width of [O III]$\lambda$4959 was fixed to that of the stronger line $\lambda$5007, the central wavelength was fixed relative to the expected separation, and the amplitude was fixed to the theoretical value of [O III]$\lambda$5007/2.98 \citep{Osterbrock}. For the H$\alpha$, $[$N II$]\lambda \lambda$6548,6584 complex, a triple Gaussian was used; the amplitude of [N II]$\lambda$6548 was fixed to be [N II]$\lambda$6584/3.071 \citep{Osterbrock}, the central wavelengths of [N II] were fixed relative to H$\alpha$, and the width of [N II]$\lambda$6548 was fixed to be that of $\lambda$6584. For Seyfert galaxies where a broad H$\alpha$ component was detected (FWHM $\gtrsim$ 1000 km/s), an additional underlying Gaussian was added.

\subsubsection{Metallicity Measurements}
We estimate the oxygen abundance\footnote{We use the terms ``oxygen abundance" and ``metallicity" interchangeably in this work.} of the ionized interstellar medium, 12+log(O/H), by using the N2 (log [N II]$\lambda$6584/H$\alpha$), O3N2 (log ([O III]$\lambda$5007/H$\beta$)/([N II]$\lambda$6584/H$\alpha$)), and O32 (log ([O III]$\lambda$5007/[O II]$\lambda\lambda$3726,3729) indicators. For the N2 and O3N2 indicators, we use the calibrations by \citet{PettiniPagel}, given by:
\begin{equation}
    \text{12 + log(O/H)}_{N2} = 8.9+0.57\times \text{N2} \\
\end{equation}
\begin{equation}
    \text{12 + log(O/H)}_{O3N2} = 8.73 - 0.32\times \text{O3N2}
\end{equation}
where the 1-$\sigma$ calibration uncertainties are 0.18 dex and 0.14 dex for N2 and O3N2, respectively. When using the N2 indicator, we limit measurements to objects with N2 $<-0.3$ to avoid saturation in nitrogen. For the O32 indicator, we use the calibration given by \citet{Jones2015}:
\begin{equation}
    \text{12+log(O/H)}_{O32} = 8.3439 - 0.4640\times \text{O32}
\end{equation}
which has an uncertainty of 0.11 dex. For reference, solar metallicity is such that 12+log(O/H) = 8.69 \citep{Asplund}.

Table \ref{tab:allobs} lists the total number of H$\alpha$ and [O II] detections with at least four photometric detections (including upper limits from 7--24 $\mu$m) and empirical metallicity measurements using either the N2 index (for H$\alpha$ and [N II]-observed sources) and/or O32 index (for [O II] sources).

\begin{deluxetable}{ccc}
    \tablenum{2}
    \tablecaption{Number of H$\alpha$ and [O II] detections from combined observations. \label{tab:allobs}}
    \tablewidth{0pt}
    \tablehead{
        \colhead{Instrument} & \colhead{$N$(H$\alpha$)} & \colhead{$N$([O II])}
        }
    \startdata
    Keck/DEIMOS & 72 & 196 \\
    Keck/MOSFIRE & 9 & 7 \\
    MMT/Hectospec, WIYN/Hydra$^{(a)}$ & 334 & 422 \\
    Subaru/FMOS$^{(b)}$ & 22 & \text{--} \\
    \hline
     & 437 & 625
    \enddata
    \tablecomments{All galaxies have at least four photometric detections (including upper limits) from 7-24 $\mu$m and empirical metallicity measurements using the emission line ratios [N II]/H$\alpha$ (for H$\alpha$ and [N II]-measured sources) and/or O32 (for [O II] sources). References: (a) \citet{Shim2013}, (b) \citet{Oi2017}.}
\end{deluxetable}

\subsection{AGN selection}
To avoid contamination from AGN in our SFR calculations, and to study the impact that the presence of an AGN has on PAH dust luminosity, we classify our source spectra. We identify AGN candidates using a combination of spectroscopic diagnostics, mid-infrared colors, and X-ray cross-matching. An object is considered to be an AGN if it meets at least one of the four following criteria:
\begin{enumerate}
	\item the spectrum has broad permitted emission lines (i.e., FWHM $\gtrsim$ 1000 km/s) consistent with a Type 1 Seyfert galaxy;
	\item the object is a Type 2 Seyfert galaxy or LINER identified with the ``BPT" line ratio diagram such that log([OIII]/H$\beta$)$>$0.61/(log([NII]/H$\alpha$)-0.02-0.1833$\times$$z$) + 1.2+0.03$\times$$z$ (\citet{Kewley2013}, Equation 1);
	\item for MIR-selected AGN, the \textit{AKARI}/IRC colors are such that \textit{N2}-\textit{N4}$>$0 and \textit{S7}-\textit{S11}$>$0 [AB magnitude] \citep{Lee2007, Shim2013}; and/or
	\item the object has a Chandra X-ray counterpart within 2.5$''$ \citep{Krumpe2015}.
\end{enumerate}

\begin{figure*}[ht!]
    \centering
	\includegraphics[scale=0.5]{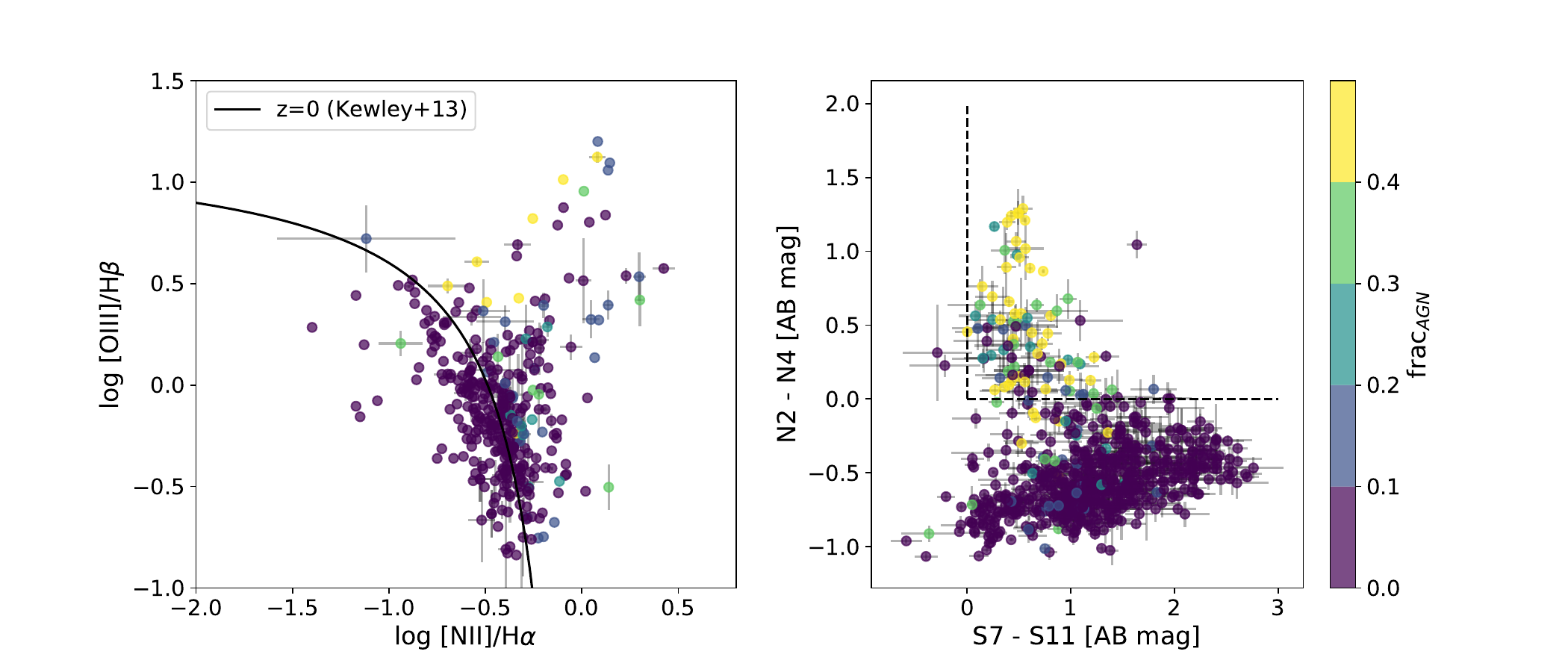}
	\caption{Left: BPT-[NII] diagram based on the redshift-dependent \citet{Kewley2013} equation. The line that separates star formation from AGN at $z=0$ is shown for reference. Right: Mid-infrared classification with \textit{AKARI} colors for objects with S/N$>$3 detections in each filter. AGN candidates are found in the region where \textit{N2}-\textit{N4}$>$0 and \textit{S7}-\textit{S11}$>$0 (dashed lines). Both panels are color-coded by their mid-infrared AGN fraction determined from CIGALE SED fits (Section \ref{sec:sed}).
	\label{bpt_mir}}
\end{figure*}

The left panel of Figure \ref{bpt_mir} shows the usual BPT-[NII] emission line ratio diagram for our sources. Based on this diagram, there are 215 star-forming galaxies (``BPT-SF") and 154 AGN candidates (``BPT-AGN"). Based on the MIR colors, there are 779 star-forming galaxies (``MIR-SF") and 157 AGN candidates (``MIR-AGN"). BPT-classified SF galaxies with \textit{N2}, \textit{N4}, \textit{S7}, and \textit{S11} filter detections are 99\% consistent with their mid-infrared color classifications; 160 BPT-SF galaxies are also MIR-SF galaxies, while only two BPT-SF galaxies are classified as MIR-AGN. However, the likelihood that a BPT-AGN is classified as a MIR-SF galaxy is high: 121 BPT-AGN are classified as MIR-SF galaxies and 8 are classified as MIR-AGN. This is probably because MIR color selection is stricter, tending to exclude objects with low AGN fraction, and many Seyfert 2's which the BPT diagram reveals. Thus, an MIR-selected AGN is powerful enough to be certainly identified as a BPT-AGN (left panel). The converse is not always true; some BPT-classified AGN have active nuclei which are too faint to contribute much mid-IR continuum.
Both panels are color-coded by their AGN fraction as determined through SED fitting (Section \ref{sec:sed}).

\section{IRC Measurement of the PAH luminosity} \label{sec:pahmeasurement}
 
 \subsection{Spectral energy distribution modeling} \label{sec:sed}
We matched our H$\alpha$ detections with their multi-wavelength photometric counterparts, using a matching radius of the full width at half maximum of the telescope point spread function. When available, we merged photometry from the following telescopes: \textit{Chandra} \citep{Krumpe2015}, \textit{GALEX} FUV and NUV \citep{Burgarella2019}, \textit{CFHT}/MegaCam \textit{u}$^{*}$, \textit{g'}, \textit{r'}, \textit{i'}, \textit{z'} \citep{Oi2014, Hwang2007}, Subaru/Hyper-Suprime Cam \textit{g/r/i/z/Y} \citep{Oi2021}, \textit{CFHT}/WIRCAM \textit{J} and \textit{K}$_{S}$ \citep{Oi2014}, \textit{KPNO}/FLAMINGOS \textit{J} and \textit{H} \citep{Jeon2014}, \textit{AKARI}/IRC \textit{N2}, \textit{N3}, \textit{N4}, \textit{S7}, \textit{S9W}, \textit{S11}, \textit{L15}, \textit{L18W}, \textit{L24} \citep{Kim2012, Murata2013}, \textit{Spitzer}/IRAC1 (3.6 $\mu$m) and IRAC2 (4.5 $\mu$m) \citep{Nayyeri2018}, \textit{WISE} 12 $\mu$m and \textit{WISE} 22 $\mu$m \citep{Wright2010, Jarrett2011, Cutri2014}, \textit{Herschel}/PACS green (100 $\mu$m) and PACS red (160 $\mu$m) \citep{Pearson2019}, and \textit{Herschel}/SPIRE 250 $\mu$m, 350 $\mu$m, 500 $\mu$m bands \citep{Pearson2017}.
We note that some of these datasets do not cover the entire \textit{AKARI} NEP-Wide field, but our multi-wavelength coverage of the NEP-Deep field is reasonably complete.

We modeled spectral energy distributions (SEDs) with the CIGALE software (Code Investigating GALaxy Emission), which is based on the assumption of energy balance between stellar radiation absorbed by dust in the UV/optical and dust emission output in the mid to far-infrared \citep{Boquien, Cigale_Yang}. 
This implicitly requires that our view of the extinction is similar to what observers along most other lines-of-sight to the galaxy would also measure.
CIGALE uses Bayesian analysis that combines stellar population models, AGN and ISM templates\footnote{We note that CIGALE includes the option to include user-input nebular emission line fluxes (e.g., H$\alpha$, [N II]), which would further constrain the metallicity and $q_{PAH}$ parameters. However, we do not include emission line fluxes in this work, but recommend this option for future studies.}, and various libraries to determine the best parameters across all possible spectra. We assumed an exponentially decaying star formation history, \citet{Chabrier} IMF, \citet{Calzetti00} dust attenuation model, \citet{Draine2014} dust emission template, and \citet{Fritz06} AGN emission model. The dust emission model characterizes PAH emission strength based on the $q_{PAH}$ parameter, which is defined as the fraction of the total dust mass in PAH grains with fewer than 10$^3$ carbon atoms. For reference, average Milky Way dust has $q_{PAH}\approx$ 4.6\% \citep{DraineLi07}. We note that the \textit{Chandra} X-ray data were only used to identify X-ray AGN and were not included in the CIGALE fits.

We define the AGN fraction, $frac_{AGN}$, as the ratio of the AGN luminosity to the total luminosity integrated between 5--10 $\mu$m. For star-forming galaxies, we assume $frac_{AGN}=0$. Otherwise, for AGN candidates and unclassified objects (i.e., objects with incomplete spectroscopic/photometric data), we allowed $frac_{AGN}$ to vary. Figure \ref{bpt_mir} is color-coded by the AGN fraction. The median AGN fraction for BPT-AGN candidates is 0.05 with interquartile range of 0.01--0.13, compared to a median of 0.33 for MIR-AGN candidates with interquartile range of 0.12--0.48. This difference demonstrates the efficacy of the BPT diagram to select AGN with lower $frac_{AGN}$ than the \textit{AKARI} color-color diagram.

\begin{figure*}[ht!]
	\centering
	\includegraphics[width=\linewidth]{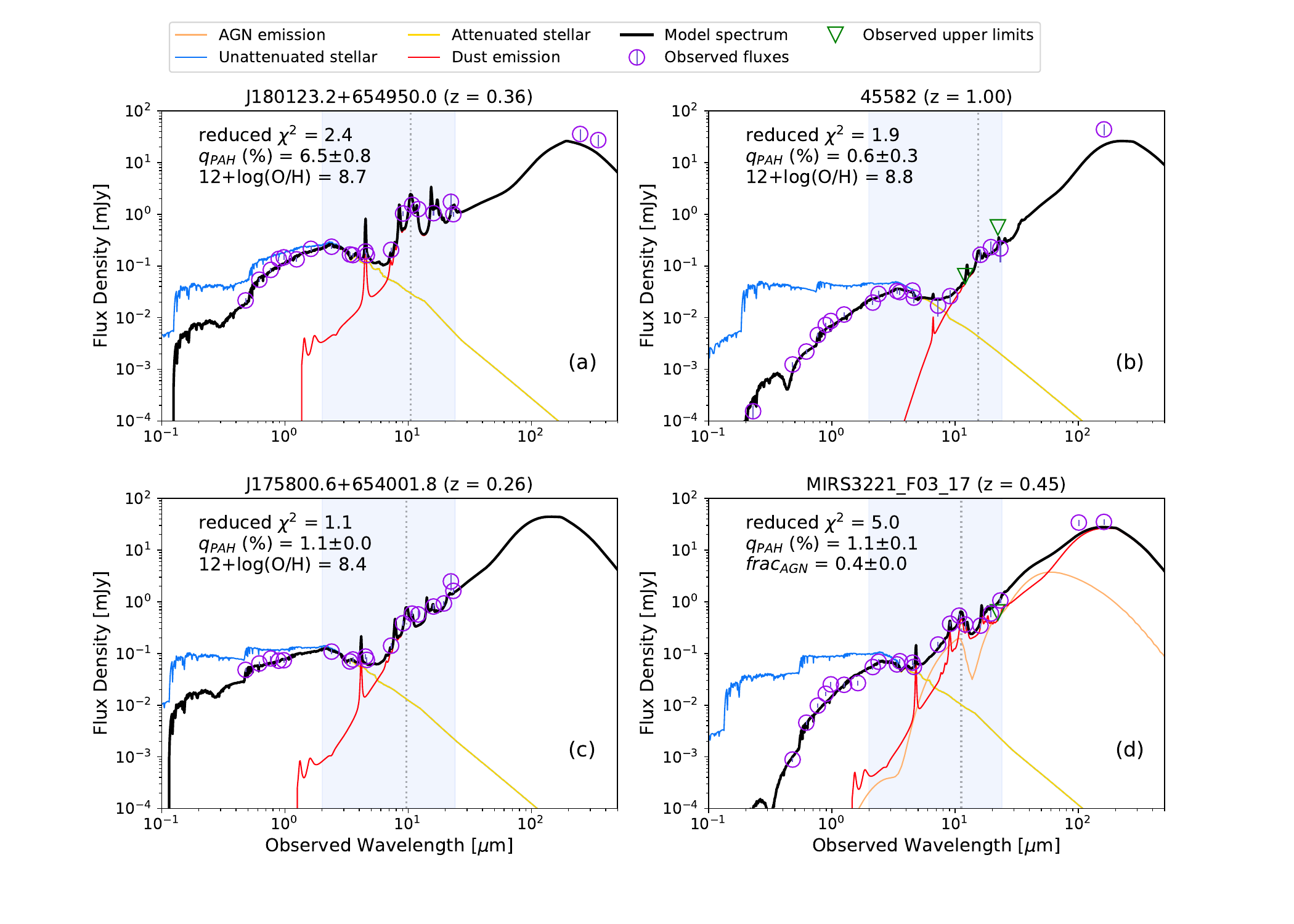}
	\vspace*{-15mm}
	\caption{Examples of typical CIGALE SED fits, including: (a) star forming ``main-sequence' galaxy with solar metallicity and strong PAH emission; (b) metal-rich starburst galaxy (R$_{SB}$=11) with weak PAH emission; (c) metal-poor starburst galaxy (R$_{SB}$=18) with weak PAH emission; (d) Type-II BPT-AGN with weak PAH emission.
	\label{fig:cigale}}
\end{figure*}

We used a 32-core processor through Amazon Web Services (AWS) cloud computing\footnote{\url{https://aws.amazon.com/}} to run our models at a rate of $\sim$30,000 models/second. For the final sample, we limit objects to those with $0.5<\text{reduced } \chi^2<10$. Objects with reduced $\chi^2<0.5$ were excluded in order to avoid models with overfitting. Table \ref{tab:cigale} summarizes our input SED parameters. In Figure \ref{fig:cigale}, we illustrate the multi-wavelength coverage of our photometric data (purple circles) and show some diverse examples of CIGALE best-fit SEDs which have a range of PAH strengths in the mid-infrared. The light blue region from 2-24 $\mu$m highlights the observed wavelength range of the \textit{AKARI}/IRC, and the dotted vertical line indicates the observed wavelength for the PAH 7.7 $\mu$m blend. The model spectrum (solid black line) represents the sum over the attenuated stellar emission, dust emission, and AGN emission components. Although we include the contribution from nebular emission when modeling SEDs, we omit it in the figure for simplicity.

The total infrared luminosity, $L(TIR)$, has been extensively used in the literature as a SFR indicator, and is known to correlate with PAH luminosity \citep{Takagi2010}. We calculate $L(TIR)$ for individual objects by shifting the best-fit SED to the rest frame and integrating from 8--1000 $\mu$m. In the Appendix, we test the dependence of FIR photometry on the integrated $L(TIR)$ and find that $L(TIR)$ measurements with and without FIR data are consistent with a root-mean-square (RMS) dispersion of 0.096 dex (Figure \ref{fig:ltirappendix}). Therefore, we estimate that our $L(TIR)$ measurements are accurate to within $\sim$20\%.

\begin{deluxetable*}{ccc}
        \rotate
	\tablenum{3}
	\tablecaption{Input parameters for CIGALE SED fitting \label{tab:cigale}}
	\tabletypesize{\footnotesize}
	\tablewidth{0pt}
	\tablehead{
		\colhead{Parameter} & \colhead{Description} & \colhead{Value}
	}
	\startdata
	\multicolumn{3}{c}{Star Formation History: $SFR(t)\propto e^{-t/\tau}$} \\
	\hline
	$\tau$ (Myr) & e-folding time of main stellar population & \makecell{100, 300, 1000, 2000, 3000, \\ 5000, 10000, 15000, 30000} \\
	$age$ (Myr) & age of main stellar population & \makecell{3, 8, 20, 50, 125, 300, 800, \\ 2000, 5000, 13000} \\
	\hline
	\multicolumn{3}{c}{Stellar Populations} \\
	\hline
	IMF & Initial Mass Function & \citet{Chabrier} \\
	$Z$ & Metallicity & 0.02 (solar) \\
	Separation age (Myr) & separation age between young and old stellar populations & 10 \\
	\hline
	\multicolumn{3}{c}{Nebular Emission} \\
	\hline
	log($U$) & ionization parameter & -2.0 \\
	$f_{esc}$ & fraction of Lyman continuum photons escaping the galaxy & 0.0 \\
	$f_{dust}$ & fraction of Lyman continuum photons absorbed by dust & 0.0 \\
	\hline
	\multicolumn{3}{c}{Dust Attenuation: \citet{Calzetti00}} \\
	\hline
	E(B-V)$_{lines}$ & color excess of nebular line emission & 0.0-1.6 in steps of 0.2 \\
    $f$ & ratio of E(B-V)$_{continuum}$ to E(B-V)$_{lines}$ & 0.44 \\
    UV extinction bump wavelength & central wavelength of UV bump in nm & 217.5 \\
    UV bump width & FWHM of UV bump in nm & 35.0 \\
	\hline
	\multicolumn{3}{c}{Dust Emission: \citet{Draine2014}} \\
	\hline
	$q_{PAH}$ (\%) & mass fraction of PAH & \makecell{0.47, 1.12, 2.50, 3.90, 4.58, \\ 5.26, 5.95, 6.63, 7.32} \\
	$U_{min}$ & minimum radiation field & 1.0, 5.0 \\
	$\alpha$ & power-law slope $dU/dM \propto U^{\alpha}$ & 2.0 \\
	$\gamma$ & fraction illuminated from $U_{min}$ to $U_{max}$ & 0.2 \\
	\hline
	\multicolumn{3}{c}{AGN Emission: \citet{Fritz06}} \\
	\hline
	$r$ ratio & ratio of the maximum to minimum radii of the dust torus & 60.0 \\
	$\tau_{9.7}$ & optical depth at 9.7 $\mu$m & 1.0, 6.0 \\
	opening angle & opening angle of dust torus & 100.0$^{\circ}$ \\
	$\psi$ & angle between equatorial axis and line of sight & 0.001$^{\circ}$, 89.990$^{\circ}$ \\
	$frac_{AGN}$ & AGN fraction & \makecell{0.0, 0.05, 0.1, 0.15, 0.2, 0.25, \\ 0.3, 0.35, 0.4, 0.45, 0.5}
	\enddata
\end{deluxetable*}
 
\subsection{Calibration sample characteristics} \label{sec:calibsample}
\begin{figure*}[ht!]
	\centering
	\includegraphics[scale=0.45]{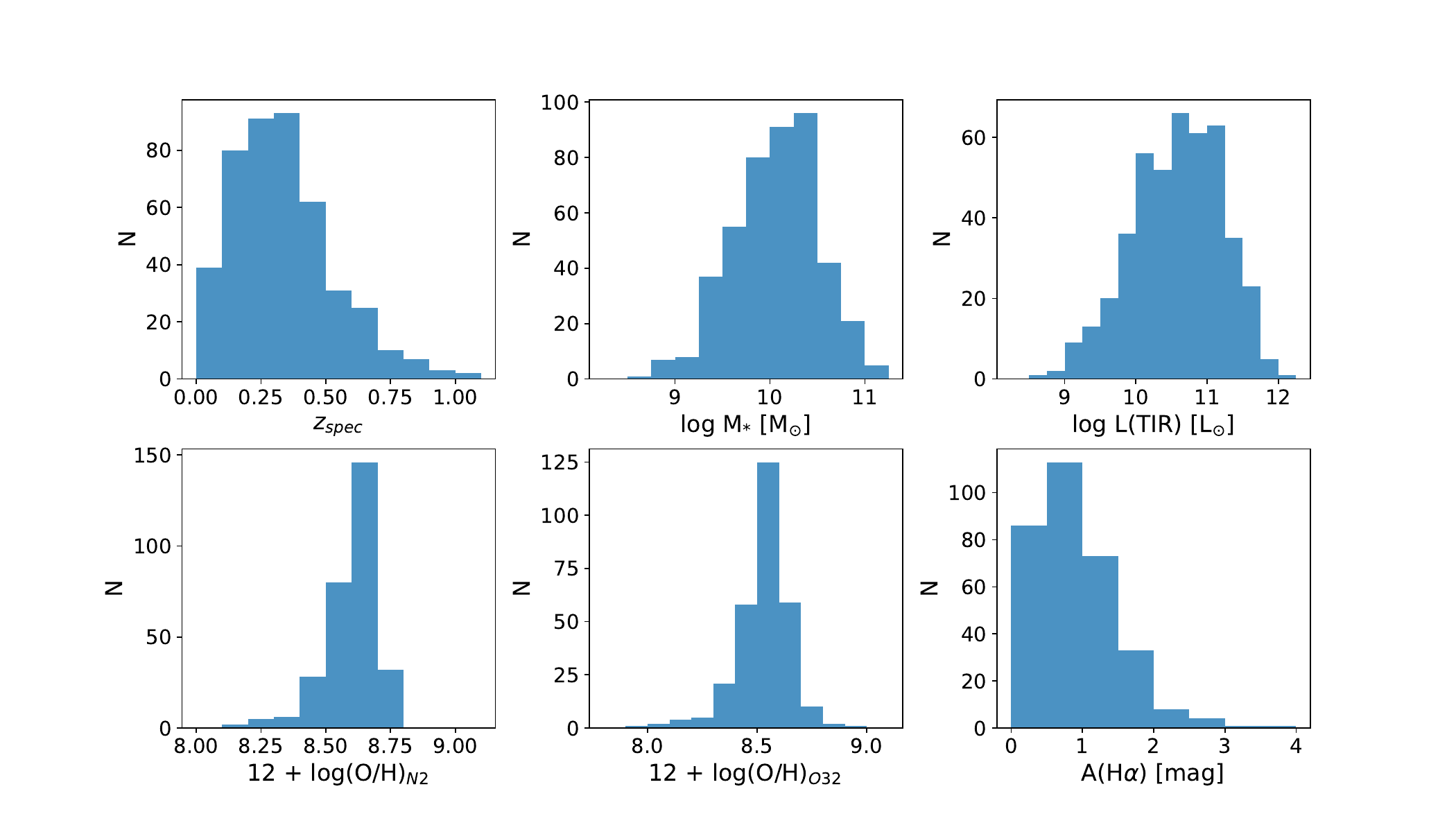}
	\caption{Distributions of main properties of the final SFR(PAH) calibration sample with H$\alpha$ and/or [O II] emission line detections. From left to right: spectroscopic redshift, stellar mass, total infrared luminosity, metallicity based on N2 index, metallicity based on O32 index, and attenuation in H$\alpha$ based on SED fitting.
	\label{hist}}
\end{figure*}

Our final PAH-derived SFR calibration sample consists of main-sequence and starburst galaxies with H$\alpha$ and/or [O II] emission lines and metallicities measured using the N2, O3N2, and/or O32 line ratio indices. The sample includes 319 galaxies with H$\alpha$ detections and 332 galaxies with [O II] detections. There are a total of 443 unique sources. We summarize the main properties of this sample in Figure \ref{hist}. The redshift, stellar mass, and total IR luminosity range from 0.048$<z<$1.025, 10$^{8.7}$ M$_{\odot}$$<$ M$_{*}$ $<$10$^{11.2}$ M$_{\odot}$, and $10^{8.7}$ L$_{\odot}$$<$ L(TIR) $<$10$^{12}$ L$_{\odot}$, respectively. Table \ref{tab:emlines} lists the emission line fluxes and physical properties for a portion of galaxies within the calibration sample. The full table is published online.

\begin{splitdeluxetable*}{ccccccccBccccccBcccc}
    \tablenum{4}
    \tablecaption{Calibration sample emission line fluxes and physical properties \label{tab:emlines}}
    \tablewidth{0pt}
    \tablehead{
        \colhead{Object} & \colhead{RA} & \colhead{Dec} & \colhead{Redshift} & \colhead{$f([O\ II])$} & \colhead{$\sigma_{f([O\ II])}$} & \colhead{$f(H\beta)$} & \colhead{$\sigma_{f(H\beta)}$} & \colhead{$f([O\ III]\lambda 5007)$} & \colhead{$\sigma_{f([O\ III]\lambda 5007)}$} & \colhead{$f(H\alpha)$} & \colhead{$\sigma_{f(H\alpha)}$} & \colhead{$f([N\ II]\lambda 6584)$} & \colhead{$\sigma_{f([N\ II]\lambda 6584)}$} & \colhead{log $\left(\frac{M_{*}}{M_{\odot}}\right)$} & \colhead{log $\left(\frac{L(TIR)}{L_{\odot}}\right)$} & \colhead{$L(PAH\ 6.2\ \mu m)$} & \colhead{$L(PAH\ 7.7\ \mu m)$} \\
        \colhead{(1)} & \colhead{(2)} & \colhead{(3)} & \colhead{(4)} & \colhead{(5)} & \colhead{(6)} & \colhead{(7)} & \colhead{(8)} & \colhead{(9)} & \colhead{(10)} & \colhead{(11)} & \colhead{(12)} & \colhead{(13)} & \colhead{(14)} & \colhead{(15)} & \colhead{(16)} & \colhead{(17)} & \colhead{(18)}
        }
    \startdata
    61017806 & 269.047 & 66.632 & 0.487 & 0.00 & 0.00 & 2.58 & 0.08 & 1.78 & 0.10 & 12.44 & 0.24 & 3.70 & 0.26 & 10.38 & 11.49 & $4.47\times10^{42}$ & $1.45\times10^{43}$ \\
61018416 & 268.719 & 66.648 & 0.395 & 1.38 & 0.11 & 0.97 & 0.06 & 0.64 & 0.11 & 6.61 & 0.15 & 2.83 & 0.16 & 10.36 & 11.11 & $3.25\times10^{42}$ & $1.14\times10^{43}$ \\
61019580 & 268.737 & 66.680 & 0.116 & 0.00 & 0.00 & 5.34 & 0.11 & 6.32 & 0.19 & 22.66 & 0.12 & 6.02 & 0.15 & 9.36 & 9.34 & $8.05\times10^{40}$ & $2.71\times10^{41}$ \\
61019692 & 268.836 & 66.685 & 0.390 & 0.00 & 0.00 & 1.59 & 0.05 & 0.37 & 0.10 & 7.73 & 0.12 & 2.63 & 0.12 & 9.92 & 10.65 & $1.12\times10^{42}$ & $3.88\times10^{42}$ \\
61020407 & 268.616 & 66.704 & 0.454 & 1.96 & 0.12 & 2.37 & 0.08 & 1.40 & 0.09 & 14.71 & 0.22 & 6.31 & 0.28 & 10.03 & 11.12 & $4.97\times10^{42}$ & $1.73\times10^{43}$ \\
    \enddata
    \tablecomments{Column names are as follows: (1) Object name; (2) Right ascension in degrees; (3) Declination in degrees; (4) Spectroscopic redshift; (5)-(14) Emission line fluxes and mean errors in $10^{-16}$ erg s$^{-1}$ cm$^{-2}$; (15) Stellar mass; (16) Total infrared luminosity; (17)-(18) PAH luminosities in erg s$^{-1}$. The errors in the PAH 6.2 $\mu$m and 7.7 $\mu$m luminosities are discussed in Section \ref{sec:spicy}. This table is published in its entirety in the machine-readable format. A portion is shown here for guidance regarding its form and content.}
\end{splitdeluxetable*}

\begin{figure*}[ht!]
	\centering
        \includegraphics[scale=0.45]{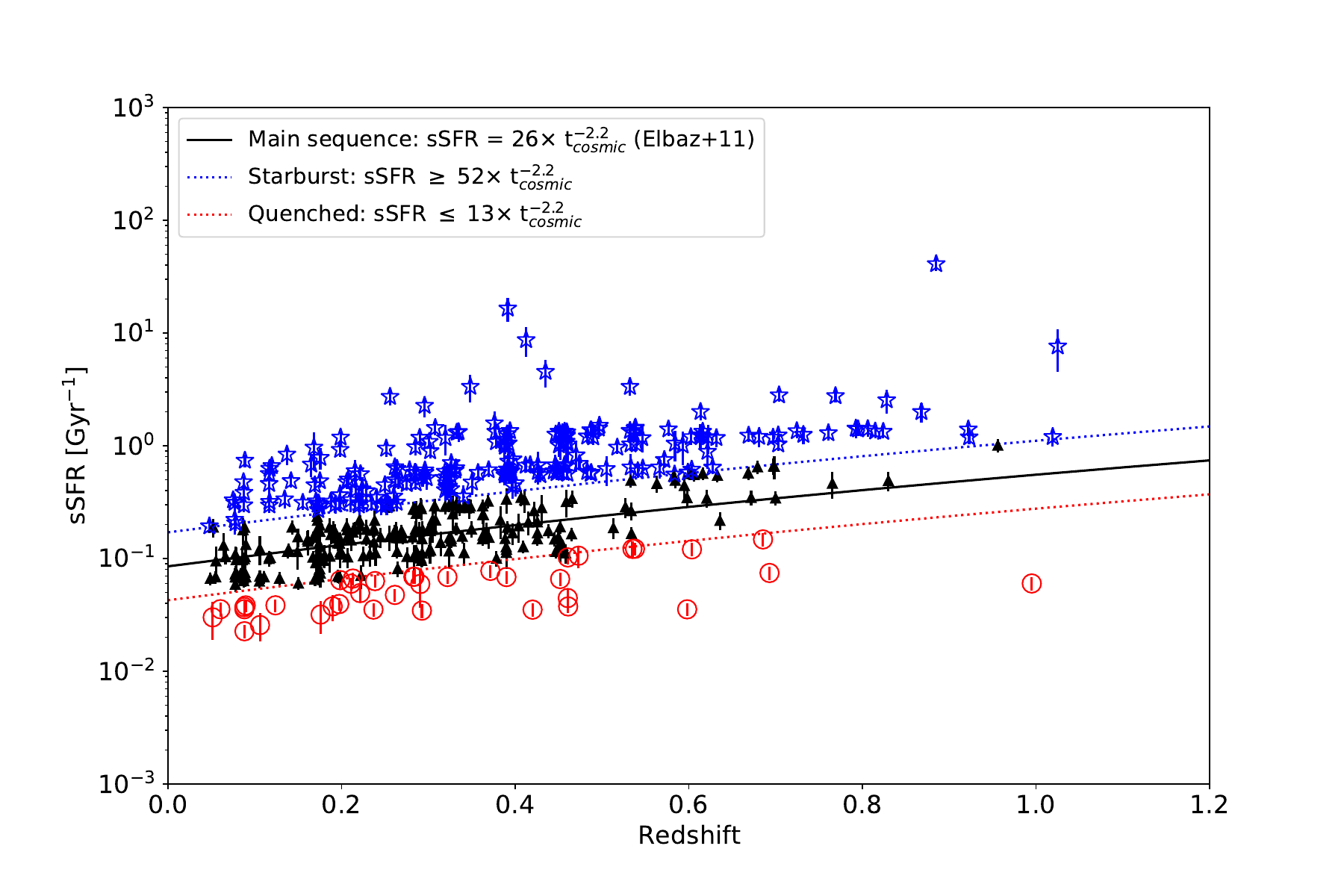}
	\caption{Specific star formation rate as a function of redshift for galaxies with H$\alpha$ and/or [O II] detections in the calibration sample. 
	\label{mainsequence}}
\end{figure*}

Figure \ref{mainsequence} shows the specific star formation rate (sSFR) as a function of redshift of all galaxy candidates with H$\alpha$ and/or [O II] detections. The sSFR, the mass-normalized star formation rate, is given by \citet{Kennicutt1998} as
\begin{equation}
sSFR\ [yr^{-1}]= [1.72\times10^{-10}\times(L(TIR)/L_{\odot})]/[(M_{*}/M_{\odot})/0.61],
\end{equation}
where $L(TIR)$ is the total IR luminosity of the best-fit SED integrated from rest-frame 8--1000 $\mu$m and (M$_{*}$/M$_{\odot}$)/0.61 is the stellar mass converted from \citet{Chabrier} to \citet{KennicuttIMF} IMF. ``Main-sequence" galaxies are defined as objects with $13\times t^{-2.2}_{cosmic}\leq$ sSFR [1/Gyr] $\leq 52\times t^{-2.2}_{cosmic}$, where $t_{cosmic}$ is the cosmic time elapsed since the Big Bang in Gyr \citep{Elbaz2011}. ``Starburst" galaxies and ``quenched" galaxies lie above and below this region, respectively. To classify starburst galaxies, we adopt the $R_{SB}$ starburst parameter from \citet{Elbaz2011}, which describes the ``starburst excess" relative to main-sequence star formation and is defined as the ratio of the sSFR to the main-sequence sSFR. With their definition, starburst galaxies have $R_{SB} \gtrsim 2$ (blue star symbols), main-sequence galaxies have $0.5\lesssim R_{SB} \lesssim 2$ (black triangle symbols), and quenched galaxies have $R_{SB}\lesssim 0.5$ (red circle symbols).

\begin{figure*}[ht!]
	\centering
        \includegraphics[width=\linewidth]{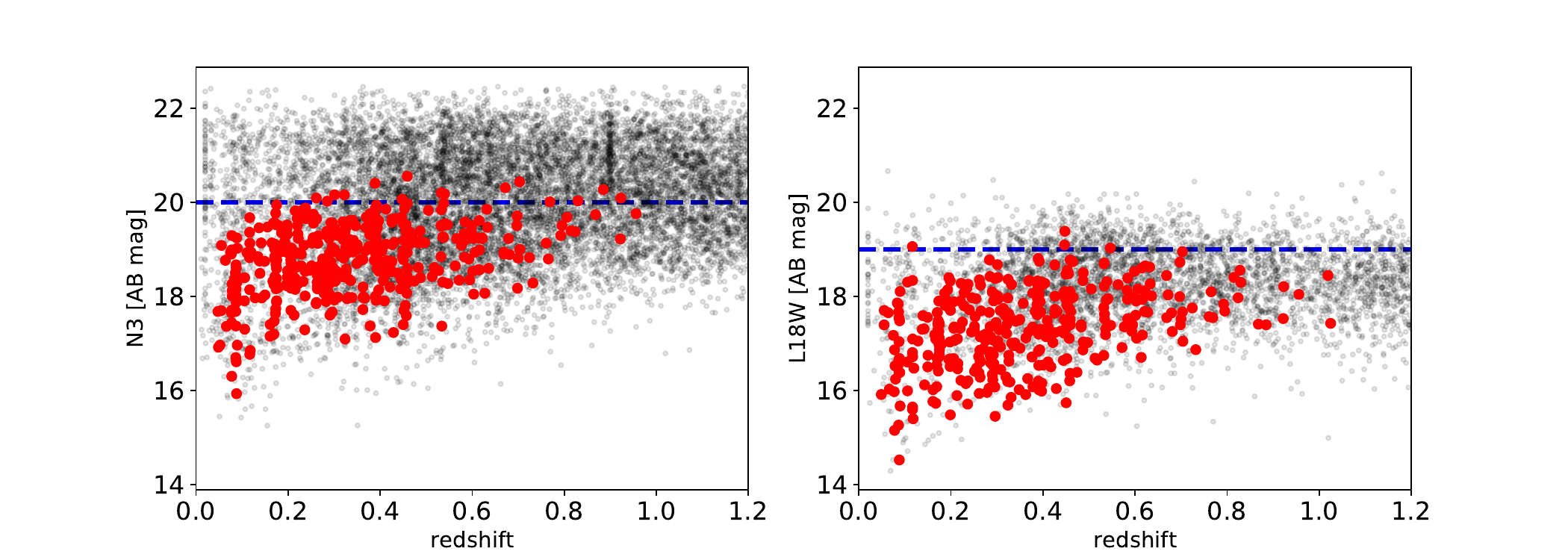}
	\caption{Observed \textit{AKARI}/IRC magnitudes at 3 and 18 $\mu$m (\textit{N3} and \textit{L18W} filters, respectively) as a function of redshift. Main-sequence and starburst galaxies from our calibration sample are shown as red circles. Observations from the \textit{AKARI} NEP-Deep field survey \citep{Oi2014} are shown as faint black dots. Blue dashed lines indicate approximate flux limits.
	\label{magnitude}}
\end{figure*}

In Figure \ref{magnitude}, we plot the observed \textit{N3} and \textit{L18W} magnitudes as a function of redshift for galaxies within our calibration sample (red circle symbols). Of the 443 star-forming galaxies in the calibration sample, 424 galaxies have detected N3 magnitudes (left panel), and 369 galaxies have detected L18W magnitudes (right). For reference, photometric detections of  galaxies in the \textit{AKARI}/IRC NEP-Deep field with photometric redshifts are shown as black dots \citep{Oi2014}. The blue dashed lines in the figure indicate that our sample is flux-limited down to $\simeq$ 20 AB mag at 3 $\mu$m and $\simeq$ 19 AB mag at 18 $\mu$m up to $z\sim0.7$, which corresponds to a stellar mass limit of $M_{*}\gtrsim 10^{9.5}$ $M_{\odot}$.

\subsection{Extinction-corrected H$\alpha$ and [O II] Luminosity} \label{extinction}
We calculate dust-corrected H$\alpha$ luminosities using three methods: the 24 $\mu$m rest-frame monochromatic luminosity ($\lambda$L$_{\lambda}$(24 $\mu$m)) correction, the Balmer decrement (H$\alpha$/H$\beta$), and the color excess E(B-V) derived with CIGALE. The empirical 24 $\mu$m correction to H$\alpha$ defined by \citet{Kennicutt1998} is given as:
\begin{equation}
    L(H\alpha)_{24\ \mu m} = L(H\alpha)_{obs} + 0.02\times \lambda L_{\lambda}(24\ \mu m)\ \text{[erg/s]}
\end{equation}
where $L(H\alpha)_{obs}$ is the observed H$\alpha$ luminosity and luminosities are in erg/s. We calculate $\lambda$L$_{\lambda}$(24 $\mu$m) as the rest-frame luminosity density of the best-fit SED model at 24 $\mu$m multiplied by the effective wavelength. The intrinsic H$\alpha$ luminosity is given by:
\begin{equation}
    L(H\alpha) = L(H\alpha)_{obs}\times 10^{0.4A(H\alpha)}
\end{equation}
where $A(H\alpha)$ is the attenuation in H$\alpha$ in magnitudes.
For star-forming galaxies with H$\alpha$ and H$\beta$ flux measurements,  $A(H\alpha)$ is given by the Balmer decrement (BD):
\begin{equation}
    A(H\alpha)_{BD}\ \text{[mag]} = 3.33\times1.97\times \text{log}\left(\frac{(F(H\alpha)/F(H\beta))_{obs}}{2.86}\right)
\end{equation}
where (F(H$\alpha$)/F(H$\beta$))$_{obs}$ is the observed flux ratio, and we assume the \citet{Calzetti00} attenuation law and an intrinsic flux ratio of 2.86 (Case B recombination and T=10$^4$ K; \citet{Osterbrock}). The SED-derived $A(H\alpha)$ is given by:
\begin{equation}
    A(H\alpha)_{CIGALE}\ \text{[mag]} = 2.45\times E(B-V)_{lines}
\end{equation}
where E(B-V)$_{lines}$ is the color excess of the nebular lines' light, and we assume a standard Milky Way extinction curve.

\begin{figure*}[ht!]
	\centering
	\includegraphics[width=\linewidth]{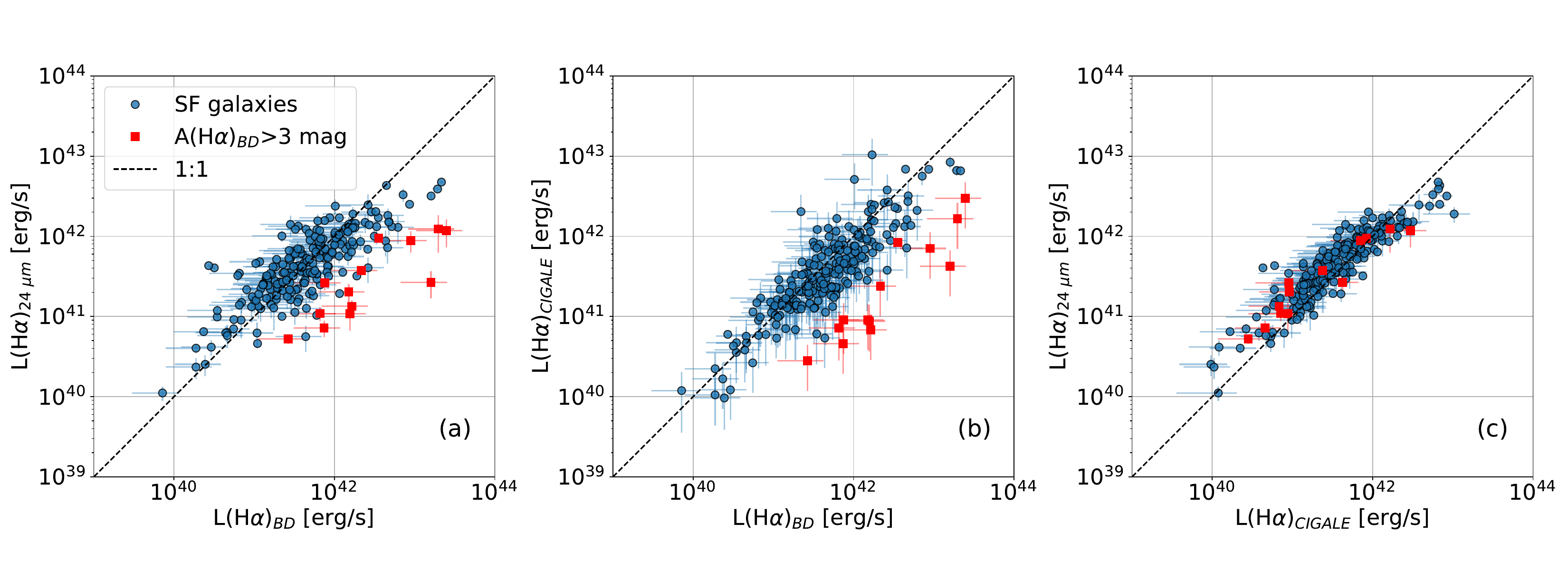}
	\vspace*{-10mm}
	\caption{Intercomparison of corrected H$\alpha$ luminosities derived from our three different de-reddening methods (Equations 6-9). The ``BD" subscript refers to the Balmer decrement (H$\alpha$/H$\beta$).
	\label{attenuation}}
\end{figure*}

Figure \ref{attenuation} shows a comparison of intrinsic H$\alpha$ luminosities for the galaxies in the calibration sample with H$\alpha$ and H$\beta$ flux measurements. Heavily dust-reddened objects with $A(H\alpha)>3$ mag as calculated via the Balmer decrement are shown by red squares. In these objects, the dust-corrected H$\alpha$ luminosity deviates significantly from those derived via the 24 $\mu$m luminosity and predicted CIGALE extinction, as shown in panels (a) and (b). However, as shown by panel (c), the corrections given by the 24 $\mu$m and CIGALE methods produce similar intrinsic H$\alpha$ luminosities, which is most likely due to their shared dependence on MIR photometry. The scatter between the Balmer decrement correction and 24 $\mu$m luminosity correction has been observed in other studies (e.g., \citealp{Boselli}, \citealp{Shipley2016}). They attribute the discrepancy as being due to uncertainty in the underlying stellar absorption correction for objects with weak H$\beta$ emission.

To correct the [O II] luminosity for dust extinction, we adopt the calibration given in \citet{Kennicutt1998}:
\begin{equation} 
    L([O\ II])_{24\ \mu m} = L([O\ II])_{obs} + 0.029\times \lambda L_{\lambda}(24\ \mu m),
\end{equation}
where the luminosities are in erg/s. The [O II] SFR is then given by \citet{Kennicutt1998} as:
\begin{equation} \label{eqn:sfroii}
    SFR\ [M_{\odot}/yr] = 8.1\times 10^{-42} L([O\ II])_{24\ \mu m}\ [erg/s]
\end{equation}
assuming a Salpeter IMF. We discuss limitations of using the [O II] luminosity to measure SFR in the Appendix. 
 
 \subsection{Measurement of PAH luminosity}
\begin{figure*}[ht!]
	\centering
	\includegraphics[width=\linewidth]{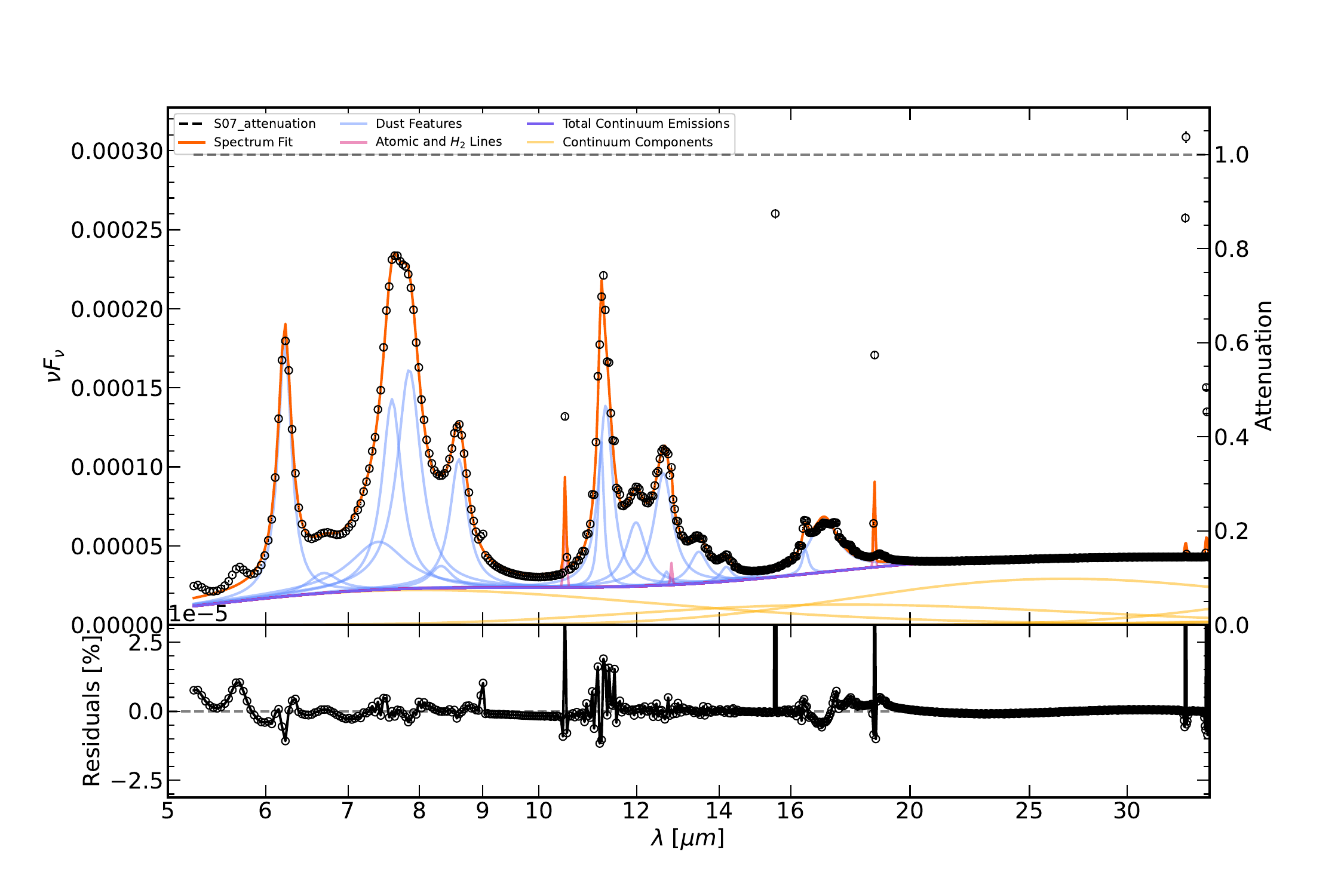}
	\vspace*{-10mm}
	\caption{Example of a typical PAHfit result for a main-sequence galaxy at $z$=0.36 (ID: J180123.2+654950.0; The SED is shown in Figure \ref{fig:cigale}(a)), fit to the model SED (circle points). The solid red line shows the best-fit result. Drude profile fits to the PAH dust emission features are shown in blue. The solid purple line underlying the emission represents the continuum, which is the summation of the individual blackbody components shown in yellow.
	\label{fig:pahfit}}
\end{figure*}

We input the rest-frame, best-fit SEDs from CIGALE into PAHFIT \citep{Smith07,Smith_PAHFIT} in order to model the dust emission and continuum from 5-35 $\mu$m, and to extract the PAH luminosities. The input best-fit SEDs include the attenuated stellar emission, nebular emission, and dust emission (i.e., modeled AGN emission is removed for AGN candidates). Then, we resample the model spectrum into the \textit{Spitzer}/IRS wavelength grid. PAHFIT uses a physical decomposition model to simulate the MIR continuum with a sum of modified blackbodies and starlight, and the PAH dust blends with Drude profiles. Because it does not take AGN emission into account, we only consider SED components from the total attenuated stellar light, nebular emission, and dust emission. For the PAH features, we fixed the central wavelengths and FWHM to their standard values and allowed the amplitudes to vary. Figure \ref{fig:pahfit} shows a typical PAHFIT result for a star-forming galaxy. The integrated intensity of the Drude profile is equal to 
\begin{equation}
    I^{(r)} = \frac{\pi c}{2}\frac{b_{r}\gamma_{r}}{\lambda_{r}},
\end{equation} where $\lambda_{r}$ is the central wavelength of the PAH feature, $\gamma_{r}$ is the fractional FWHM, and $b_{r}$ is the central amplitude \citep{Smith07}. For example, the luminosities for the PAH 6.2 $\mu$m and 7.7 $\mu$m emission features are given by:
\begin{equation}
L(\text{PAH}\ 6.2\ \mu m)\ [erg/s] = 10^7 \times 4\pi d_{L}^2  \frac{\pi c}{2} \times \left(\frac{0.030\ b_{6.22 \mu m}}{6.22\ \mu m}\right)
\end{equation}
\begin{eqnarray}
L(\text{PAH}\ 7.7\ \mu m)\ [erg/s] = 10^7 \times 4\pi d_{L}^2  \frac{\pi c}{2} \times \nonumber \\ \left(\frac{0.126\ b_{7.42 \mu m}}{7.42\ \mu m}
+ \frac{0.044\ b_{7.60\mu m}}{7.60\ \mu m} + \frac{0.053\ b_{7.85 \mu m}}{7.85\ \mu m}\right)
\end{eqnarray}
with amplitudes $b_{r}$ in units of  W/m$^2$/Hz, luminosity distance $d_{L}$ in meters, and $c$ in $\mu$m/s.

\subsubsection{Validation of PAH luminosity measurements} \label{sec:spicy}
\begin{figure*}[ht!]
	\centering
	\includegraphics[width=\linewidth]{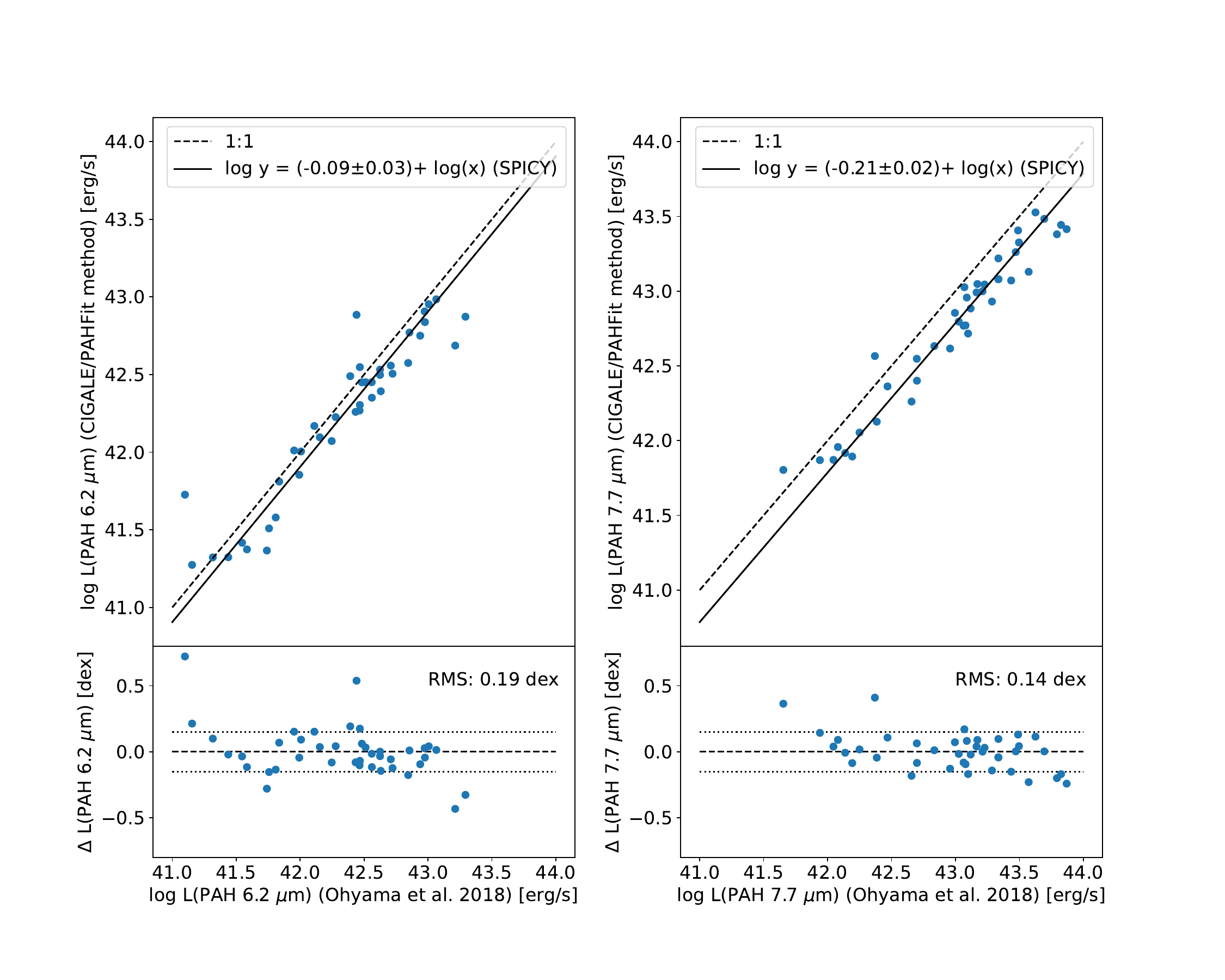}
	\vspace*{-10mm}
	\caption{Top panel: Comparison between measured PAH luminosity from Ohyama et al. (2018) SPICY sample and inferred luminosity from our CIGALE and PAHFIT method for 6.2 $\mu$m (left) and 7.7 $\mu$m (right) blends. The solid line represents the least-squares linear fit to the data with slope fixed to unity. Bottom panel: Residuals as a function of the measured PAH luminosity with the dotted lines indicating the RMS dispersion in dex.
	\label{fig:spicy_error}}
\end{figure*}

Figure \ref{fig:spicy_error} compares our PAH luminosity measurements for the 6.2 $\mu$m and 7.7 $\mu$m features to those by \citet{Ohyama2018} in their SPICY spectra of the same galaxies. The figure demonstrates the validity of our method for measuring $L(PAH\ 6.2\ \mu m)$ and $L(PAH\ 7.7\ \mu m)$, with systematic offsets of only $\sim0.1$ dex and $\sim0.2$ dex, respectively. The small root-mean-square (RMS) dispersions, $\sigma_{L(PAH\ 6.2\ \mu m)} =$ 0.19 dex and $\sigma_{L(PAH\ 7.7\ \mu m)} =$ 0.14 dex, represent the approximate errors in our PAH luminosity measurements. Our measurements of $L(PAH\ 7.7\ \mu m)$ have a larger systematic offset from Ohyama et al.'s measurements compared to those of $L(PAH\ 6.2\ \mu m)$, which is likely due to the difficulty in constraining the underlying continuum beneath the wide 7.7 $\mu$m blend in noisy spectra. Their method uses a single broad Lorentzian to fit the 7.7 $\mu$m blend from low resolution spectra, which may overestimate the flux somewhat. We note that although \citet{Ohyama2018} initially fit some IRC spectra of brighter sources with PAHFIT, they adopted a simpler model that could better detect faint PAH features in low-resolution IRC spectra ($R\simeq 50$), compared to the \textit{Spitzer}/IRS spectra ($60<R<130$) for which PAHFIT was designed.

We further tested the importance of mid-infrared photometry on SED fitting by re-running CIGALE on the sample of 48 SPICY galaxies without the MIR \textit{AKARI}, \textit{WISE}, and \textit{IRAC} bands. We then compared the output $q_{PAH}$ and SFR values with and without the MIR data. When comparing the $q_{PAH}$ parameter, the results produced a scatter diagram, highlighting the necessity of including MIR data in determining $q_{PAH}$ (and PAH luminosity). The SFRs derived from CIGALE without MIR data were comparable to the original results for about 50\% of the galaxies; however, for the remaining half, the SFRs calculated without MIR data were highly uncertain because these galaxies also lacked UV and FIR photometry.

In addition to the PAH luminosity, we also calculated the peak luminosity at 7.7 $\mu$m, $\nu L \nu (7.7 \mu$m), which is sometimes used as a proxy for the PAH luminosity in the literature because it can be estimated photometrically (e.g., \citealp{Takagi2010}). The relation for the peak luminosity is:
\begin{equation}
 \nu L_{\nu}(7.7 \mu m) = F_{\nu}(7.7 \mu m) \times 4\pi d_{L}^2\nu(7.7 \mu m),
\end{equation}
where $F_{\nu}(7.7 \mu m)$ is the best-fit spectrum from PAHFIT evaluated at 7.7 $\mu$m (i.e., the sum of the blackbody continuua and Drude profiles at 7.7 $\mu$m) and $\nu(7.7 \mu m)$ is the frequency at 7.7 $\mu$m in Hz. We use these peak luminosities in Section \ref{sec:disc} to compare our results to \citet{Takagi2010}.

The convenience of estimating PAH flux from measurements in just a few broadband filters, particularly the IRAC-4 8 $\mu$m band, has been studied by many authors. For example, \citet{Figueira2022} attempted this with a limited set of IRAC and MIPS photometry. However, they found that the rest-frame 8 $\mu$m band luminosity did not correlate well with SFR, and the logarithmic slope they found, 0.81, was much less than linear. They speculated that their difficulty may have stemmed from their inability to separate the PAH 7.7 $\mu$m flux from the underlying continuum components in that wavelength region.

\begin{figure*}[ht!]
    \centering
    \includegraphics[scale=0.5]{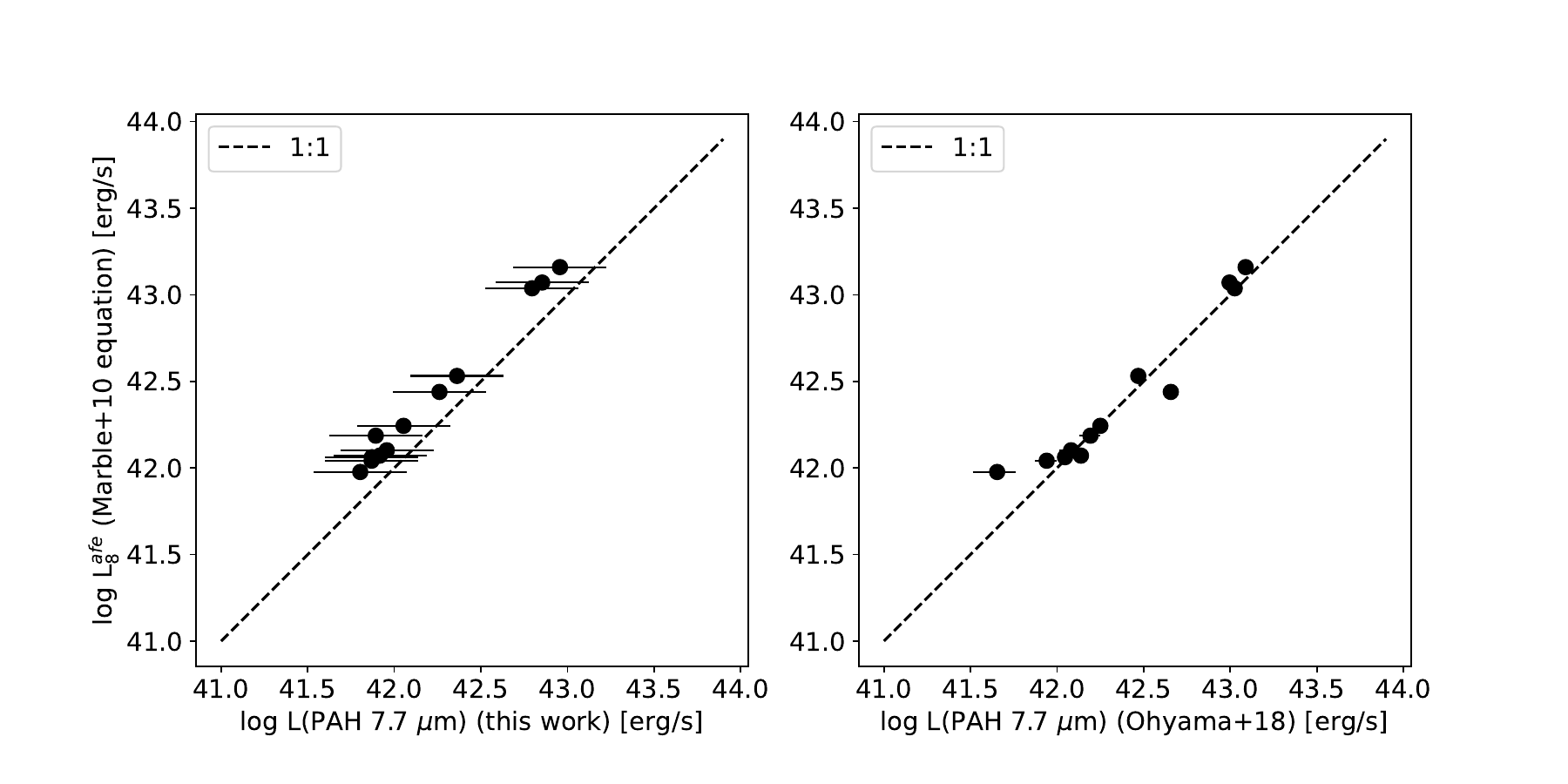}
    \vspace*{0mm}
    \caption{Comparison between the aromatic feature emission luminosities at 8 $\mu$m from \citet{Marble2010} and our PAH 7.7 $\mu$m luminosities (left panel) and those of \citet{Ohyama2018} (right panel) for SPICY galaxies at $z<0.1$. The RMS scatter in the left panel is 58\% (or, approximately 10\% after a 0.2 dex offset is applied to the residuals), while the RMS scatter in the right panel is 32\%.
    \label{marble}}
\end{figure*}

\citet{Marble2010} derived PAH fluxes for a local sample of SINGS galaxies based on the \textit{Spitzer}/IRAC 3.6, 4.5, 5.8, and 8 $\mu$m bands and MIPS 24 $\mu$m. We compare our 7.7 $\mu$m PAH luminosity to their photometric-based PAH luminosity, known as the ``aromatic feature emission" of the 8 $\mu$m blend ($L_{8}^{afe}$), by selecting SPICY galaxies with $z<0.1$ and estimating the observed \textit{Spitzer} flux densities with CIGALE. The resulting sub-sample contains 13 objects. Figure \ref{marble} shows the comparison between $L_{8}^{afe}$ and our measured $L(PAH\ 7.7\ \mu m)$ (left panel) and $L(PAH\ 7.7\ \mu m)$ from \citet{Ohyama2018}. Our results are consistent with \citet{Marble2010}, with a slight offset of $\sim 0.15-0.2$ dex. This offset is likely due to $L_{8}^{afe}$ including the 8.33 and 8.61 $\mu$m features in addition to the 7.7 $\mu$m blend, leading to an overestimation of the 7.7 $\mu$m luminosity. The result from \citet{Ohyama2018} is likely more consistent with \citet{Marble2010} due to the simplified Lorentzian profile that was used to fit the 7.7 $\mu$m feature, again leading to a slight overestimate in the flux density.

\subsubsection{Limitations on constraining PAH line ratios}
Throughout this work, we present our results using both the 6.2 $\mu$m and 7.7 $\mu$m luminosities. It is important to note that these luminosities and their resulting SFR calibrations are \textit{not} completely independent of each other due to our use of the CIGALE templates. In all
of these templates, $L(PAH\ 6.2\ \mu m)/L(PAH\ 7.7\ \mu m) \simeq 0.3$. A separate photometric approach would be needed to constrain the variability in 6.2 $\mu$m/7.7 $\mu$m. However, assuming that the SED templates give descriptions of the PAH dust emission in star-forming galaxies, which we tested with the SPICY sample, we believe that it is valuable to present both indicators for future research, with this caveat. For example, the 6.2 $\mu$m feature is narrower than the 7.7 $\mu$m blend and may be easier to measure in some cases.

\section{Analysis} \label{sec:analysis}

\subsection{Effects of starburst intensity, metallicity, and AGN fraction}

\begin{figure*}[ht]
	\centering
        \includegraphics[width=\linewidth]{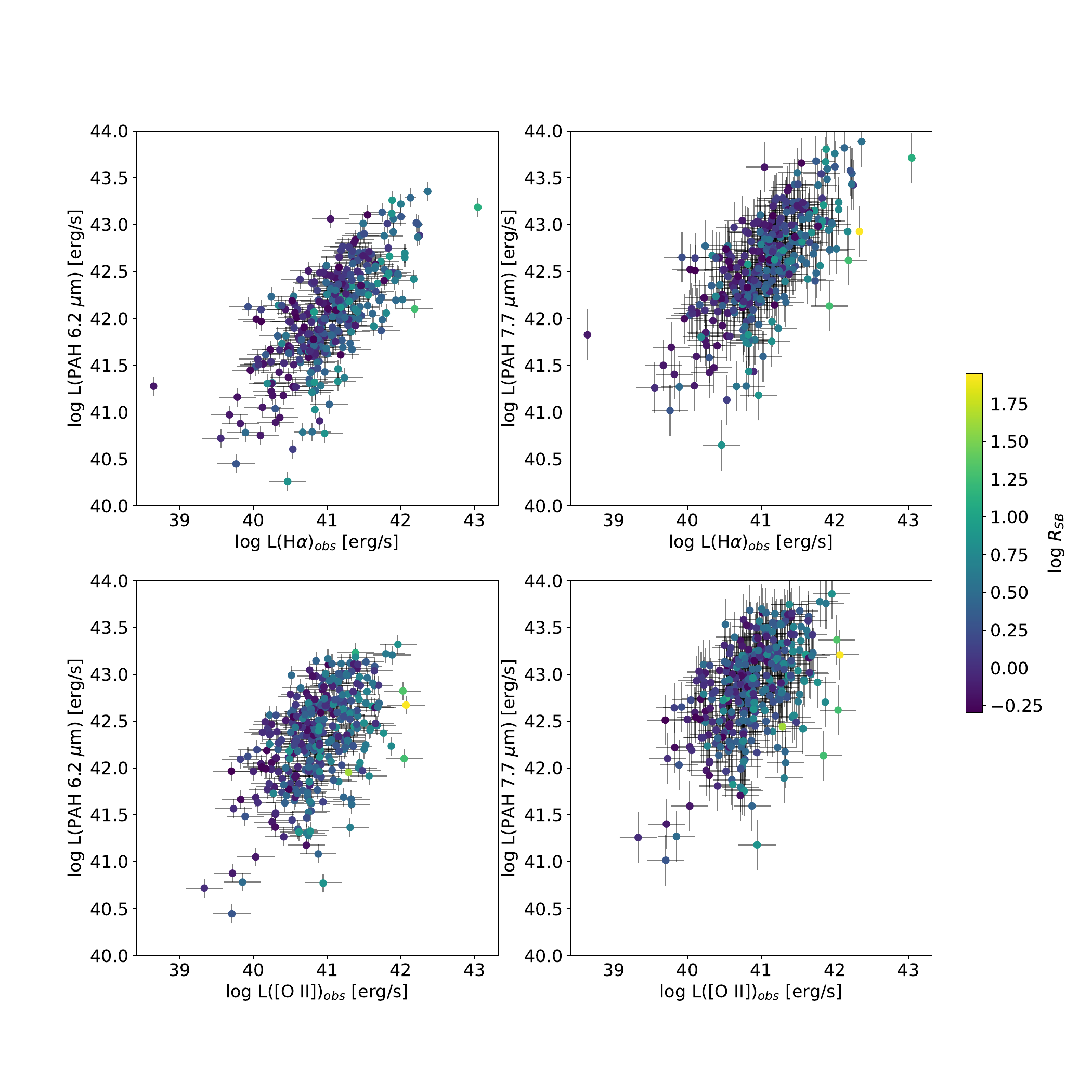}
	\vspace*{-18mm}
	\caption{PAH 6.2 $\mu$m (left column) and 7.7 $\mu$m (right column) luminosities vs. {\it observed} H$\alpha$ and [O II] luminosities in star-forming galaxies, colored by the logarithm of the starburst intensity, $R_{SB}$.
	\label{fig:lpahlhaobs}}
\end{figure*}

To determine how well PAH luminosities predict SFR, we correlate the PAH 6.2 $\mu$m and 7.7 $\mu$m luminosities with the reddening-corrected H$\alpha$ and [O II] luminosities. Figure \ref{fig:lpahlhaobs} shows the PAH 6.2 $\mu$m and 7.7 $\mu$m luminosity as a function of observed (reddened) H$\alpha$ and [O II] luminosity. The four panels show a positive correlation between PAH luminosity and observed H$\alpha$ and [O II] luminosity, with considerable scatter due to reddening effects which have not yet been taken into account. For reference, the symbols are colored by the logarithm of the starburst intensity, $R_{SB}$. Despite the large scatter, there is a visible trend such that galaxies with higher $R_{SB}$ generally have depreciated PAH luminosity for a given observed H$\alpha$ or [O II] luminosity (or enhanced H$\alpha$ or [O II] for a given PAH luminosity).

\begin{figure*}[ht!]
	\centering
	\includegraphics[width=\linewidth]{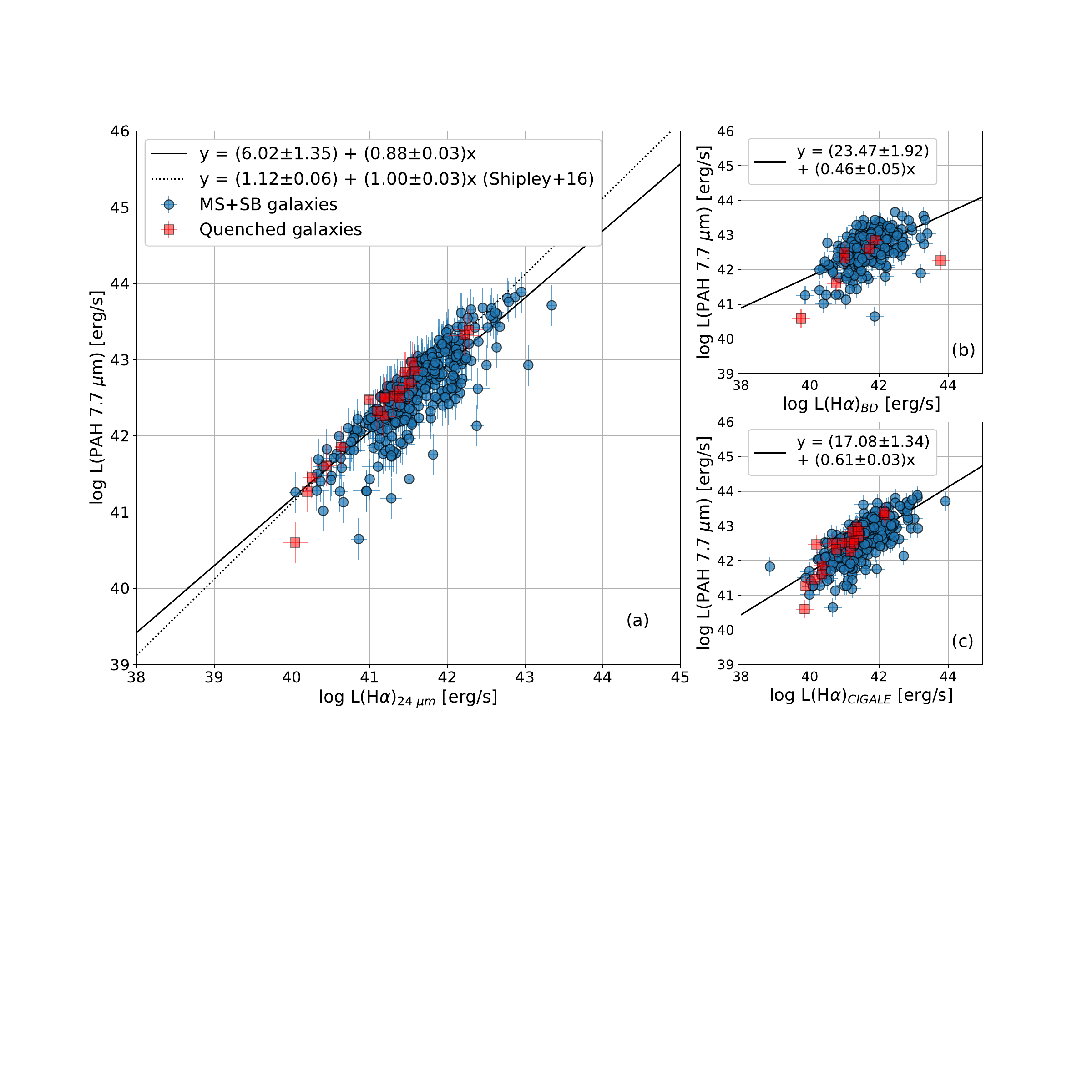}
	\vspace*{-65mm}
	\caption{Comparison of PAH 7.7 $\mu$m luminosity vs. intrinsic H$\alpha$ luminosity of main-sequence and starburst galaxies for different dust extinction correction methods: (a) rest-frame 24 $\mu$m luminosity correction from \citet{Kennicutt_2009}; (b) Balmer decrement; (c) SED fitting. The solid lines represent linear fits.
	\label{lpah_lhalpha_corr}}
\end{figure*}

Figures \ref{lpah_lhalpha_corr} shows the PAH 7.7 $\mu$m luminosity as a function of the de-reddened H$\alpha$ luminosity. We fit linear relations of the form log $L(PAH\ 7.7\ \mu m)$ = $c_{0}$ + $c_{1}$ $\times$log $L(H\alpha)$ to the star-forming sample. Quenched galaxies (i.e., $R_{SB}<0.5$; see Section \ref{sec:calibsample}) are shown for reference as red square symbols but are excluded from the fits. The three panels in Figure \ref{lpah_lhalpha_corr} show results for different methods of correcting the observed H$\alpha$ luminosity as discussed in Section \ref{extinction}. Compared to Figure \ref{fig:lpahlhaobs}, the scatter between $L(PAH\ 7.7\ \mu m)$ and $L(H\alpha)_{24\ \mu m}$ is significantly reduced; the RMS dispersion of the fit is 0.31 dex. The dotted line in panel (a) represents the linear relation from \citet{Shipley2016} derived from \textit{Spitzer}/IRS galaxies with solar metallicity at $z<0.4$. For the majority of galaxies, $L(PAH\ 7.7\ \mu m)$ is linearly correlated with $L(H\alpha)_{24\ \mu m}$ (i.e., slope $\sim$ 1). However, we find that a small minority, 47/319 (15\%), of the galaxies deviate from the fit by -0.4 dex, contributing to the uncertainty in the normalization and non-linear slope. This deviation is consistent with \citet{Calzetti07}, who attribute it to low metallicity and decline to provide a calibration for 8 $\mu$m emission as a local star formation rate indicator on this basis. Panels (b) and (c) show that the correlations between $L(PAH\ 7.7\ \mu m)$ and $L(H\alpha)$ are non-linear with the Balmer decrement and SED fitting corrections, with the Balmer decrement method resulting in the most scatter.

The majority of the quenched galaxies shown in panel (a) of Figure \ref{lpah_lhalpha_corr} (red square symbols) lie along or slightly above the linear correlation between $L(PAH\ 7.7\ \mu m)$ and dust-corrected $L(H\alpha)$, which is consistent with the physical scenario where intense starbursts lead to the destruction of PAH dust grains due to UV dissociation \citep{Draine07}; in an environment with low $R_{SB}$, there are significantly fewer young, massive O and B stars compared to actively star-forming galaxies. Based on a sample of local SDSS galaxies, \citet{Smercina2018} found that the dominant PAH emission in post-starburst galaxies can overestimate the SFR relative to traditional ionized gas tracers in the mid-infrared that are typically present in H II regions (e.g., [Ne II]+[Ne III]). \citet{French2023} similarly found that the TIR luminosity overestimates the SFR relative to H$\alpha$, [Ne II]+[Ne III], and other ionized gas tracers when the star formation history is abruptly truncated, which is believed to be the scenario in post-starburst galaxies.

\begin{figure*}[ht!]
	\centering
        \includegraphics[width=\linewidth]{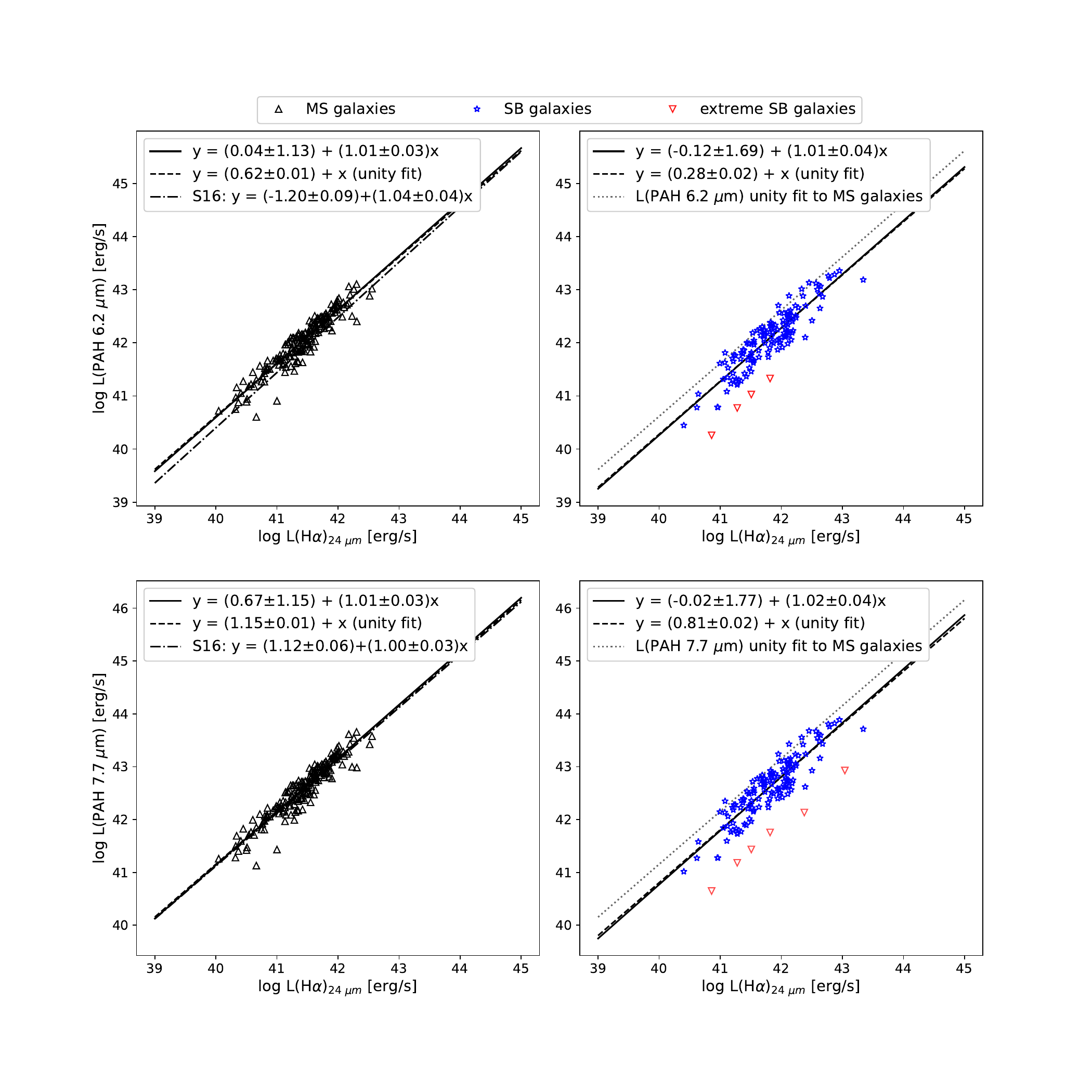}
	\vspace*{-20mm}
	\caption{Relation between PAH 6.2 $\mu$m and 7.7 $\mu$m luminosities and intrinsic H$\alpha$ luminosity for star-forming galaxies. Main-sequence galaxies (black triangles) are shown in the left column, and starburst galaxies (blue stars) are shown in the right column. ``Extreme" starburst galaxies (i.e., $L(TIR)/L(8\ \mu m)>20$) are shown as red triangles. The solid and dashed lines represent linear and fixed unity fits, respectively. The dash-dotted line represents the linear relation given by \citet{Shipley2016}. 
	\label{lpah_ms_sb}}
\end{figure*}

Figure \ref{lpah_ms_sb} shows the correlations between the PAH 6.2 $\mu$m and 7.7 $\mu$m luminosities with the dust-corrected H$\alpha$ luminosity for the star-forming sample, with main-sequence and starburst galaxies fit separately. For main sequence galaxies, the linear and unity slope fits for the PAH 7.7 $\mu$m luminosities are consistent with the relations found by Shipley et al. The linear relations for $L(PAH\ 6.2\ \mu m)$ in main-sequence galaxies are also consistent within $\sim$0.25 dex. The figure demonstrates that starburst galaxies have lower PAH luminosities for a given intrinsic H$\alpha$ luminosity. If we assume unity slope in the linear fits, the trend for starburst galaxies has a systematic offset of $\sim$0.35 dex relative to that of main-sequence galaxies for both PAH 6.2 $\mu$m and 7.7 $\mu$m features, suggesting that starburst galaxies are either PAH-deficient for fixed H$\alpha$ luminosity or H$\alpha$-enhanced for fixed PAH luminosity.

\begin{figure*}[ht!]
	\centering
        \includegraphics[width=\linewidth]{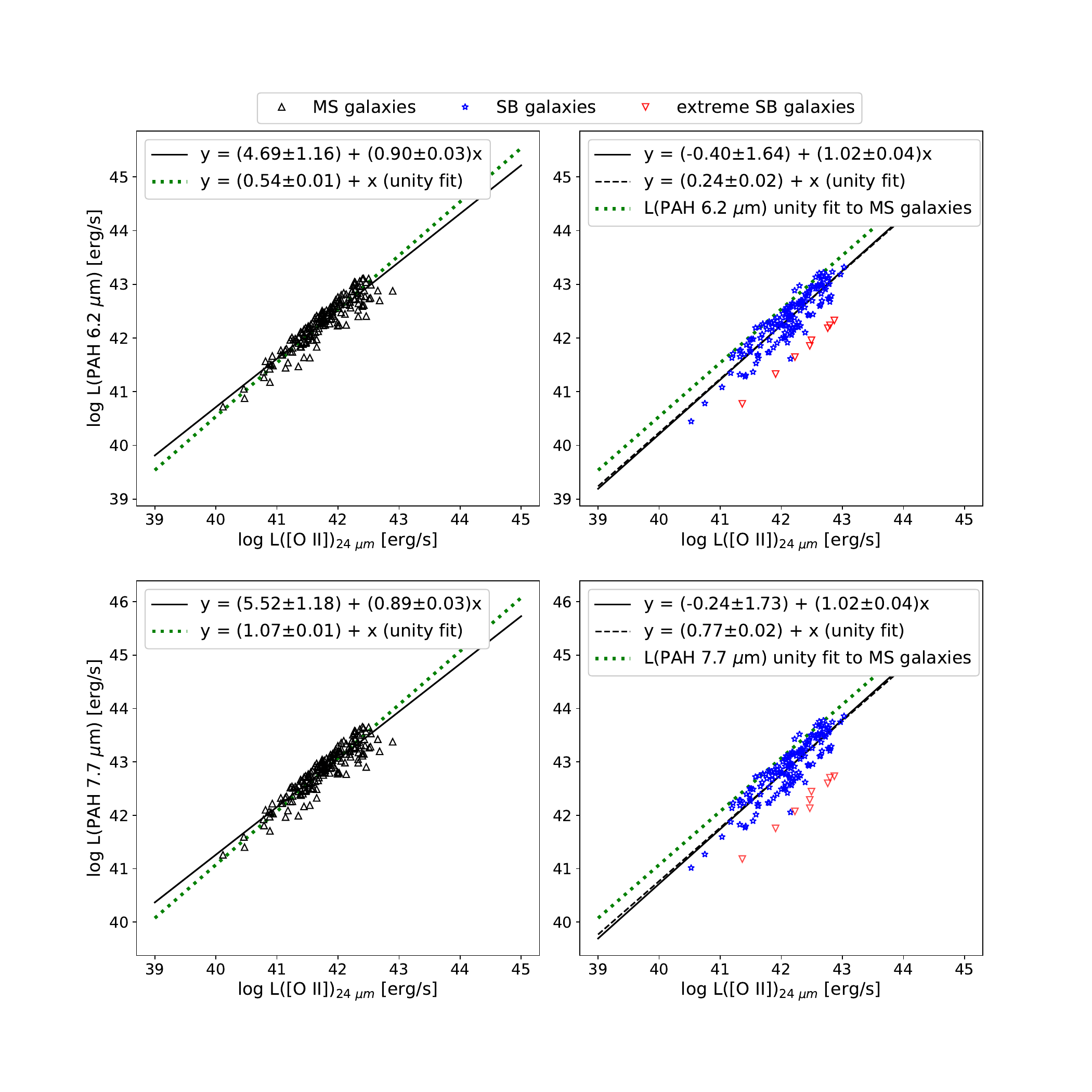}
	\vspace*{-15mm}
	\caption{Relations between PAH 6.2 $\mu$m and 7.7 $\mu$m luminosities and intrinsic (de-reddened) [O II] luminosities for star-forming galaxies. Symbols and lines are the same as for Figure \ref{lpah_ms_sb}.
	\label{lpah_ms_sb_oii}}
\end{figure*}

The offset between starburst galaxies and main-sequence galaxies is also present in the correlation between PAH luminosity and dust-corrected [O II] luminosity, as shown in Figure \ref{lpah_ms_sb_oii}. Starburst galaxies are systematically lower by a factor of 0.3 dex. We also find that linear fits to the main-sequence sample of the form log $L(PAH) = c_{0} + c_{1} \times$ log $L([O II])$ result in a shallower slope compared to fits with respect to $L(H\alpha)$ such that $L(PAH) \propto L([O II])^{0.9}$.

\begin{figure*}[ht!]
	\centering
        \includegraphics[width=\linewidth]{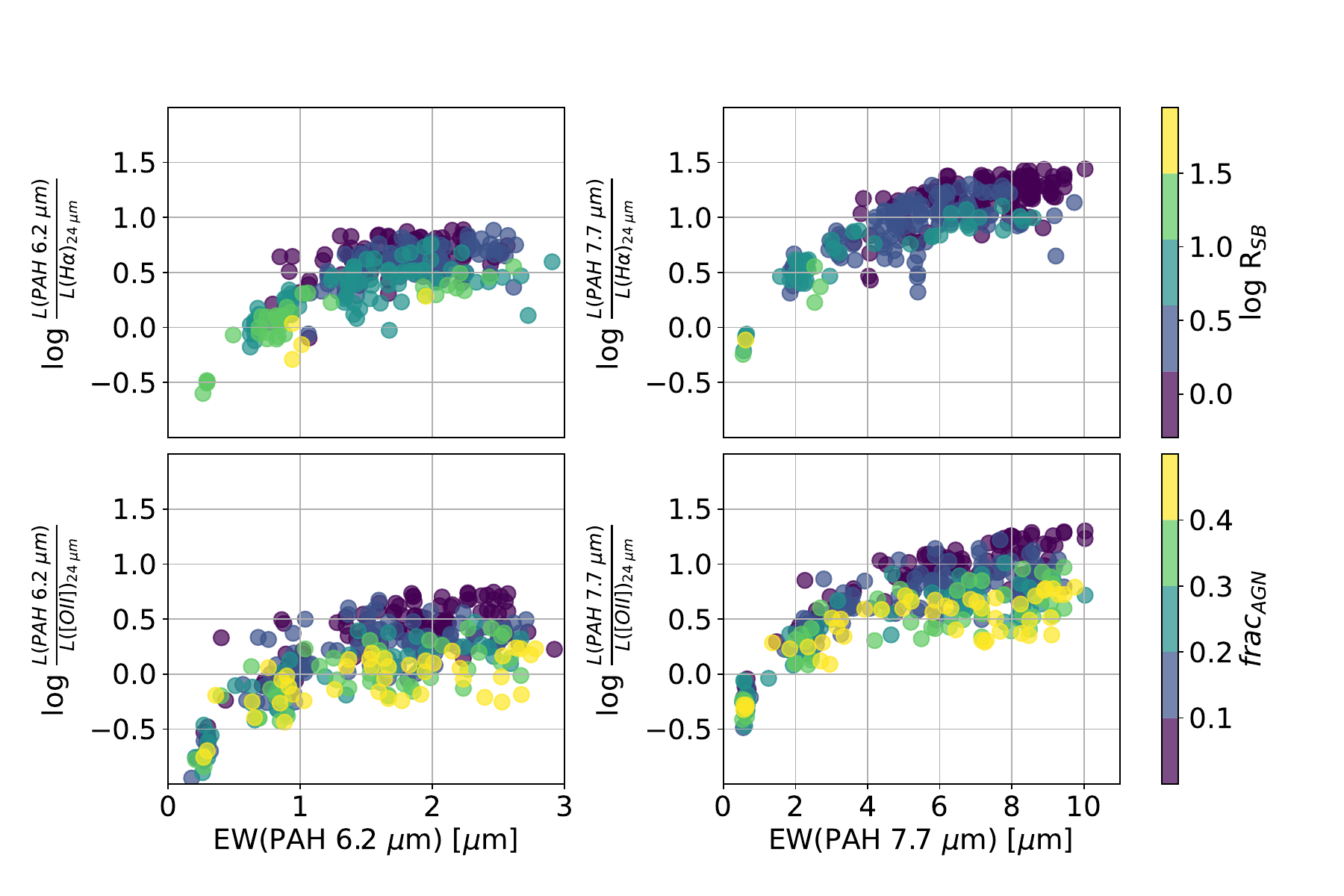}
	\vspace*{-5mm}
	\caption{Effects of starburstiness (log ratio of SFR to main sequence, $R_{SB}$, shown with color bar on top row) and AGN fraction (bottom row) on the correlation between PAH luminosity and intrinsic H$\alpha$ and [O II] luminosity.}
	\label{fig:lpahew}
\end{figure*}

We further examine the dependence of PAH luminosity on starburst intensity and AGN strength by plotting the log ratio of PAH luminosity to intrinsic H$\alpha$ and [O II] luminosity as a function of PAH equivalent width (EW) in Figure \ref{fig:lpahew}. The equivalent width is calculated as the flux of the PAH emission feature (given by PAHFIT) divided by the flux density of the continuum components, which includes the starlight, evaluated at the central wavelength of the feature. We note that it is difficult to constrain the values for $EW(PAH\ 7.7\ \mu m)$ for individual galaxies because the calculation relies heavily on assumptions for the underlying continuum shape and blend of Drude profiles. However, the exact equivalent widths do not affect our overall interpretations. The top row of the figure shows the correlations for star-forming galaxies, color-coded by starburst parameter. The bottom row shows the correlations for AGN candidates, color-coded by AGN fraction. Objects with low PAH equivalent width (i.e., $EW(PAH\ 6.2\ \mu m)\lesssim 1\ \mu$m or $EW(PAH\ 7.7\ \mu m)\lesssim$ $4\ \mu$m) scatter downwards from the $L(PAH\ 6.2/7.7\ \mu m)$ and dust-corrected $L(H\alpha)$/$L([O\ II])$ fits. The figure demonstrates that galaxies with higher starburst intensity ($R_{SB}$) have lower $L(PAH\ 6.2\ \mu m)$ and $L(PAH\ 7.7\ \mu m)$ for a given $L(H\alpha)_{24\ \mu m}$ and that AGN with higher AGN fraction have lower $L(PAH\ 6.2\ \mu m)$ and $L(PAH\ 7.7\ \mu m)$ for a given $L([O II])_{24\ \mu m}$. The offset such that galaxies with high $frac_{AGN}$ have low log $L(PAH)/L([O\ II])$ across all equivalent widths is likely due to the destruction of PAH dust grains by hard radiation from AGN. Given that ionization from AGN limits the efficacy of using the intrinsic H$\alpha$ and [O II] luminosities to trace star formation, we exclude AGN from all SFR calibrations.

\begin{figure*}[ht!]
	\centering
        \includegraphics[width=\linewidth]{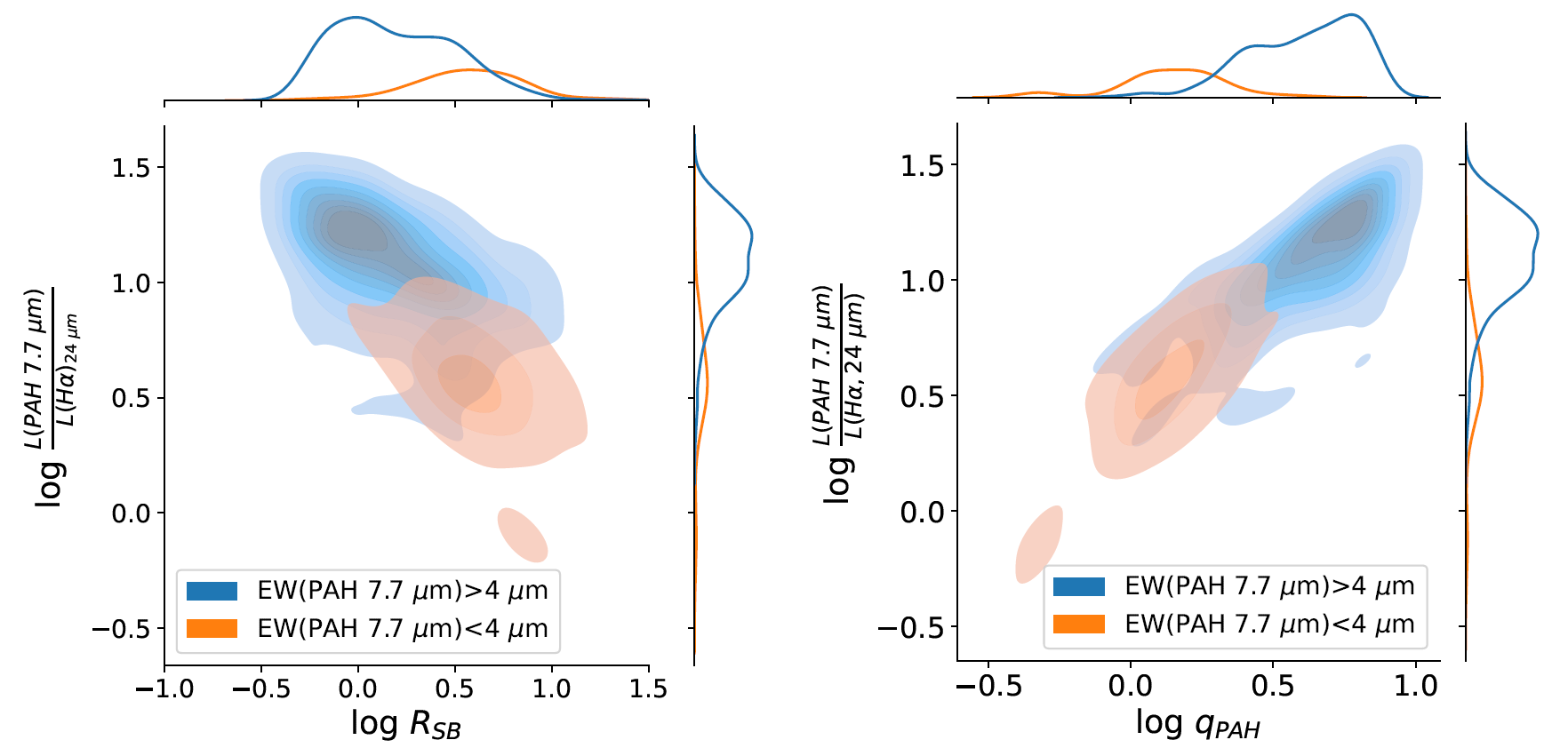}
	\caption{Kernel density estimate of the PAH luminosity to intrinsic H$\alpha$ luminosity as a function of starburstiness (left) and $q_{PAH}$ (right) for main-sequence and starburst galaxies. The data are grouped according to the estimated PAH 7.7 $\mu$m equivalent width. Marginal distributions are shown on the top and right sides of each plot.}
	\label{fig:marginal2}
\end{figure*}

To further illustrate the effect of starburstiness on PAH intensity, we show the kernel density estimate and marginal distributions of the PAH 7.7 $\mu$m luminosity to intrinsic H$\alpha$ luminosity as a function of $R_{SB}$ on the left column of Figure \ref{fig:marginal2}. The log $L(PAH\ 7.7\ \mu m)$/$L(H\alpha)_{24\ \mu m}$ ratio is anti-correlated with $R_{SB}$. In addition, galaxies with higher $R_{SB}$ are associated with lower PAH 7.7 $\mu$m equivalent width ($\lesssim 4$ $\mu$m), as shown by the orange contours. The Spearman rank-order correlation coefficient between $EW(PAH\ 7.7\ \mu m)$ and $R_{SB}$ is -0.48 with p-value of $2\times10^{-26}$, indicating a moderate monotonically decreasing correlation. The anti-correlation between log $L(PAH)/L(H\alpha)$ and $R_{SB}$ is consistent with the scenario in which PAH dust grains are destroyed by UV dissociation in compact star-forming regions \citep{Peeters, Murata2014}. For example, \citet{Murata2014} observe a deficit in $\nu L(8)/\nu L(4.5)$ for star-forming galaxies with log $R_{SB}>0.5$, which they interpret as a deficit of PAH emission in starburst galaxies.

The right column of Figure \ref{fig:marginal2} illustrates the correlation between log $L(PAH\ 7.7\ \mu m)$/$L(H\alpha)_{24\ \mu m}$ and $q_{PAH}$ parameter derived from best-fit SED models. The $q_{PAH}$ parameter is highly correlated with PAH equivalent width; the Spearman rank-order correlation coefficient between $EW(PAH\ 7.7\ \mu m)$ and $q_{PAH}$ is 0.8 with a p-value of $1.6\times10^{-98}$. We note, however, that the strong linear correlation between log $L(PAH\ 7.7\ \mu m)$/$L(H\alpha)_{24\ \mu m}$ and $q_{PAH}$ was to be expected given the correlation between $L(PAH)$ and $q_{PAH}$ based on assumptions in the SED modeling.

\begin{figure*}[ht!]
	\centering
        \includegraphics[width=\linewidth]{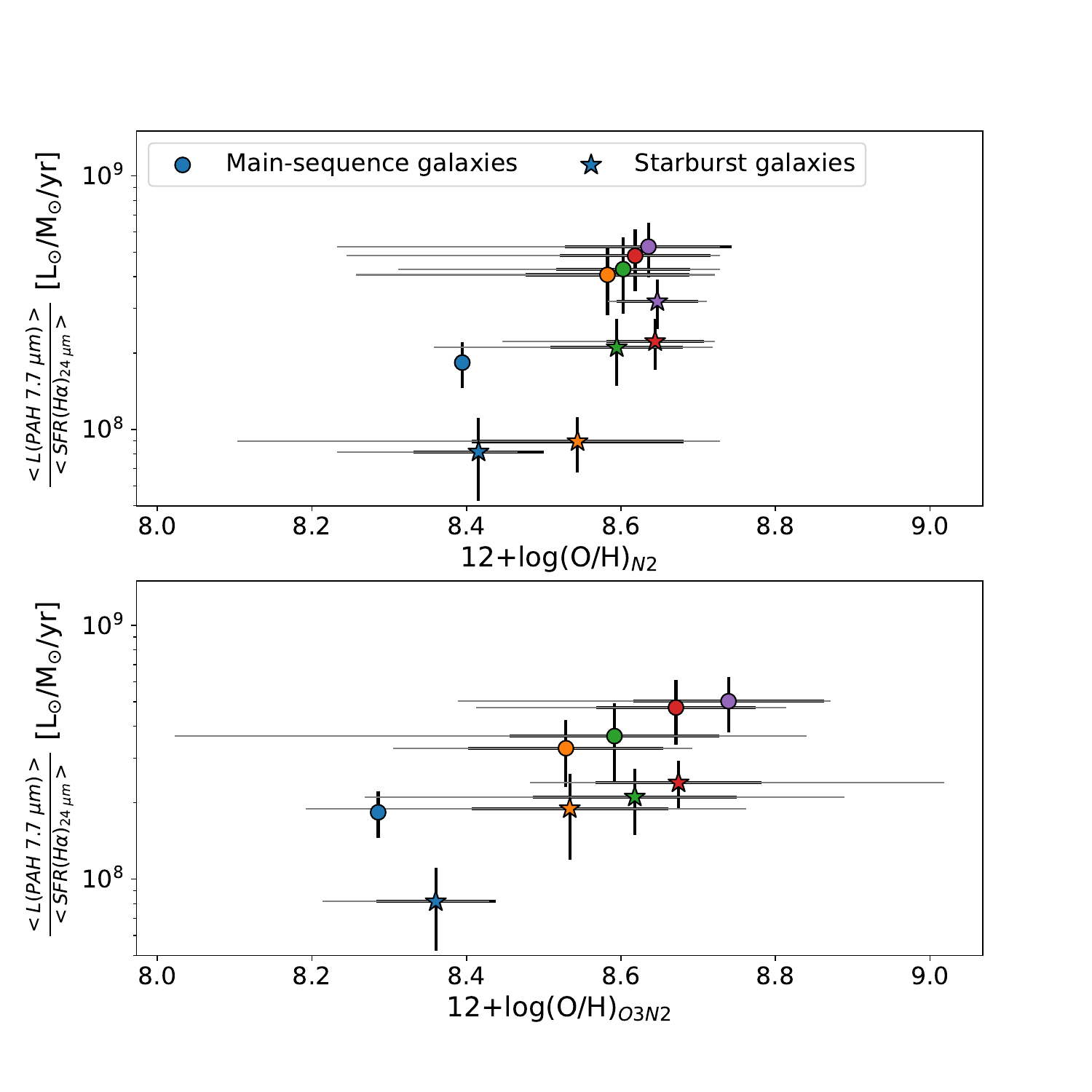}
	\vspace*{-15mm}
	\caption{Ratio of the PAH 7.7 $\mu$m luminosity to the dust-corrected H$\alpha$ SFR, both averaged over stellar mass, as a function of average metallicity using the N2 (top) and O3N2 (bottom) indices. Average metallicity is calculated as $<$12+log(O/H)$>$. Symbols are color-coded based on stellar mass bins as follows, with metallicity monotonically increasing with stellar mass: $M/M_{\odot}<10^{9}$ (blue), $10^{9} < M/M_{\odot} < 10^{9.5}$ (orange), $10^{9.5} < M/M_{\odot} < 10^{10}$ (green), $10^{10} < M/M_{\odot} < 10^{10.5}$ (red), $ M/M_{\odot} > 10^{10.5}$ (purple). Light grey horizontal lines indicate the metallicity ranges per bin.
	\label{fig:lpahmetal}}
\end{figure*}

In Figure \ref{fig:lpahmetal}, we investigate the dependence of PAH intensity on metallicity. Here, we calculate the PAH intensity as the ratio of $L(PAH\ 7.7\ \mu m)$ to $SFR(H\alpha)_{24\ \mu m}$, both averaged over stellar mass. The stellar masses range from $M/M_{\odot}<10^{9}$ to $M/M_{\odot}>10^{10.5}$ in bins of 0.5 dex. The figure shows the correlation between PAH intensity and average metallicity, $<$12+log(O/H)$>$, calculated using either the N2 (top panel) or O3N2 (bottom panel) index. Main-sequence galaxies (i.e., $0.5<R_{SB}<2$) and starburst galaxies (i.e., $R_{SB}>2$) are shown as circle and star symbols, respectively. Table \ref{table:bin_numbers} lists the number of main-sequence and starburst galaxies included in each bin. There is a positive correlation between PAH 7.7 $\mu$m intensity and metallicity in both main-sequence and starburst galaxies, with starburst galaxies having systematically lower PAH 7.7 $\mu$m intensity per stellar mass. 

There has been some concern that these ``strong-line" ratios
might give systematically incorrect metallicities. One
possibility, for example, is that the N/O ratio 
in strongly star-forming galaxies might deviate from the solar value \citep{Masters2014, Henry2021, Spinoglio2022}. However, we do not find any clear systematic differences between the metallicities estimated with or without the [OIII] line in Figure \ref{fig:lpahmetal}. Another concern is that electron-temperature based metallicities (using the
[OIII]4363 emission line which is too weak to measure in our spectra) might yield systematically lower metallicities than our strong-line methods \citep{Shin2021, Ly2016a, Ly2016b}. However, even if a 
systematic offset of a few tenths of a dex were applied to all of our [O/H] estimates, our finding of a relative increase of PAH strength with metallicity is not significantly changed, other than in the normalization.

\begin{deluxetable}{cllll}
\tablenum{5}
\tabletypesize{\footnotesize}
\tablecolumns{6}
\tablewidth{0pt}
\tablecaption{Number of star-forming galaxies in each stellar mass bin to calculate PAH intensity as a function of average metallicity in Figure \ref{fig:lpahmetal}. \label{table:bin_numbers}}
\tablehead{
\colhead{stellar mass bin} & \multicolumn{2}{c}{N2 index} & \multicolumn{2}{c}{O3N2 index} \\
\cline{2-5}
\colhead{} &  \colhead{$N_{MS}$} & \colhead{$N_{SB}$} & \colhead{$N_{MS}$} & \colhead{$N_{SB}$}}
\startdata
$M/M_{\odot}<10^9$ & 2 & 6 & 2 & 6 \\
$10^9 < M/M_{\odot} < 10^{9.5}$ & 15 & 28 & 7 & 23 \\
$10^{9.5} < M/M_{\odot} < 10^{10}$ & 62 & 52 & 37 & 48 \\
$10^{10} < M/M_{\odot} < 10^{10.5}$ & 69 & 44 & 37 & 31 \\
$M/M_{\odot}>10^{10.5}$ & 18 & 3 & 14 & 0
\enddata
\end{deluxetable}

\begin{figure}[ht!]
	\centering
	\includegraphics[scale=0.5]{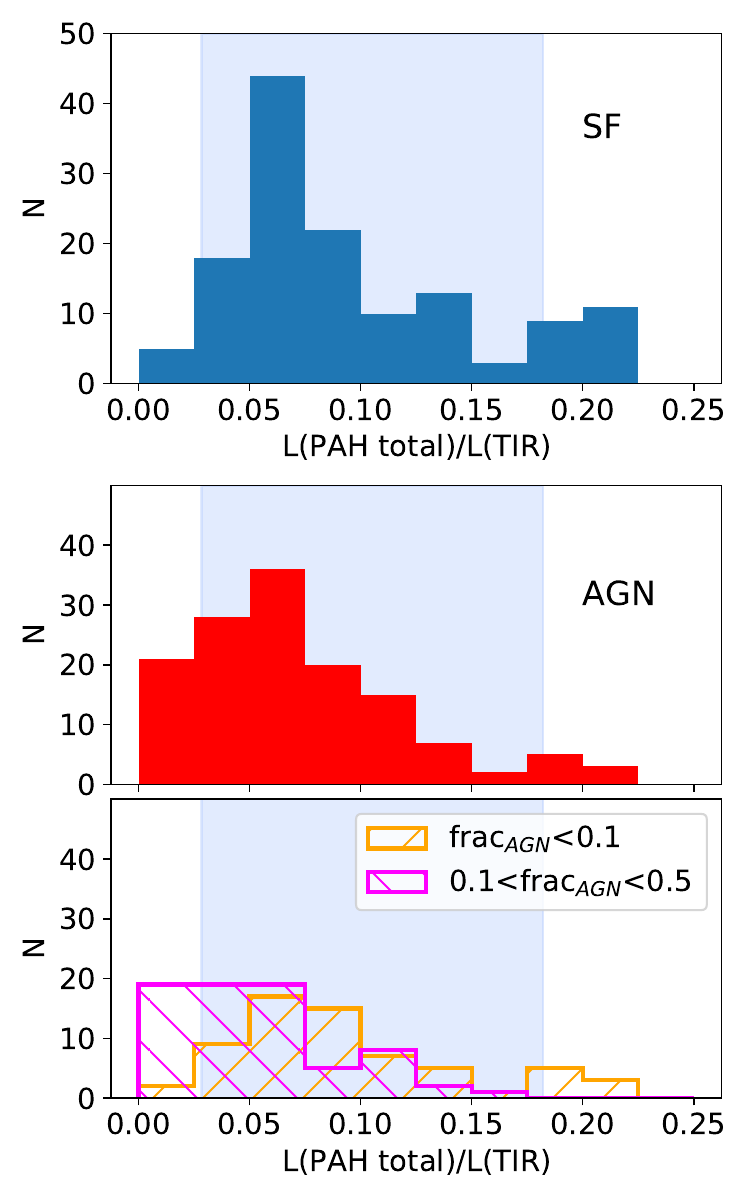}
	\caption{Distributions of the ratio of total PAH luminosity to the total infrared luminosity for star-forming galaxies (top panel) and AGN (middle panel). The bottom panel shows the histograms of PAH/TIR separately for the ``strong" and ``weak" AGN that were shown together in the middle panel. For reference, the pale blue shaded region in all three panels represents the 10-90\% range in $L(PAH)/L(TIR)$ for the star-forming galaxies.
	\label{lpahltir_hist}}
\end{figure}

In Figure \ref{lpahltir_hist}, we compare the PAH contribution to the total IR luminosity in star-forming galaxies and AGN by showing the distributions of the total PAH luminosity, $L(PAH)$, to $L(TIR)$ for objects with at least one \textit{Herschel} FIR detection. We define $L(PAH)$ as the sum of luminosities from PAH emission features at 6.2, 7.7, 8.6, 11.3, 12.6, and 17 $\mu$m. Star-forming galaxies are shown in the top panel and AGN are shown in the bottom panels, with the bottom-most panel separating out ``weak" AGN (i.e., $frac_{AGN}<0.1$) from stronger AGN (i.e., $frac_{AGN}>0.1$). We find that $L(PAH\ 7.7\ \mu m)$ contributes a median $\sim45$\% of the total PAH luminosity, and that the total PAH luminosity in turn can contribute up to $\sim20$\% of $L(TIR)$. The median $L(PAH)/L(TIR)$ for star-forming galaxies (main-sequence and starburst galaxies) is 0.08 with an interquartile range of 0.05 to 0.14. The median in the AGN sample is lower at 0.06 with an interquartile range of 0.03 to 0.10. Objects that are more AGN-dominant with $frac_{AGN}>0.1$ have lower $L(PAH)/L(TIR)$ with a median 0.04. For star-forming galaxies, the 10-90\% range in $L(PAH)/L(TIR)$ is 0.03 to 0.18, and is plotted as the pale blue shaded region in all three panels in Figure \ref{lpahltir_hist}. Our results are consistent with those of \citet{Shipley2013}. A simple interpretation is that up to half of the TIR luminosity in AGN may be produced by the AGN itself, rather than from star formation. Previous detailed studies of the far-IR suggest that most of this AGN luminosity emerges at wavelengths shortward of 100 $\mu$m \citep{Spinoglio}.

\subsection{The PAH SFR calibration}
In this section, we develop our PAH SFR calibration based on the dust-corrected H$\alpha$ and [O II] luminosities. Given that the AGN fraction is a significant driver of $L(PAH)/L(H\alpha)$ and $L(PAH)/L([O II])$, as demonstrated in Figure \ref{fig:lpahew}, our calibration sample excludes AGN and is limited to the 443 unique star-formation-dominated galaxies (Section \ref{sec:calibsample}). Therefore, $frac_{AGN}$ is not included in the following fits.

Assuming that PAH luminosity corrections for metallicity and starburst intensity are both simply linear, we perform a multi-linear regression to the PAH luminosity with the dust-corrected H$\alpha$ luminosity, R$_{SB}$ parameter, and metallicity as the independent variables in the form of: log L(PAH) [erg/s] = $c_{0}$ + $c_{1}\times$ log R$_{SB}$ + $c_{2}\times$(12+log(O/H)$_{index}$ - 8.6) + $c_{3}\times$[log $L(H\alpha)_{24\ \mu m}$ - 42] [erg/s], where the constant 8.6 is the median metallicity and 42 is the median logarithm of the intrinsic H$\alpha$ luminosity. The relation for predicting the 7.7 $\mu$m PAH luminosity as the dependent variable is:
\begin{eqnarray} \label{eqn:sfr_pah7.7}
\text{log}\ L(PAH\ 7.7\ \mu m)\ [erg/s] = (43.18 \pm 0.02) \nonumber \\
- (0.67 \pm 0.04)(\text{log}\ R_{SB}) \nonumber \\
+ (0.42 \pm 0.12)(12+\text{log}(O/H)_{N2} - 8.6) \nonumber \\
+ (1.07 \pm 0.03)[\text{log}\ L(H\alpha)_{24\ \mu m}-42]\ [erg/s]
\end{eqnarray}
where -0.30 $\leq$ log $R_{SB}$ $\leq$ 1.95, 8.1 $\leq$ 12+log(O/H)$_{N2}$ $\leq$ 8.7, $40.0 \leq \text{log}\ L(H\alpha)_{24\ \mu m}\ [erg/s] \leq 43.3$. The redshift range is 0.05 $\leq$ $z$ $\leq$ 1.03. Table \ref{table:constants} lists the results of the linear fits for L(PAH 6.2 $\mu$m) and L(PAH 7.7 $\mu$m) for both metallicity indicators. We note that the maximum threshold of log $R_{SB}=1.95$ is a result of the outliers with upper limits on the PAH luminosity; only nine objects in the SFR calibration sample have $R_{SB}>10$. The median value of $R_{SB}$ is 1.94 with an interquartile range of 1.01--3.48. We note that \citep{Whitcomb2020} also found that increasing metallicity correlates with an increase in the PAH luminosity for a given star formation rate. Similarly,
\citep{Kouroumpatzakis2021} found that PAH luminosity for a given star formation rate increases with metallicity and decreases with the strength of star formation normalized by stellar mass (similar to our $R_{SB}$ parameter).

To calculate the H$\alpha$ SFR calibrations, we substitute the SFR equation given by Kennicutt et al. into the multi-linear fit results:
\begin{equation} \label{eqn:k98}
    SFR\ [M_{\odot}/yr] = 7.9\times10^{-42}L(H\alpha)_{24\ \mu m}\ [erg/s]
\end{equation}
where the normalization factor assumes a \citet{Salpeter} IMF. We note that more modern SFR calibrations assume a \citet{Kroupa} or \citet{Chabrier} IMF; based on equation 2 in \citet{Speagle2014}, equation \ref{eqn:k98} can be converted to a Kroupa or Chabrier IMF by dividing by a factor of 1.6 and 1.7, respectively. The form of the PAH SFR calibration is then:
\begin{eqnarray} \label{eqn:sfr_calibration}
log\ SFR\ [M_{\odot}/yr] = c_{0} + c_{1}\times log\ R_{SB} \nonumber \\
+ c_{2}\times(12+log(O/H)_{index} - 8.6) \nonumber \\
+ c_{3}\times[log\ L(PAH_{n}) - 42]\ [erg/s]
\end{eqnarray}
where the index $n$ in $L(PAH_{n}$) is equal to either 6.2 $\mu$m or 7.7 $\mu$m. To calculate the [O II] SFR calibrations, we similarly substitute Equation \ref{eqn:sfroii} into the multi-linear fit results, where the H$\alpha$ luminosity is replaced with $L([O\ II])_{24\ \mu m}$. The coefficients for the SFR calibrations are listed in Tables \ref{table:constants} and \ref{table:constants_oii}. The results show that the SFR is linearly predicted by the PAH luminosity (i.e., $c_{3}\sim1$) when corrected for starburst intensity and metallicity.

\begin{deluxetable*}{lccccc}
\tablenum{6}
\tabletypesize{\footnotesize}
\tablecolumns{6}
\tablewidth{0pt}
\tablecaption{Multi-linear fit results for the PAH 6.2 $\mu$m and 7.7 $\mu$m luminosities and H$\alpha$ star formation rates. \label{table:constants}}
\tablehead{
\colhead{y} & \colhead{$c_{0}$} & \colhead{$c_{1}$ (starburstiness)} & \colhead{$c_{2}$ (metallicity)} & \colhead{$c_{3}$} & \colhead{metallicity calibration}}
\startdata
log L(PAH 6.2 $\mu$m) & 42.63$\pm$0.02 & -0.61$\pm$0.04 & 0.38$\pm$0.12 & 1.07$\pm$0.03 & N2\\
log SFR(PAH 6.2 $\mu$m) & -39.12$\pm$0.02 & 0.57$\pm$0.04 & -0.35$\pm$0.12 & 0.94$\pm$0.03 & N2 \\
\hline
log L(PAH 6.2 $\mu$m) & 42.61$\pm$0.03 & -0.60$\pm$0.06 & 0.27$\pm$0.12 & 1.09$\pm$0.04 & O3N2\\
log SFR(PAH 6.2 $\mu$m) & -38.15$\pm$0.03 & 0.55$\pm$0.05 & -0.25$\pm$0.12 & 0.92$\pm$0.04 & O3N2\\
\hline
log L(PAH 7.7 $\mu$m) & 43.18$\pm$0.02 & -0.67$\pm$0.04 & 0.42$\pm$0.12 & 1.07$\pm$0.03 & N2\\
log SFR(PAH 7.7 $\mu$m) & -39.32$\pm$0.02 & 0.62$\pm$0.04 & -0.39$\pm$0.12 & 0.93$\pm$0.03 & N2\\
\hline
log L(PAH 7.7 $\mu$m) & 43.15$\pm$0.03 & -0.65$\pm$0.05 & 0.32$\pm$0.12 & 1.09$\pm$0.04 & O3N2\\
log SFR(PAH 7.7 $\mu$m) & -38.59$\pm$0.03 & 0.60$\pm$0.05 & -0.29$\pm$0.12 & 0.92$\pm$0.04 & O3N2
\enddata
\tablecomments{Luminosity equations are in the form of y [erg/s] = $c_{0}$ + $c_{1}\times$log $R_{SB}$ + $c_{2}\times$(12+log(O/H)$_{index}$-8.6) + $c_{3}\times$(log $L(H\alpha)_{24\ \mu m}$ - 42) [erg/s]. SFR equations are in the form of y [M$_{\odot}$/yr]= $c_{0}$ + $c_{1}\times$log R$_{SB}$ + $c_{2}\times$(12+log(O/H)$_{index}$-8.6) + $c_{3}\times$log L(PAH$_{n}$) [erg/s], where L(PAH$_{n}$) refers to either 6.2 $\mu$m or 7.7 $\mu$m PAH luminosities. The gas-phase metallicity 12+log(O/H) is calculated via the N2 or O3N2 index \citep{PettiniPagel}. The SFR equations assume a Salpeter IMF; to convert to a Kroupa or Chabrier IMF, the constant $c_{0}$ would be subtracted by log(1.61) $\approx 0.21$ (Kroupa) or log(1.71) $\approx 0.23$ (Chabrier).}
\end{deluxetable*}

\begin{deluxetable*}{lccccc}
\tablenum{7}
\tabletypesize{\footnotesize}
\tablecolumns{6}
\tablewidth{0pt}
\tablecaption{Multi-linear fit results for the PAH 6.2 $\mu$m and 7.7 $\mu$m luminosities and [O II] star formation rates. \label{table:constants_oii}}
\tablehead{
\colhead{y} & \colhead{$c_{0}$} & \colhead{$c_{1}$ (starburstiness)} & \colhead{$c_{2}$ (metallicity)} & \colhead{$c_{3}$} & \colhead{metallicity calibration}}
\startdata
log L(PAH 6.2 $\mu$m) & 42.59$\pm$0.02 & -0.61$\pm$0.04 & 0.33$\pm$0.14 & 1.06$\pm$0.04 & N2\\
log SFR(PAH 6.2 $\mu$m) & -39.19$\pm$0.02 & 0.57$\pm$0.05 & -0.31$\pm$0.14 & 0.94$\pm$0.04 & N2 \\
\hline
log L(PAH 6.2 $\mu$m) & 42.51$\pm$0.02 & -0.51$\pm$0.04 & 0.39$\pm$0.12 & 1.00$\pm$0.03 & O32\\
log SFR(PAH 6.2 $\mu$m) & -41.53$\pm$0.02 & 0.50$\pm$0.04 & -0.39$\pm$0.12 & 1.00$\pm$0.03 & O32\\
\hline
log L(PAH 7.7 $\mu$m) & 43.13$\pm$0.03 & -0.66$\pm$0.05 & 0.40$\pm$0.15 & 1.06$\pm$0.04 & N2\\
log SFR(PAH 7.7 $\mu$m) & -39.82$\pm$0.03 & 0.63$\pm$0.05 & -0.38$\pm$0.15 & 0.94$\pm$0.04 & N2\\
\hline
log L(PAH 7.7 $\mu$m) & 43.05$\pm$0.02 & -0.54$\pm$0.04 & 0.43$\pm$0.13 & 1.00$\pm$0.03 & O32\\
log SFR(PAH 7.7 $\mu$m) & -41.97$\pm$0.02 & 0.54$\pm$0.04 & -0.43$\pm$0.13 & 1.00$\pm$0.03 & O32
\enddata
\tablecomments{Luminosity equations are in the form of y [erg/s] = $c_{0}$ + $c_{1}\times$log $R_{SB}$ + $c_{2}\times$(12+log(O/H)$_{index}$-$d$) + $c_{3}\times$(log $L([O\ II])_{24\ \mu m}$ - 42) [erg/s]. SFR equations are in the form of y [M$_{\odot}$/yr]= $c_{0}$ + $c_{1}\times$log R$_{SB}$ + $c_{2}\times$(12+log(O/H)$_{index}$-$d$) + $c_{3}\times$log L(PAH$_{n}$) [erg/s], where L(PAH$_{n}$) refers to either 6.2 $\mu$m or 7.7 $\mu$m PAH luminosities, and $d=8.6$ for the N2 index and $d=8.5$ for the O32 index. The gas-phase metallicity 12+log(O/H) is calculated via the N2 or O32 index \citep{PettiniPagel, Jones2015}. The SFR equations assume a Salpeter IMF; to convert to a Kroupa or Chabrier IMF, the constant $c_{0}$ would be subtracted by log(1.61) $\approx 0.21$ (Kroupa) or log(1.71) $\approx 0.23$ (Chabrier).}
\end{deluxetable*}

We additionally tested for dependence on redshift and total infrared luminosity by including them as parameters in the multi-linear fit. However, we found that there was {\textit no significant dependence on either parameter}; 
the coefficients for both parameters were close to zero with large uncertainties. To investigate the relative contributions of starburst intensity vs. metallicity in the multi-linear fit, we tested fits by removing either parameter, resulting in fits in the form of log L(PAH) = $c_{0}$ + $c_{1}\times$ log R$_{SB}$ + $c_{3}\times$[log $L(H\alpha)_{24\ \mu m}$ - 42] [erg/s] for the ``starburst-only" fit and log L(PAH) = $c_{0}$ + $c_{2}\times$(12+log(O/H)$_{index}$ - 8.6) + $c_{3}\times$[log $L(H\alpha)_{24\ \mu m}$ - 42] [erg/s] for the ``metallicity-only" fit. We found that for the starburst-only fit, the $c_{1}$ coefficients and uncertainties were consistent with those reported in our complete multi-linear fit results, whereas for the metallicity-only fit, the $c_{2}$ coefficients were $\sim 2.5-3$ times higher than the original results, indicating that the correction for the $R_{SB}$ parameter is more dominant than that for 12+log(O/H). However, both starburst-only and metallicity-only fits resulted in less linearity between L(PAH) and L(H$\alpha$) (i.e., $c_{3}$ deviated more from 1). Therefore, we conclude that the complete multi-linear fits, that correct for both starburst intensity and metallicity as parameters, provide the most reliable results.

Our methodology implicitly assumes that all of our galaxies have the same ratio of 7.7/6.2 $\mu$m PAH luminosity as in the CIGALE templates. This ratio can be determined from the $c_{0}$ coefficients in Table 6 to be about 3.4--3.6. This is an {\it assumption} that is not derivable from our \textit{AKARI} mid-IR photometry. We are not able to independently determine the Table 6 correlations for the 6.2 $\mu$m PAH luminosity. There is evidence that stronger 6.2 $\mu$m emission is produced by more highly ionized PAHs. However, since the same trend is seen for the 7.7 $\mu$m emission \citep{Maragkoudakis2022}, it turns out that the 7.7/6.2 $\mu$m {\it ratio} has little variation with PAH ionization or other galaxy properties. Larger PAH molecules are thought to emit slightly higher 7.7/6.2 $\mu$m ratios.  However, in LIRGs and ULIRGs which are the main targets of this study, the observed 7.7/6.2 $\mu$m ratios are all close to 3.6. The small scatter in this ratio, of $\sim 0.05$ dex \citep{McKinney2021}, indicates that CIGALE is indeed giving us reasonably accurate 6.2 $\mu$m luminosities. We therefore include the 6.2 $\mu$m PAH / SFR correlations in Tables 6 and 7 as a convenient aid for future researchers who will measure that emission feature.

The PAH SFR calibrations highlight four main results:
\begin{enumerate}
    \item The PAH luminosity per intrinsic H$\alpha$ or [O II] luminosity is increasingly deficient (or the H$\alpha$ and [O II] luminosities per PAH luminosity is enhanced) as $R_{SB}$ increases or metallicity decreases.
    \item The PAH SFR calibration does not depend on the total infrared luminosity.
    \item There is no apparent redshift evolution in the PAH SFR calibration from the local universe out to $z\sim 1.2$.
    \item Although starburst intensity, $R_{SB}$, has a stronger effect than metallicity on the correlation between PAH luminosity and intrinsic H$\alpha$ or [O II] luminosity, both are important for reliably calibrating the PAH SFR.
\end{enumerate}
  
\subsection{Possible physical interpretation}
Although this paper has focused on empirical analysis of observational data, the results described above may shed light on possible physical mechanisms that can suppress PAH emission. In particular, we can offer two speculations about why the PAH luminosity--for a given SFR--is significantly reduced in galaxies with active star-bursts ($R_{SB}$), and in those having active galactic nuclei ($frac_{AGN} \ge 0.05$). PAHs are very large molecules, and are therefore vulnerable to destruction by various energetic processes that can occur in the interstellar medium. In the nearby Orion star-forming complex there is direct evidence for PAH destruction by sputtering, and/or by photo-dissociation \citep{Giard1994}. Observations and models of H II region spectra indicates that PAH destruction occurs in regions of
elevated strength and hardness of the interstellar radiation field \citep{lebou2011}, both of which are expected in AGN and extreme starbursts.

\section{Star formation rate density from 0 $<$ $z$ $<$ 1.2} \label{sec:sfrd}
The largest source of uncertainty in UV-based cosmic star formation rate density (SFRD) measurements arises from correcting for dust attenuation \citep{Burgarella2005, Kobayashi}. For example, \citet{Salim2007} show that there is significant scatter between $A_{FUV}$ and the rest-frame FUV-NUV color for their sample of $\sim$50,000 local, optically-selected galaxies with \textit{GALEX} photometry. The scatter is especially prominent for starburst galaxies, which can deviate from the trend found for normal star-forming galaxies by $\gtrsim 1$ mag. SFRD indicators based on IR observations are particularly invaluable towards the epoch of peak cosmic star formation when much of this activity was heavily dust-obscured. \citet{Takeuchi} estimate that the FIR luminosity density is approximately 15 times higher than that of the FUV at $z\sim1$. However, studies have shown the FIR SFR to be less certain for galaxies whose UV to optical emission is dominated by old stellar populations or AGN \citep{Kennicutt1998}.

We apply our extinction-independent PAH 7.7 $\mu$m SFR calibrations to estimate the star formation rate density to $z\sim1.2$ for star-forming galaxies with metallicity detections. We use the rest-frame 8 $\mu$m and 12 $\mu$m luminosity functions (LF) from \citet{Goto2010}, which were derived using \textit{AKARI}/IRC sources in the NEP-Deep field. The luminosity functions assume a double power law given by:
\begin{equation}
\Phi(L)dL/L^{*} = \Phi^{*}\left(\frac{L}{L^{*}}\right)^{1-\alpha}dL/L^{*},\ (L<L^{*})
\end{equation}
\begin{equation}
\Phi(L)dL/L^{*} = \Phi^{*}\left(\frac{L}{L^{*}}\right)^{1-\beta}dL/L^{*},\ (L>L^{*})
\end{equation}
where $\Phi^{*}$ is the normalization in Mpc$^{-3}$ dex$^{-1}$, $L^{*}$ is the characteristic luminosity or ``knee" of the luminosity function in L$_{\odot}$, $L$ is the monochromatic luminosity $\nu L_{\nu}$(8 $\mu$m) or $\nu L_{\nu}$(12 $\mu$m), and $\alpha$ and $\beta$ are the slopes of the luminosity function at the low- and high-luminosity sides. We adopt the best-fit parameters given in Table 2 of \citet{Goto2010}. The luminosity density is then given as the integral:
\begin{equation} \label{eqn:ld}
    \Omega = \int_{0.1L^{*}}^{10L^{*}} L\Phi(L)\,dL
\end{equation}
where we define the lower and upper limits to be within an order of magnitude of $L^{*}$. We calculate the 8 $\mu$m luminosity density in redshift bins of 0.38$<$$z$$<$0.58 (\textit{S11}) and 0.65$<$$z$$<$0.90 (\textit{L15}), and the 12 $\mu$m LF in bins of 0.15$<$$z$$<$0.35 (\textit{L15}), 0.38$<$$z$$<$0.62 (\textit{L18W}), and 0.84$<$$z$$<$1.16 (\textit{L24}), where the corresponding \textit{AKARI}/IRC filters are given in parentheses. At 0$<$$z$$<$0.3, we use the 8 $\mu$m luminosity density given by \citet{Huang07}
where $\Omega_{8 \mu m}$=3.1$\times$10$^{7}$ L$_{\odot}$ Mpc$^{-3}$.

\begin{figure*}[ht!]
	\centering
	\includegraphics[width=\linewidth]{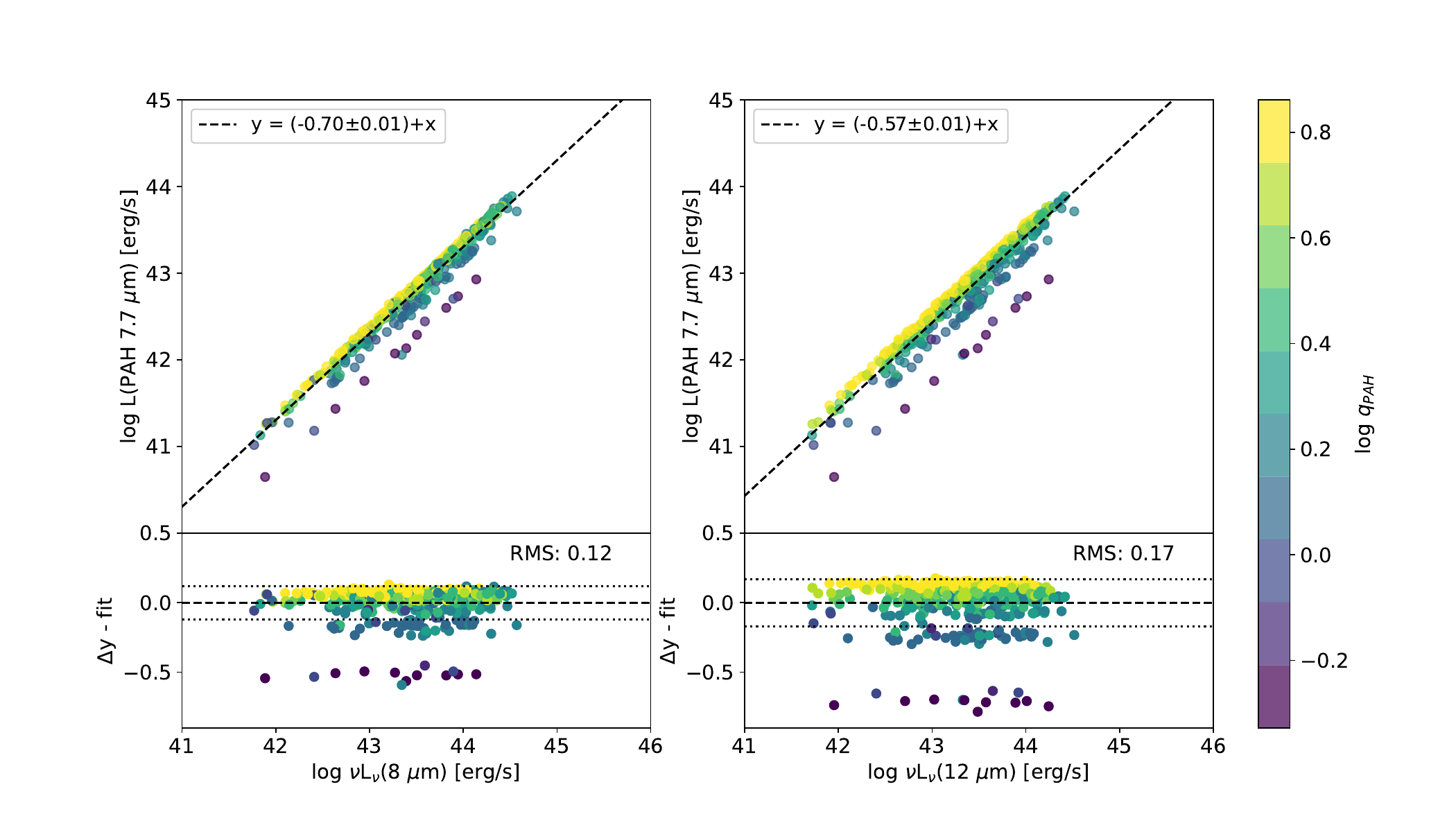}
	\vspace*{-10mm}
	\caption{Top panels: Correlations between PAH 7.7 $\mu$m luminosity and monochromatic luminosity at 8 $\mu$m (left) and 12 $\mu$m (right) for star-forming galaxies. Dashed lines represent linear fits with fixed unity slope. Bottom panels: Residuals of the fit as a function of monochromatic luminosity with dotted lines indicating the RMS dispersion in dex. Each galaxy is color-coded by the relative abundance of its small PAH dust grains, shown by the vertical scale on the right. The outliers correspond to low $q_{PAH}$ sources and represent upper limits.
	\label{fig:lpah_l8_l12}}
\end{figure*}

We use linear fits with fixed unity slope to convert PAH luminosity to monochromatic luminosity at 8 and 12 $\mu$m, as shown in Figure \ref{fig:lpah_l8_l12}. The correlations are given by:
\begin{equation} \label{eqn:l8}
    \text{log}\ L(PAH\ 7.7\ \mu m)= (-0.70\pm 0.01)
    + \text{log}\ \nu L_{\nu}(8\ \mu m)
\end{equation}
\begin{equation} \label{eqn:l12}
    \text{log}\ L(PAH\ 7.7\ \mu m)= (-0.57\pm 0.01)
    + \text{log}\ \nu L_{\nu}(12\ \mu m)
\end{equation}
where $\nu L_{\nu}$(8 $\mu$m) and $\nu L_{\nu}$(12 $\mu$m) are in erg/s. The scatter in the linear fits represented by the RMS dispersion in the residuals is approximately 0.13 and 0.18 dex for the 8 $\mu$m and 12 $\mu$m relations, respectively (bottom panels). These relations suggest that the PAH 7.7 $\mu$m luminosity may be estimated to within 30\% by multiplying the rest-frame broadband luminosity at 8 $\mu$m by a factor of 0.2 (unless the galaxy is an extreme starburst). The outliers with residuals $\lesssim -0.5$ are starburst galaxies with exceptionally high IR8 (=$L(TIR)/L(8\ \mu m)\gtrsim20$) and represent only 7\% of the overall starburst galaxy sample. However, as shown by the color bar, these PAH-weak outliers also have SEDs which CIGALE fitted with the minimum $q_{PAH}$ value ($q_{PAH}\approx0.47$), and therefore represent upper limits due to modeling constraints.

Then, we substitute equations \ref{eqn:l8} and \ref{eqn:l12} for $L(PAH\ 7.7\ \mu m)$ into our SFR calibrations to derive a function of SFR($R_{SB}$, 12+log(O/H), $\nu L_{\nu}$(8 $\mu$m)) or SFR($R_{SB}$, 12+log(O/H), $\nu L_{\nu}$(12 $\mu$m)). The SFRD ($\rho_{SFR}$) is calculated in each redshift bin as:

\begin{equation}
\rho_{SFR} = c_{0} + c_{1}\times \langle \text{log}\ R_{SB} \rangle 
+ c_{2}\times \langle (12+\text{log(O/H)}-8.6)\rangle
+ c_{3}\times \left(c_{4} +\Omega_{8\mu m} \right)
\end{equation}

\begin{equation}
    \rho_{SFR} = c_{0} + c_{1}\times \langle \text{log}\ R_{SB} \rangle
    + c_{2}\times \langle(12+\text{log(O/H)}-8.6)\rangle
    + c_{3}\times \left(c_{5} +\Omega_{12\mu m} \right)
\end{equation}
in M$_{\odot}$yr$^{-1}$Mpc$^{-3}$, where the constant coefficients $c_{0},...,c_{3}$ are given by the SFR calibrations in Table \ref{table:constants}, $c_{4}$ and $c_{5}$ are the best-fit intercepts from equations \ref{eqn:l8} and \ref{eqn:l12}, respectively. $\Omega_{8\mu m}$ and $\Omega_{12\mu m}$ are the luminosity densities in L$_{\odot}$Mpc$^{-3}$ calculated from equation \ref{eqn:ld} based on $\nu L_{\nu}(8\ \mu m)$ and $\nu L_{\nu}(12\ \mu m)$ and converted to L$_{\odot}$ units. The averages $\langle \text{log}\ R_{SB} \rangle$ and $\langle$12+log(O/H)$\rangle$ are averaged over each redshift bin. Table \ref{table:sfrd_numbers} shows the number of star-forming galaxies integrated in each bin for a given PAH 7.7 $\mu$m and optical emission line calibrator.

\begin{deluxetable*}{clllll}
\tablenum{8}
\tabletypesize{\footnotesize}
\tablecolumns{6}
\tablewidth{0pt}
\tablecaption{Number of star-forming galaxies integrated in each redshift bin to calculate the star formation rate density. \label{table:sfrd_numbers}}
\tablehead{
\colhead{LF} & \colhead{redshift bin} & \multicolumn{4}{c}{$N_{tot}$ ($N_{MS}$, $N_{SB}$)} \\
\cline{3-6}
\colhead{} & \colhead{} & \colhead{H$\alpha$ N2} & \colhead{H$\alpha$ O3N2} & \colhead{[O II] N2} & \colhead{[O II] O32}}
\startdata
8 $\mu$m & 0.02$<$$z$$<$0.3 & 203 (129, 74) & 131 (70, 61) & 120 (77, 43) & 98 (59, 39) \\
8 $\mu$m & 0.28$<$$z$$<$0.47 & 105 (51, 54) & 89 (38, 51) & 85 (46, 39) & 135 (66, 69) \\
8 $\mu$m & 0.65$<$$z$$<$0.90 & 10 (1, 9) & 3 (1, 2) & 1 (1, 0) & 17 (8, 9) \\
\hline
12 $\mu$m & 0.15$<$$z$$<$0.35 & 183 (106, 77) & 132 (67, 65) & 131 (80, 51) & 112 (67, 45) \\
12 $\mu$m & 0.38$<$$z$$<$0.62 & 30 (10, 20) & 29 (7, 22) & 23 (8, 15) & 120 (39, 81) \\
12 $\mu$m & 0.84$<$$z$$<$1.16 & 7 (1, 6) & 1 (0, 1) & 1 (0, 1) & 0
\enddata
\end{deluxetable*}

\begin{figure*}[ht!]
	\centering
	\includegraphics[scale=0.5]{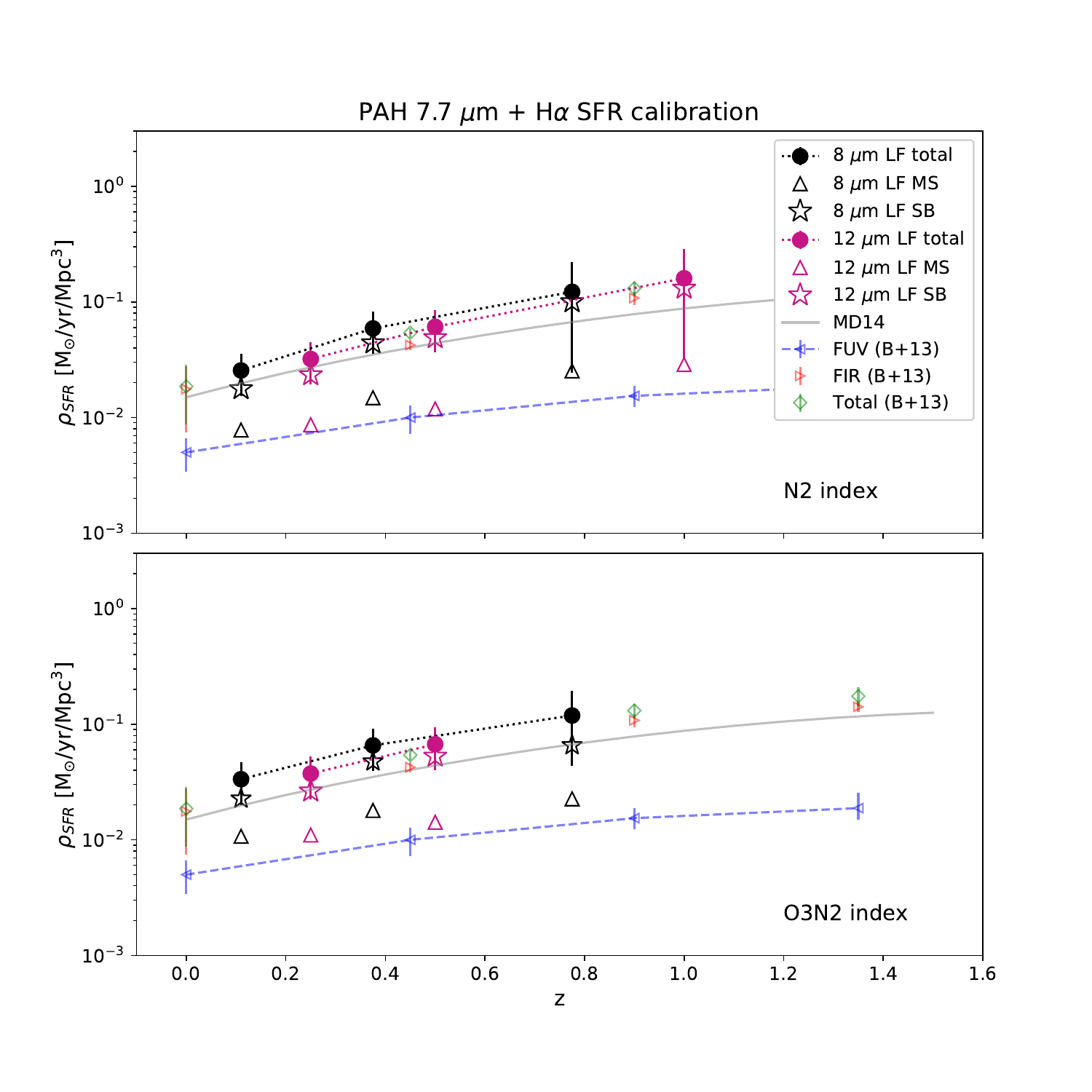}
	\vspace*{-10mm}
	\caption{Star formation rate density as a function of redshift for the N2 (top) O3N2 (bottom) metallicity indicators, using the PAH 7.7 $\mu$m calibration tied to metallicity, $R_{SB}$, and dust-corrected H$\alpha$. Three dotted lines of different colors connect the estimates based on the 8 and 12 $\mu$m and FUV luminosity functions. The best-fit function from \citet{Madau} is shown in light grey.
	\label{sfrd_halpha}}
\end{figure*}

Figure \ref{sfrd_halpha} shows the redshift evolution of the PAH 7.7 $\mu$m-derived SFRD based on the H$\alpha$ calibration and N2/O3N2 metallicity indicators. We calculate the SFRD for the combined main-sequence and starburst galaxy sample in each redshift bin (filled circles) and show the separate contributions from main-sequence galaxies only (triangle symbols) and starburst galaxies only (star symbols). The redshifts plotted are the median redshifts per bin. The SFRDs derived from either the 8 $\mu$m and 12 $\mu$m LFs are consistent with each other within the uncertainties.

For reference, we plot the best-fit cosmic star formation history from \citet{Madau} (MD14; grey line). In addition, we plot the SFRD calculated from FUV and FIR observations from \citet{Burgarella2013} (B+13) who use calibrations from \citet{Kennicutt1998}. We find that the PAH 7.7 $\mu$m star formation rate density includes significant contribution from dust-obscured star formation absorbed in the UV. In fact, at $z\sim1$, the PAH SFRD is an order of magnitude higher than the FUV SFRD. Our results are consistent with the total (UV+FIR) SFRD, and suggest that when corrected for metallicity and starburst intensity, the PAH 7.7 $\mu$m luminosity traces the total star formation (i.e., obscured and unobscured) in actively star-forming galaxies. In addition, the contribution from starburst galaxies to the SFRD becomes more significant at higher $z$ relative to main-sequence galaxies. Our results are consistent with past attempts to estimate the SFR based on \textit{Herschel}/PACS detections. The SFR calibration with the O3N2 index (bottom panel) is consistent with that of the N2 index. The MD14 cosmic star formation rates are $\sim$ 20 -- 30\%  
lower than ours. 

\begin{figure*}[ht!]
	\centering
	\includegraphics[scale=0.5]{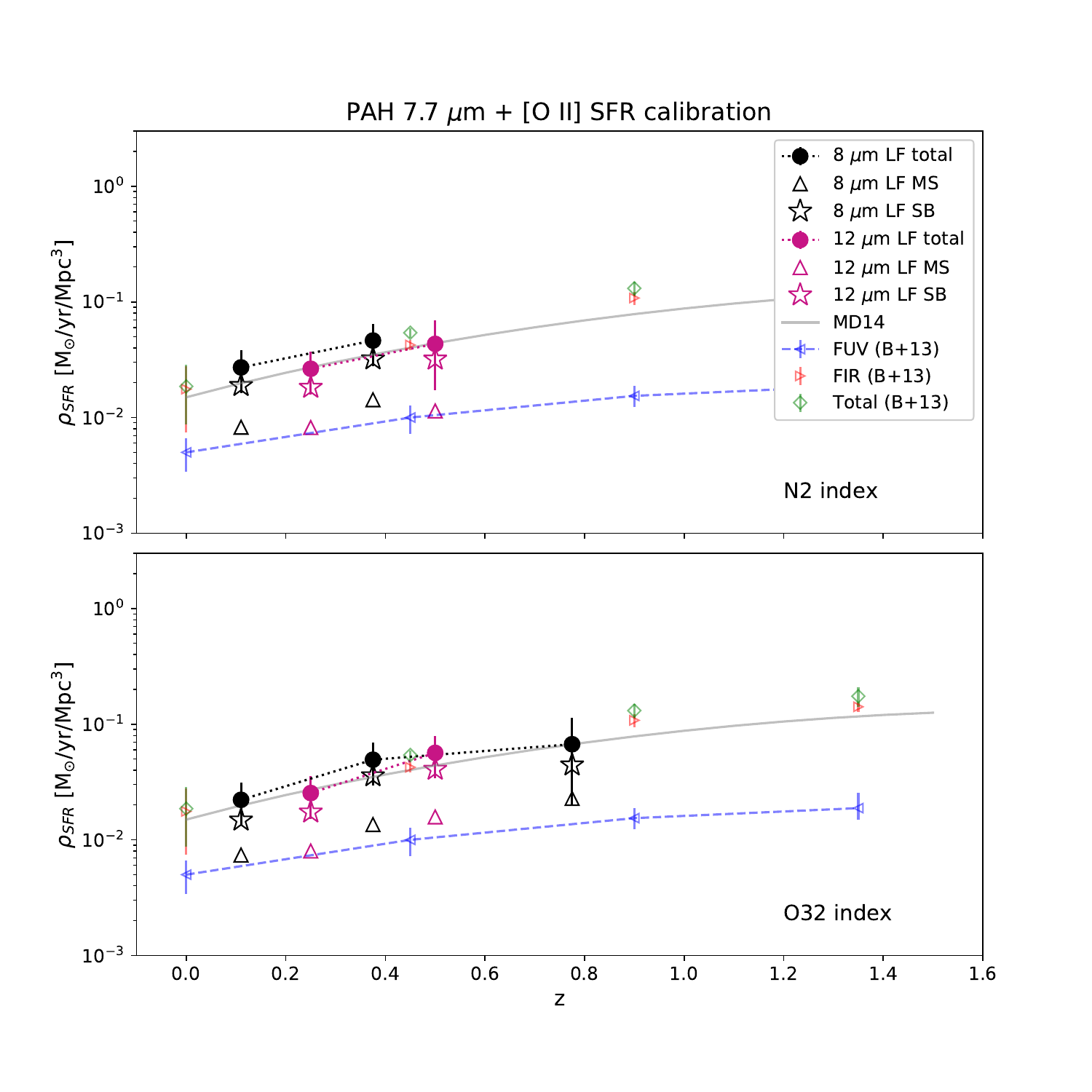}
	\vspace*{-10mm}
	\caption{ Star formation rate density as a function of redshift for the N2 (top) O32 (bottom) metallicity indicator using the PAH 7.7 $\mu$m calibration tied to metallicity, $R_{SB}$, and dust-corrected [O II]. Symbols and lines are the same as in the previous figure.
	\label{sfrd_oii}}
\end{figure*}

We also show the redshift evolution for the PAH 7.7 $\mu$m-derived SFRD, based on the de-reddened [O II] luminosity calibration and N2 (top panel) and O32 (bottom panel) metallicity indicators in Figure \ref{sfrd_oii}. Although more data are needed at $z\gtrsim 0.8$ to constrain the shape of the SFRD, the PAH SFRD derived from the $L[O\ II]$ calibration is consistent with that of H$\alpha$.

\section{Discussion} \label{sec:disc}

\subsection{Comparison with previous studies}
To our knowledge, this study is the first to derive PAH SFR calibrations from \textit{AKARI}/IRC 2--24 $\mu$m photometry. We extend the relationship between PAH luminosity and SFR by determining the dependence of the SFR calibration on metallicity and starburst intensity. \citet{Shipley2016} derived PAH SFR calibrations using a sample of \textit{Spitzer}/IRS galaxies at $z<0.4$. We find that their linear fits to $L(PAH)$ vs. $L(H\alpha)$ are consistent with our results for main-sequence galaxies up to $z\sim1.2$, but must be corrected in starburst galaxies, which have systematically higher $L(TIR)/L(8\ \mu m)$ ratios. This discrepancy is most likely because starburst galaxies were excluded from their sample due to having low S/N ratios. Based on Figure \ref{fig:lpahew}, we estimate that Shipley et al.'s calibration sample is limited to main-sequence galaxies with $EW(PAH\ 6.2\ \mu m)\gtrsim 1$ $\mu$m and $EW(PAH\ 7.7\ \mu m)\gtrsim4$ $\mu$m.

\begin{figure*}[ht!]
    \centering
    \includegraphics[scale=0.4]{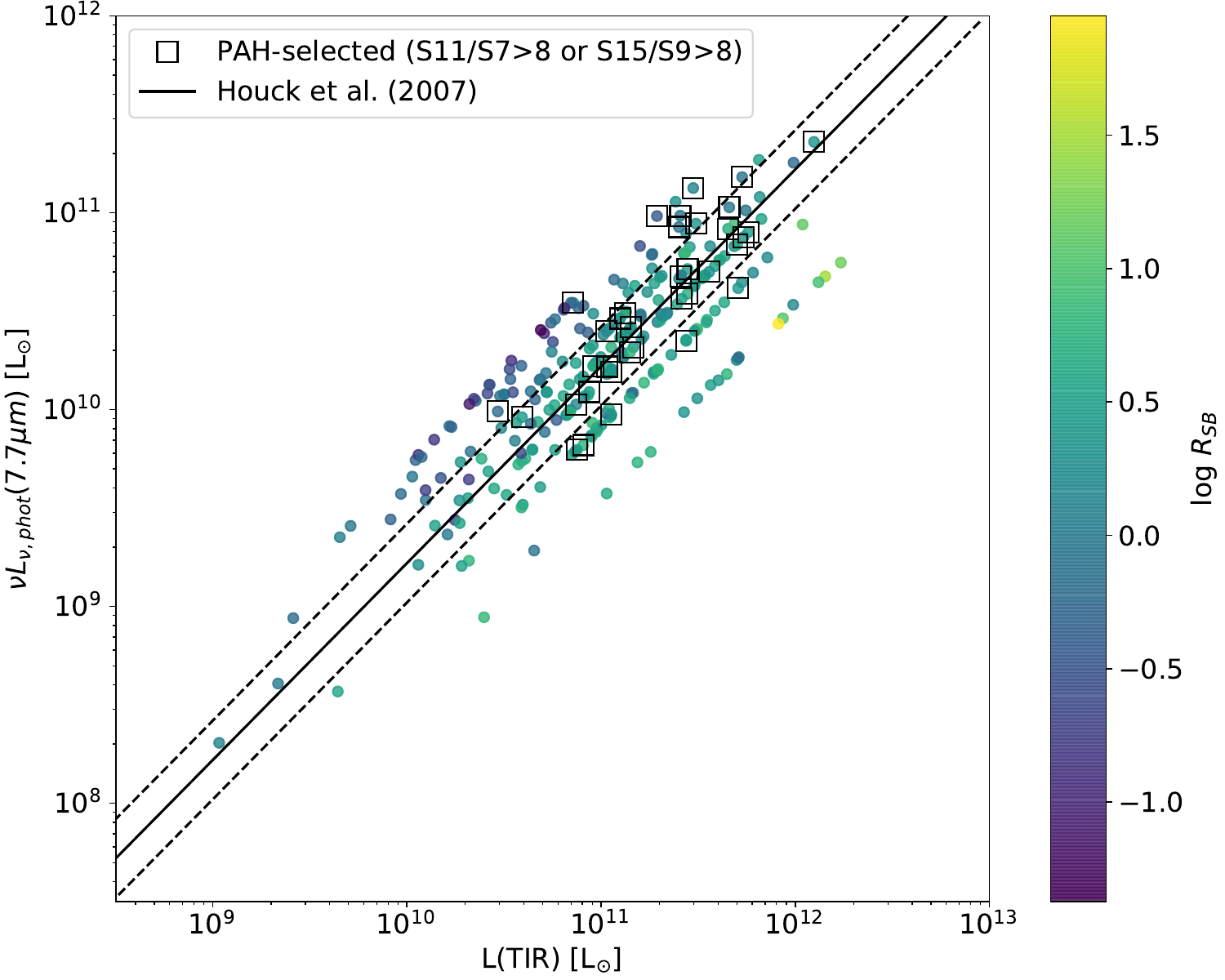}
    \vspace*{0mm}
    \caption{Peak photometric monochromatic PAH 7.7 $\mu$m luminosity as a function of total infrared luminosity for star-forming galaxies in the NEP-Deep region. The points in boxes meet the selection criteria of ``PAH enhancement" from Takagi et al. The solid line represents the relation found by \citet{Houck} for local starburst galaxies. Symbols are color-coded by the logarithm of the starburst intensity.
    \label{takagi}}
\end{figure*}

\citet{Takagi2010} studied ``PAH-luminous" starburst galaxies at $z\sim0.5$ and $z\sim1$ in the \textit{AKARI} NEP-Deep survey that were selected using color cuts based on the 15 $\mu$m/9 $\mu$m and 11 $\mu$m/7 $\mu$m flux ratios. Applying the same color criteria to our star-forming galaxies in the NEP-Deep region, we select 22 such galaxies at $0.39<z<0.62$ ($\tilde{z}=0.47$) and 16 galaxies at $0.76<z<1.23$ ($\tilde{z}=1.01$). Figure \ref{takagi} shows the correlation between the photometric monochromatic luminosity at 7.7 $\mu$m, $\nu L_{\nu,phot}(7.7\ \mu m)$ and $L(TIR)$, color-coded by $R_{SB}$. Galaxies shown by square symbols represent ``PAH-selected galaxies" that satisfy Takagi et al.'s criteria. To calculate the photometric monochromatic luminosity, we scaled our peak luminosity measurements by a factor of 10$^{-0.24}$ (see Appendix). Most objects lie along the local relation for less luminous starbursts, indicating that MIR-selected galaxies up to 10$^{12}$ L$_{\odot}$ are ``PAH-normal" rather than ``PAH-enhanced" for a given TIR luminosity.

\subsection{Applications with JWST}
The \textit{James Webb Space Telescope} Mid-Infrared Instrument (\textit{JWST}/MIRI) has an observing wavelength range of 5--28 $\mu$m, which includes spectroscopy that resolves the PAH 6.2 $\mu$m and 7.7 $\mu$m features to $z\sim3$. To apply our PAH SFR calibrations to \textit{JWST} observations, follow-up spectroscopy in the near-IR such as with \textit{Keck I}/MOSFIRE \citep{Shivaei2017} or \textit{JWST}/NIRSPEC would also be useful to determine the gas metallicity with rest-frame optical emission lines. Far-infrared photometry would be needed to measure the total infrared luminosity and infer the starburst intensity/compactness of star formation, although extrapolations from multi-wavelength SED fitting may be used provided there are sufficient data in the observed optical to near-infrared wavelengths. In addition, joint observations with \textit{Hubble Space Telescope} will be able to constrain the sSFR. PAH-based SFR calibrations that do not correct for metallicity or starburstiness would underestimate the true SFR in metal-poor and starburst galaxies.

\section{Summary} \label{sec:summary}
In this work, we explore PAH dust emission as an extinction-independent star formation rate indicator, a major scientific goal of the \textit{AKARI} mission. We combine ground-based, optical and near-infrared spectroscopy with \textit{AKARI}/Infrared Camera multi-band photometry to measure the infrared spectra of $\sim500$ galaxies in the \textit{AKARI} NEP field. We present the first results of PAH 6.2 and 7.7 $\mu$m SFR calibrations with corrections for metallicity and starburst intensity, applicable to star-forming galaxies. Our calibration sample consists of 443 main-sequence and starburst galaxies from the \textit{AKARI}/IRC NEP survey at $0.05\leq z\leq 1.03$, with a broad range of stellar masses ($M_{*}\sim10^{8.7} - 10^{11.2}$ $M_{\odot}$), total infrared luminosities ($L(TIR)\sim 10^{8.7}-10^{12}\ L_{\odot})$, and SFRs ($\sim0.06-500\ M_{\odot}$yr$^{-1}$). These mid-infrared-selected galaxies were observed to have strong emission lines with optical/NIR spectroscopy, including \textit{Keck II}/DEIMOS and \textit{Keck I}/MOSFIRE observations, which we newly presented in this work. A summary of our main conclusions are as follows:
\begin{itemize}
\item To measure the PAH luminosity of \textit{AKARI}/IRC galaxies, we first derive best-fit spectral energy distributions with CIGALE to model the mid-IR dust emission. Then from the best-fit, rest-frame SEDs, we decompose the PAH emission features by using PAHFIT. Our photometrically derived PAH 6.2 and 7.7 $\mu$m luminosities are consistent with \textit{AKARI}/IRC slitless spectroscopic measurements to within 20-40\%. 
\item The PAH luminosity per dust-corrected H$\alpha$ and [O II]$\lambda\lambda3726,3729$ luminosity increases as a function of metallicity, and decreases as a function of starburst intensity ($R_{SB}$) and AGN fraction. Starburst galaxies (i.e., galaxies with $R_{SB}>2$) are systematically deficient in $L(PAH)$ per dust-corrected $L(H\alpha)$ and $L([O\ II])$ relative to main-sequence galaxies by a factor of 0.34 and 0.3 dex, respectively. Based on multi-linear fits, we derive, for the first time, corrections for metallicity and starburst intensity to the PAH luminosity and SFR. Due to the effect of AGN on $L(PAH)$ per dust-corrected $L(H\alpha)$ and $L([O\ II])$, our PAH SFR calibrations may not be applicable to AGN systems. In addition, we find that the PAH SFR calibration is independent of total infrared luminosity and redshift (at least up to $z\sim1$).
\item We apply our PAH SFR calibrations to the extensive dataset of \textit{AKARI} to study the dust-obscured cosmic star formation rate density per comoving volume. We combine our correlations between $L(PAH\ 7.7\ \mu m)$ vs. $\nu L_{\nu}(8\ \mu m)$ and $\nu L_{\nu}(12\ \mu m)$ with luminosity functions at 8 and 12 $\mu$m to derive the SFRD as a function of PAH luminosity. The SFRD as predicted by our PAH 7.7 $\mu$m SFR calibration is a factor of 5 higher at $z\sim0.15$ and a factor of $\sim 10$ higher at $z\sim1$ than observed FUV estimates due to dust attenuation, with significant contribution from starburst galaxies at $z\gtrsim 0.6$. Compared to infrared-based SFR indicators, our PAH SFRD is consistent with FIR and TIR estimates from $0.15 \lesssim z \lesssim 1$ \citep{Burgarella2013}.
\item Future studies that involve PAH luminosity as a SFR indicator, such as those conducted with \textit{JWST}, would need to correct for the effects of metallicity and starburst intensity; otherwise, the PAH SFR would be underestimated in metal-poor or starburst galaxies.
\end{itemize}

\section*{Acknowledgements}
We wish to thank the anonymous referee whose helpful comments greatly improved this work. Some of the data presented herein were obtained at the W. M. Keck Observatory, which is operated as a scientific partnership among the California Institute of Technology, the University of California and the National Aeronautics and Space Administration. The Observatory was made possible by the generous financial support of the W. M. Keck Foundation. This research has made use of the Keck Observatory Archive (KOA), which is operated by the W. M. Keck Observatory and the NASA Exoplanet Science Institute (NExScI), under contract with the National Aeronautics and Space Administration. The authors wish to recognize and acknowledge the very significant cultural role and reverence that the summit of Maunakea has always had within the indigenous Hawaiian community. We are most fortunate to have the opportunity to conduct observations from this mountain. The analysis pipeline used to reduce the DEIMOS data was developed at UC Berkeley with support from NSF grant AST-0071048. H. K. gratefully acknowledges financial support from the S.O.S. program of the National Radio Astronomy Observatory, grant number SOSPA6-024. T. M. is supported by UNAM-DGAPA PAPIIT (IN111319, IN114423) and CONACyT Grant Ciencias B\'{a}sicas 252531.

\bibliography{HelenKim_PAH_final_R2_arxiv}

\appendix
\section{Subaru/FMOS H$\alpha$ detections}
Table \ref{tab:fmos} lists the 10 additional H$\alpha$ sources from \citet{Oi2017} that are newly identified as secure detections based on other emission line detections that confirm the spectroscopic redshift.
\begin{deluxetable}{lllcc}
    \tablenum{9}
    \tablecaption{Updated additional secure H$\alpha$ detections from \textit{Subaru}/FMOS. \label{tab:fmos}}
    \tablewidth{0pt}
    \tablehead{
        \colhead{AKARI ID} & \colhead{RA (deg)} & \colhead{Dec (deg)} & \colhead{Redshift} & \colhead{F(H$\alpha$) [$10^{-16}$ erg/s/cm${^2}$]}
        }
    \startdata
    61021137 & 269.24127 & 66.72676 & 0.949 & 0.34$\pm$0.13 \\
    61014814 & 269.13637 & 66.55578 & 1.030 & 0.22$\pm$0.11 \\
    61016430 & 269.00271 & 66.59490 & 1.003 & 1.45$\pm$0.27 \\
    61009352 & 268.98370 & 66.40349 & 0.902 & 0.82$\pm$0.14 \\
    61023314 & 268.83595 & 66.80321 & 0.925 & 0.78$\pm$0.23 \\
    61010363 & 268.77946 & 66.43416 & 0.902 & 0.58$\pm$0.12 \\
    61023651 & 268.71468 & 66.81580 & 0.715 & 0.68$\pm$0.09 \\
    61023133 & 268.64746 & 66.79599 & 0.714 & 0.55$\pm$0.12 \\
    61012206 & 268.34379 & 66.48484 & 1.026 & 0.63$\pm$0.22 \\
    61015448 & 268.91175 & 66.57122 & 1.003 & 1.02$\pm$0.23 \\
    \enddata
\end{deluxetable}

\section{Measurement of total infrared luminosity}
\begin{figure*}[ht!]
	\centering
	\includegraphics[scale=0.5]{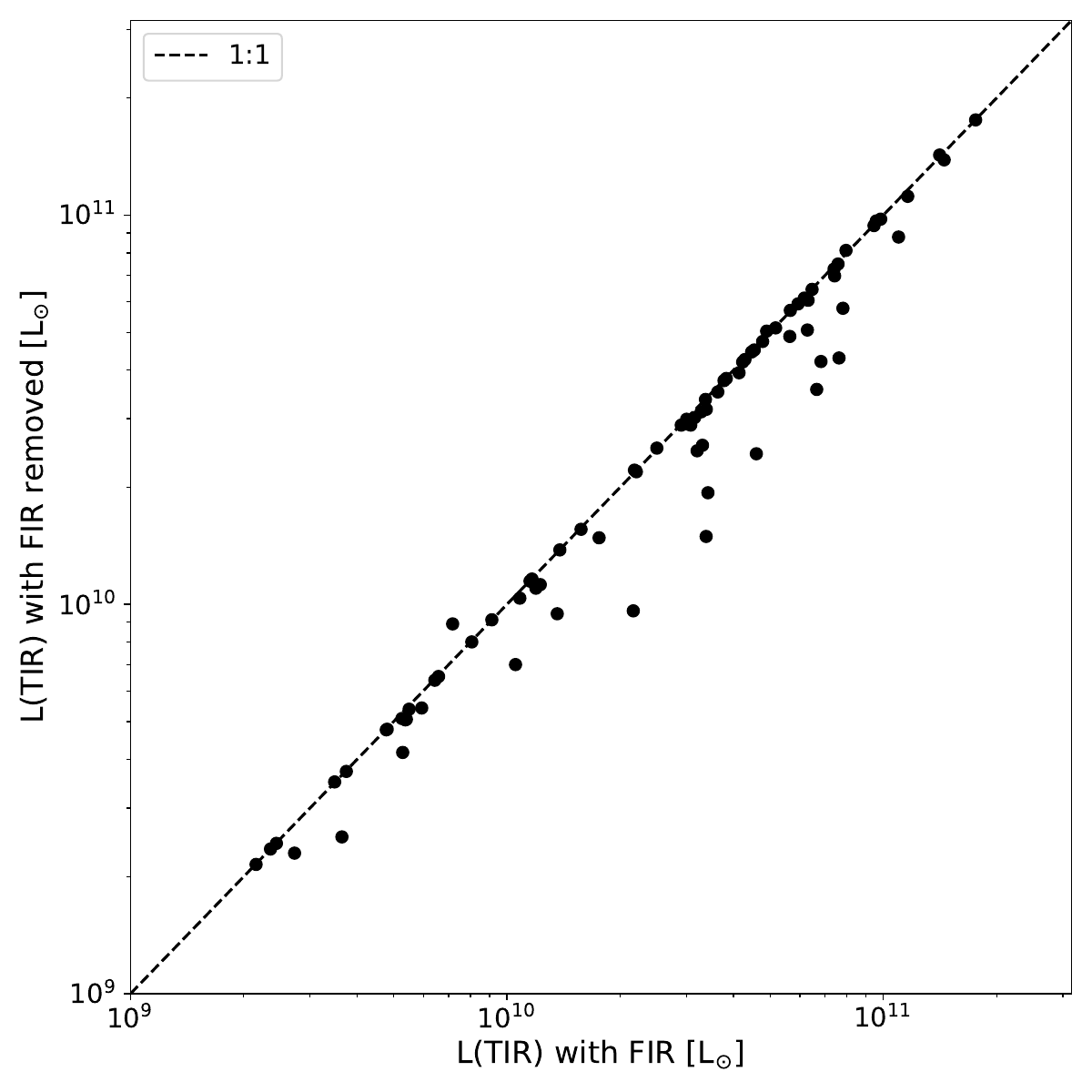}
	\caption{Comparison between L(TIR) measurements when including and removing \textit{Herschel} FIR photometry in SED fitting. The dashed line represents a perfect 1:1 correlation, and has not been fitted to the data.
	\label{fig:ltirappendix}}
\end{figure*}

In this section, we describe our procedure for testing the dependence of our L(TIR) measurements on FIR photometry. We selected 82 star-forming galaxies from the NEP-Wide survey with one to five FIR flux density measurement(s) from \textit{Herschel}/PACS and/or SPIRE. Then, using the same input parameters described in Table \ref{tab:cigale}, we modeled SEDs using CIGALE twice -- including and excluding the FIR data. Figure \ref{fig:ltirappendix} shows the resulting comparison between L(TIR) when the FIR data are included and excluded. For 78\% of galaxies, the L(TIR) measurements agree to within $\lesssim$ 12\%. In a minority of galaxies, removing the FIR data results in significant under-estimation of the TIR luminosity.  There was no correlation between the scatter and the number of FIR filter detections. We conclude that our estimates of L(TIR) are generally accurate, even for those galaxies which lack \textit{Herschel} detections.

\section{Measurement of peak PAH 7.7 $\mu$m luminosity}
The PAH luminosity measurements presented in \citet{Ohyama2018} are determined by the integrated Lorentzian fits to the SPICY spectra. To compare their measurements to other works that rely on rest-frame SEDs, \citet{Ohyama2018} define the ``photometric monochromatic luminosity," $\nu L_{\nu, photo}(7.7\ \mu m)$, and the ``spectroscopic monochromatic luminosity," $\nu L_{\nu, spec}(7.7\ \mu m)$. Both the photometric and spectroscopic monochromatic luminosity include contributions from the underlying continuum and measure the peak PAH 7.7 $\mu$m luminosity. However, the photometric monochromatic luminosity at 7.7 $\mu$m is the peak PAH 7.7 $\mu$m luminosity based on broad-band SED fitting \citep{Takagi2003}, while the spectroscopic monochromatic luminosity is based on the SPICY spectra. According to the scaling relation in their Section 3.3.1, $\nu L_{\nu, photo}(7.7\ \mu m)$ is a factor of 0.6 dex higher than their integrated PAH 7.7 $\mu$m luminosity due to the different luminosity definitions and continuum contribution. Based on our $L(PAH\ 7.7\ \mu m)$ measurements, $\nu L_{\nu, photo}(7.7\ \mu m)$ as measured by Ohyama et al. is $\sim0.7$ dex higher than our integrated PAH 7.7 $\mu$m luminosities, which is consistent with the errors.

\begin{figure*}[ht!]
	\centering
	\includegraphics[width=\linewidth]{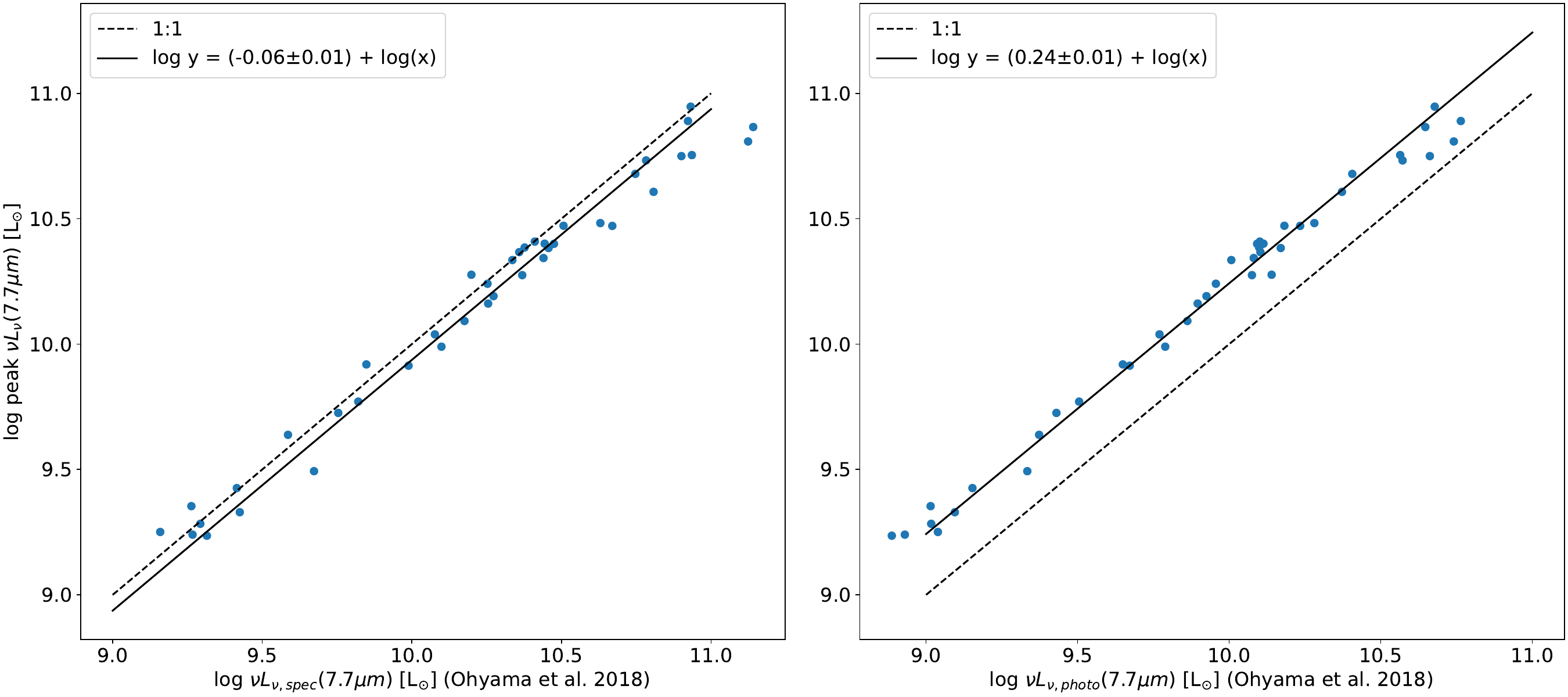}
	\vspace*{-5mm}
	\caption{Comparison between our measured peak PAH 7.7 $\mu$m luminosity vs. spectroscopic monochromatic luminosity (left) and photometric monochromatic luminosity (right) by \citet{Ohyama2018} for SPICY galaxies. Linear fits are shown as solid lines.  The dotted lines show exact agreement between the two luminosity estimates.
	\label{fig:spicy_nulnu}}
\end{figure*}

Figure \ref{fig:spicy_nulnu} shows a comparison of our method for measuring the peak 7.7 $\mu$m luminosity vs. the monochromatic spectroscopic and photometric luminosities given by \citet{Ohyama2018} in 41 SPICY galaxies in our sample. We find that our peak $\nu L_{\nu}(7.7\ \mu m)$ measurements are consistent with Ohyama et al.'s spectroscopic monochromatic 7.7 $\mu$m luminosities within 13\% (left panel). In contrast, there is a 0.24 dex offset between our peak $\nu L_{\nu}(7.7\ \mu m)$ measurements and the photometric monochromatic 7.7 $\mu$m luminosities (right panel), indicating that our values should be multiplied by a factor of 10$^{-0.24}$ when comparing to photometric-based methods that rely on broad-band SED fitting. 

\section{Comparison between [O II] and H$\alpha$ SFR}
\begin{figure*}[ht!]
	\centering
	\includegraphics[scale=0.45]{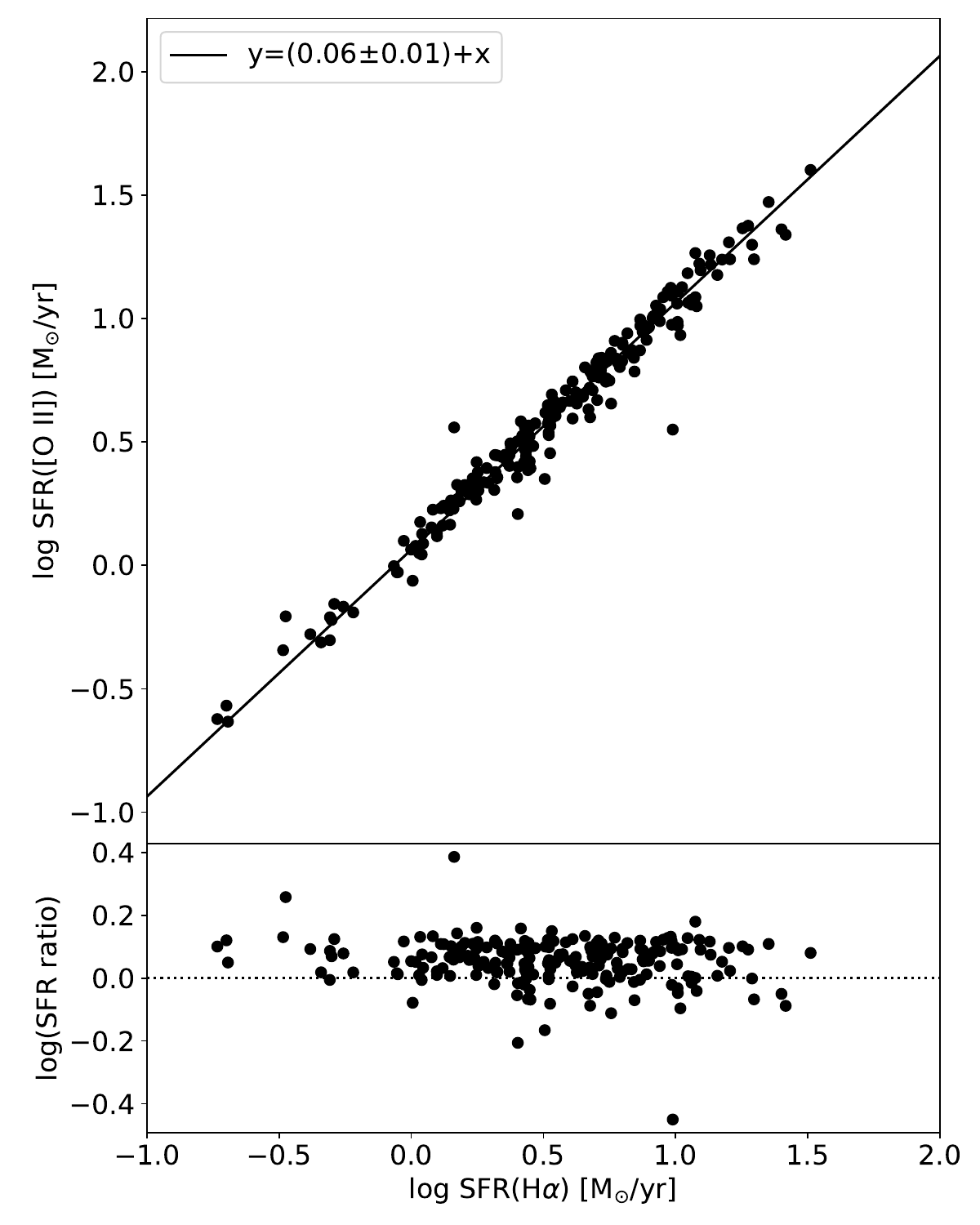}
	\caption{Comparison between [O II] and H$\alpha$ SFR for calibration sample, dust-corrected assuming the monochromatic 24 $\mu$m luminosity correction given by \citet{Kennicutt1998} equations. A linear fit to the data is shown by the solid line.
	\label{fig:sfr_ha_oii}}
\end{figure*}

Although the [O II] luminosity calibration from \citet{Kennicutt1998} is commonly used as an alternative SFR diagnostic in galaxies where H$\alpha$ is redshifted out of the visible range ($0.4\lesssim z \lesssim 1.5$), the [O II] emission line is sensitive to reddening and metallicity, which can cause disagreement between SFR([O II]) and SFR(H$\alpha$) \citep{Kewley2004}. In light of these effects, we compare SFR([O II]) and SFR(H$\alpha$) in Figure \ref{fig:sfr_ha_oii}, where we correct for reddening using the rest-frame 24 $\mu$m luminosity calibration given in \citet{Kennicutt1998}. For the majority of galaxies, the SFR calibrations are consistent with a 0.06 dex offset and dispersion of 20\%. 

\end{document}